\documentclass[12pt]{iopart}

\usepackage{iopams}  

\expandafter\let\csname equation*\endcsname\relax 
\expandafter\let\csname endequation*\endcsname\relax 

\usepackage{graphicx}
\usepackage{amsmath}
\usepackage{overpic}

\def\bra#1{\mathinner{\langle{#1}}}
\def\ket#1{\mathinner{{#1}\rangle}}

\newcommand{\Braket}[2]{\langle #1|#2\rangle}
\def\Bra#1{\left<#1\right|}
\def\Ket#1{\left|#1\right>}

\DeclareMathOperator*{\argmin}{arg\,min}

\def\bc{\begin{center}}
	\def\ec{\end{center}}
\def\bea{\begin{eqnarray}}
	\def\eea{\end{eqnarray}}
\newcommand{\avg}[1]{\langle{#1}\rangle}

\makeatletter
\def\@mkboth#1#2{}
\makeatother

\begin{document}

\title{Complex Quantum Networks: a Topical Review}

\author{Johannes Nokkala}
\address{Department of Physics and Astronomy, University of Turku, FI-20014, Turun Yliopisto, Finland}
\ead{jsinok@utu.fi}

\author{Jyrki Piilo}
\address{Department of Physics and Astronomy, University of Turku, FI-20014, Turun Yliopisto, Finland}

\author{Ginestra Bianconi}
\address{School of Mathematical Sciences, Queen Mary University of London, London, United Kingdom\\
The Alan Turing Institute, The British Library, London, United Kingdom}
\ead{ginestra.bianconi@gmail.com}

\begin{abstract}
These are exciting times for quantum physics as new quantum technologies are expected to soon transform computing at an unprecedented level. Simultaneously network science is flourishing proving an ideal mathematical and computational framework to capture the complexity of large interacting systems. Here we provide a comprehensive and timely review of the rising field of complex quantum networks.  On one side, this subject is key to harness the potential of complex networks in order to provide design principles to boost and enhance quantum algorithms and quantum technologies. On the other side this subject can provide a new generation of quantum algorithms to infer significant complex network properties. 
The field  features  fundamental research questions as diverse as designing networks to shape Hamiltonians and their corresponding phase diagram, taming the complexity of many-body quantum systems with network theory, revealing how quantum physics and quantum algorithms can predict novel network properties and phase transitions, and studying the interplay between architecture, topology and performance in quantum communication networks. Our review covers all of these multifaceted aspects in a self-contained presentation  aimed both at network-curious quantum physicists and at quantum-curious network theorists. We provide a framework that  unifies the field of quantum complex networks along  four main research lines: network-generalized, quantum-applied, quantum-generalized and quantum-enhanced. Finally we  draw attention to the connections between these research lines, which can lead to new opportunities and new discoveries at the interface between quantum physics and network science.
\end{abstract}

%
%
%
%
%

\tableofcontents

\section{Introduction and motivation}

Quantum physics emerged in the 20th century to explain phenomena not accounted for by classical physics, such as the spectrum of black body radiation and the photoelectric effect, and has since been developed into a mature and highly successful theory of Nature. Its deviations from its classical counterpart have more recently been recognized as an opportunity especially in computing \cite{montanaro2016quantum}, sensing \cite{giovannetti2011advances}, communication \cite{wehner2018quantum} and simulation \cite{georgescu2014quantum}. In this context,  regimes or circumstances have been identified where quantumness can provide an advantage or indeed, facilitate an otherwise impossible task. Discovery and pursuit of these new applications has led to the creation of several specialized subfields  such as quantum enhanced approaches to classical tasks or generalizing purely classical concepts to the quantum case, further fueling both theoretical and experimental progress towards realizing the envisioned technology. Today we already enjoy the fruits of the so called first quantum revolution which gave us the transistor, the laser and the atomic clock. The second revolution is generally considered to mean that deeply quantum phenomena such as entanglement move from laboratories to the field and their applications are commercialized, meaning in particular that one has to deal with states, systems and architecture of increasing complexity---among the many formidable hurdles to be overcome, this complexity must be tamed \cite{bianconi2015interdisciplinary,biamonte2019complex}.

While physics in the past centuries has followed mostly a reductionist direction, in this last century we have witnesses the recognition that ``more is different" \cite{anderson1972more}, i.e. new physics arises from large complex interacting systems. In particular starting from the late nineties, complexity has flourished  thanks to the increased understanding of complex interacting systems in terms of their underlying network structure \cite{SW,BA}. Network theory \cite{barabasi,newman2018networks} is now pivotal to characterize complexity across domains, ranging from the Internet to the brain \cite{bullmore2009complex}. Specifically, a complex system 
is formed by a set of  interacting elements, where typically these interactions are considered  pairwise. Examples of networks representing complex systems are not only communication, transportation networks, and power grids, but also protein-protein interaction networks in the cell and neural networks in the brain. More in general networks can be considered as representation of both classical and quantum data. For example networks can be used as a mathematical representation of quantum statistics~\cite{bianconi2001bose} or they can capture the complexity of entangled spin chains~\cite{sundar2018complex,mendes2023wave}. Networks encode the information  of complex systems in their topology, hence a fundamental goal of network science has been to mine network structures finding the key statistical  and topological properties. Interestingly while some properties are very specific of some complex systems other properties such as the small world~\cite{SW} property and the  scale-free~\cite{BA} degree distribution, are ubiquitous and define universality classes.  A key result of network science is that the topology of the network strongly affects dynamical processes defined on these structures \cite{dorogovtsev2008critical}. For instance  scale-free networks as different  as the Internet or the biological transcription networks respond to random and targeted damage \cite{cohen2000resilience,cohen2001breakdown,dorogovtsev2008critical} of their nodes in a similar way, which is very distinct from the response of lattices and random graphs to similar perturbations. 

\begin{figure}[t]
    \centering
    \includegraphics[trim=0cm 5.25cm 0.5cm 0cm,clip=true,width=0.95\textwidth]{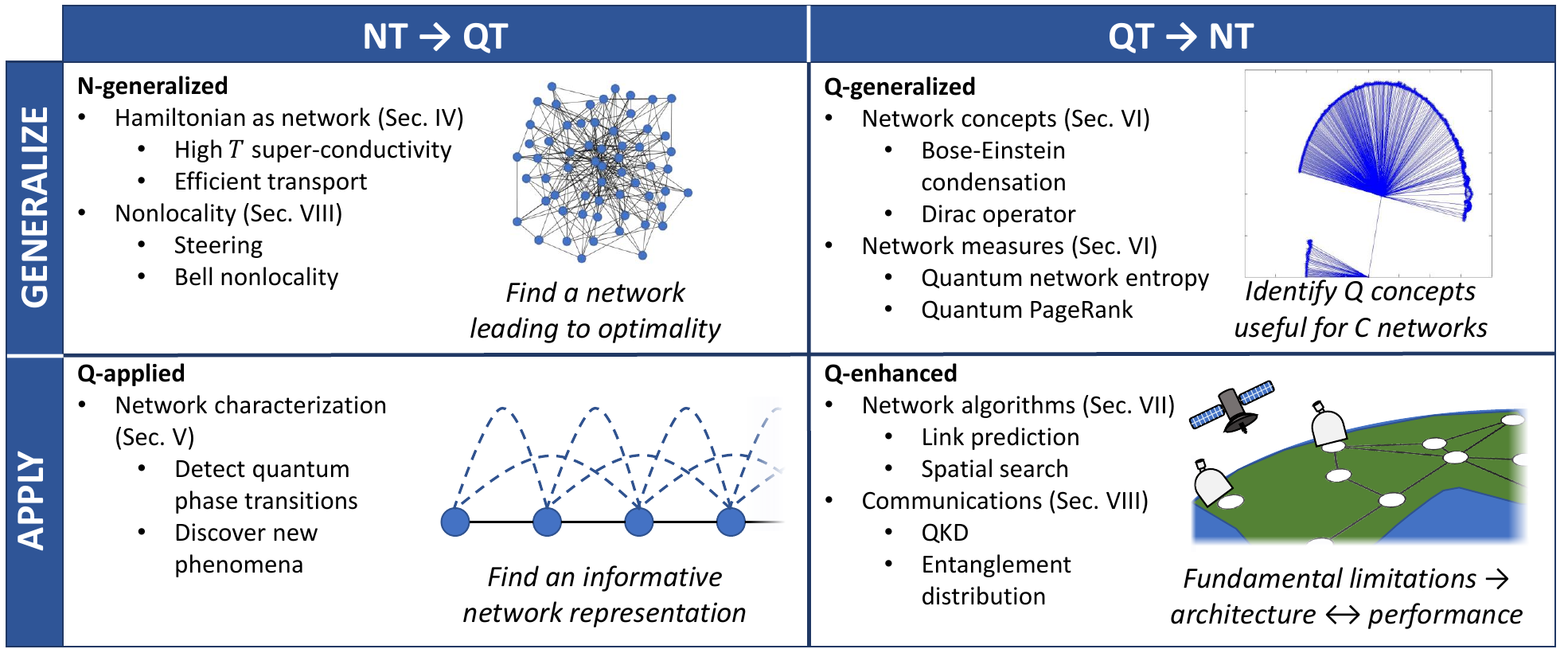}
    \caption{We distinguish between four distinct research lines on complex quantum networks: network-generalized, quantum-applied, quantum-generalized and quantum-enhanced. Here these research lines are schematically presented with a figure and with  illustrative research questions. These different research lines  will be presented in the indicated Sections.}
    \label{fig:intro}
\end{figure}

This review focuses on the intersection between quantum physics and network theory and therefore on cases where there is both a quantum and a network aspect. Although such research has seen steadily increasing interest since 2000s, the term complex quantum  network still does not have a stringent definition and the various research lines have developed independently. However the field is now developing further and 
we strongly believe that with the impressive advances in the pursuit of quantum technology it is now very timely to cover these multi-faceted 
complex quantum networks and introduce the emerging field in a pedagogical, self-contained and comprehensive review. This topical review is intended both for quantum physicists and network scientists to serve as an entry point to the literature{, complementing and in some ways extending the treatment of previous reviews \cite{bianconi2015interdisciplinary,biamonte2019complex}}.

We distinguish between four main directions of investigation of complex quantum networks. In a similar vein to V. Dunjko's and P. Wittek's categorization \cite{dunjko2020non} of quantum machine learning \cite{ventura2000quantum,trugenberger2001probabilistic,trugenberger2002quantum}, we call these different research lines  quantum-applied, network-generalized, quantum-generalized and quantum-enhanced approaches, respectively. We summarize them in Fig.~\ref{fig:intro}. Quantum physics research generalizing to or taking inspiration from networks (the network-generalized research line) includes optimizing excitation transport in networks  \cite{mulken2016complex} or studying synchronization in open networks of interacting quantum harmonic oscillators \cite{manzano2013synchronization}. In general, in this line of research the network structure is typically encoded into the system Hamiltonian. 
Thus important developments have considered quantum critical phenomena \cite{sachdev2011quantum} on lattices, graphs and complex networks and quantum walks \cite{kempe2003quantum,aharonov1993davidovich}. This latter are playing an important role for quantum computation \cite{montanaro2016quantum} providing significant speedups of quantum algorithm with respect to classical algorithms \cite{childs2009universal,childs2003exponential} in some network topology.
Moreover,  research  investigating network nonlocality in terms of Bell inequalities \cite{tavakoli2022bell} or quantum steering \cite{jones2021network} also follows in this research line. The quantum-applied research direction includes adopting a network approach to address a quantum problem, such as predicting a quantum phase transition \cite{valdez2017quantifying} or discovering a previously unknown collective phenomenon \cite{sokolov2022emergent} from a judiciously chosen network representation of the considered state. {The} very active quantum generalized research direction identifies  quantum concepts that are useful for modelling and characterizing complex   networks. This line of research includes the formulation of the Bose-Einstein condensation \cite{bianconi2001bose} in complex networks, the  definition the von Neumann entropy \cite{passerini2008neumann} and of the topological Dirac operator \cite{bianconi2021topological} of a network. Finally, the quantum-enhanced research line  includes quantum enhanced communication such as quantum key distribution \cite{mehic2020quantum} or entanglement distribution \cite{kimble2008quantum} and its various applications. 

The review is structured as follows. The basics of quantum theory and network theory are covered in Secs.~\ref{sec:qtheory} and \ref{sec:ntheory}, respectively. They provide a relatively broad selection of topics to account for the variety of ways they can and have been combined with a special emphasis on content relevant to reviewed material, and are intended primarily for readers unfamiliar with either field. The following Sections each focus on a different aspect of {complex quantum networks}. They are mostly independent and may be read in any order or individually as per interest---for the sake of compactness, the related examples primarily highlight some relevant contemporary research, while technical details are often left for more specialized reviews which are suggested where necessary. Sec.~\ref{sec:physical} focuses on network-generalized research and specifically on cases where the network is embedded in the Hamiltonian, often such that the interaction terms play the role of links and the systems the role of nodes. The following Sec.~\ref{sec:properties} focuses on quantum-applied research where a judicious network representation is sought to simplify, predict, understand or discover properties of interest. {Network theory oriented} research is covered from two complementary points of view with network models exhibiting emergent quantumness presented in Sec.~\ref{sec:emergence} and quantum algorithms for both conventional and novel properties of classical networks given in Sec.~\ref{sec:propertiesII}. Sec.~\ref{sec:comms} focuses on quantum-enhanced communication networks including quantum key distribution and entanglement distribution networks but also briefly introduces state transfer in networks interacting quantum systems as well as network generalized nonlocality. Finally, conclusions are drawn in Sec.~\ref{sec:discus} where we discuss the overall state of the field and outlook, as well as the connections {between the research lines}, which we hope encourages cross-fertilization and fosters new research in this promising field.

\section{\label{sec:qtheory}Basics of quantum mechanics}

\subsection{Basic concepts}

This Section presents briefly some relevant background starting from basic concepts and definitions. Some familiarity with the topics is assumed, and for the sake of compactness the text is not self-contained. We recommend Refs.~\cite{heinosaari2011mathematical} and \cite{nielsen2010quantum}, \cite{breuer2002theory} as further reading; experts may wish to skip this Section. We set the Planck constant $\hbar=1$ and the Boltzmann constant $k_B=1$.

Quantum mechanics is a probabilistic theory concerning outcome probabilities of measurements performed on physical systems. In particular, given the state of a physical system and a measurement there must be a rule to arrive at ordinary probabilities: a set of real non-negative numbers that add up to one such that the probability of a union of mutually exclusive outcomes is just the sum of their probabilities \cite{kolmogorov2018foundations}. The theory must also be able to account for evolution and composition of states as physical systems typically evolve in time---think for example of a swinging pendulum in a grandfather clock---and experiments can involve joint systems. Such a probabilistic framework does in fact leave some leeway, however what sets quantum physics apart is its remarkable predictive power. This makes applications and quantum technology possible; because of fundamental differences with classical mechanics, they have different limitations and advantages.

Specifically, a physical system is associated with a Hilbert space $\mathcal{H}$, a complex vector space complete with respect to the norm induced by its inner product. Then a vector of this space is indicated by the ket $\Ket{\psi}$. Any non-zero vector is a possible state for the system, and two vectors that differ only by a constant represent the same state. The inner product between some vectors $\Ket{\phi},\Ket{\psi}\in\mathcal{H}$ is indicated by $\Braket{\phi}{\psi}$ and maps them to a scalar. Additionally, it must be linear in its second argument, have conjugate symmetry and be positive definite provided that the argument is not the zero vector. The induced norm $\lVert\psi\rVert$ of some vector $\Ket{\psi}$ reads
$
\lVert\psi\rVert=\sqrt{\Braket{\psi}{\psi}}.
$
The vector is normalized if $\lVert\psi\rVert=1$. Such unit vectors are also called state vectors. On the other hand, if $\Braket{\psi}{\phi}=0$ the vectors are orthogonal, and if this holds for any distinct pair of elements in some set $S\subset\mathcal{H}$ then $S$ is orthogonal. If the elements of $S$ are also unit vectors, then $S$ is also orthonormal. The dimension of the Hilbert space $\mathcal{H}$ is determined by the largest possible size of such a set: if the size is limited by some positive integer $d$ then $\mathcal{H}$ is $d$-dimensional, and otherwise infinite dimensional. 
Omitting some details, any orthogonal set $S$ with $d$ elements is a possible basis for $\mathcal{H}$, meaning that any of its vectors can be expressed by a linear combination of the elements of $S$.  Conversely, any linear combination $a\Ket{\psi}+b\Ket{\phi}$ where $a,b\in\mathbb{C}$ is a valid vector and therefore a valid state; this is also known as the superposition principle.

It turns out that dynamics, measurements and even more general states will all be accounted for by linear operators acting in the relevant Hilbert space; therefore from now on whenever we say operator we mean a linear operator. 

Starting from dynamics, suppose that for some operator $U$ and some state vector $\Ket{\psi}$ we have $\Ket{\rho}=U\Ket{\psi}$. If $U$ describes a physical transformation then also $\Ket{\rho}$ should be normalized. This requirement is fulfilled when $U$ preserves the inner product between vectors and then we say that it is a unitary operator. As a side note, this also ensures a unitary operator can be used to change from one orthogonal basis to another. A paradigmatic example of a unitary operator is the one obtained as the solution to the well-known Schr{\"o}dinger equation. If the system is in state $\Ket{\psi(t)}$ at time $t\in\mathbb{R}$ and its Hamiltonian is $H$, then
$
\mathrm{i}\frac{\partial}{\partial t}\Ket{\psi(t)}=H\Ket{\psi(t)}
$
from which one gets $\Ket{\psi(t)}=e^{-\mathrm{i}Ht}\Ket{\psi(0)}$ where $\Ket{\psi(0)}$ is some initial state and the anticipated unitary operator reads $U(t)=e^{-\mathrm{i}Ht}$. A relevant property of unitary operators is that they are reversible, implying particularly that given $\Ket{\psi(t)}$, $t$ and $H$ the initial state can always in principle be recovered. This seemingly simple property of unitary evolution has deep implications that we will return to momentarily.

Focusing on the Hamiltonian, it is not only an operator but also a Hermitian operator. In short, Hermitian operators have real eigenvalues and eigenvectors corresponding to different eigenvalues are orthogonal; suppose $\Ket{\varphi_i}$ are the eigenvectors and the associated eigenvalues are $\lambda_i$. Thanks to Hermiticity, the numbers $p_{\lambda_i}=\Bra{\psi}\ket{\varphi_i}\!\bra{\varphi_i}\Ket{\psi}$ satisfy the requirements for probabilities and indeed can be interpreted as outcome probabilities of so called projective measurements of the observable with Hermitian operator $H$, the energy, where $\lambda_i$ are the corresponding outcomes that in this case are the possible energies of the system. The outcome $\lambda_i$ also indicates that the state was projected into a corresponding eigenvector. The eigenvector with the lowest energy is called the ground state, whereas the rest are called excited states. An important quantity is the expected value $\langle H\rangle$, which can be recovered just from $\langle H\rangle=\sum_ip_{\lambda_i}\lambda_i=\Bra{\psi}H\Ket{\psi}$. More generally, any observable $O$, be it the polarization of a photon or the momentum of a nanomechanical oscillator, is associated with a Hermitian operator.

We make a brief remark about mixed states which are statistical mixtures of state vectors, also called pure states. Suppose we prepare a pure state drawn from some given set according to some fixed probabilities such that state vector $\Ket{\phi_i}$ appears with probability $p_i$. We introduce the density operator $\varrho=\sum_ip_i\Ket{\phi_i}\!\Bra{\phi_i}$ where $\Ket{\phi_i}\!\Bra{\phi_i}$ is the projector to the one dimensional subspace spanned by $\Ket{\phi_i}$ and its action on some vector $\Ket{\psi}$ is just $\Braket{\phi_i}{\psi}\Ket{\phi_i}$. Then $p_{\lambda_i}=\mathrm{Tr}\left(\varrho \Ket{\varphi_i}\!\Bra{\varphi_i}\right)$, $\langle O\rangle=\mathrm{Tr}\left(\varrho O\right)$ and $\varrho(t)=\sum_ip_i\Ket{\varphi_i(t)}\!\Bra{\varphi_i(t)}$ where $\Ket{\varphi_i(t)}=U(t)\Ket{\varphi_i(0)}$. Here $\mathrm{Tr}$ evaluates the trace of an operator which is equal to the sum of its eigenvalues and is therefore basis independent.

Quantum mechanics is inherently linear, therefore a very important problem is to characterize the quantum signature of chaos. This has lead to the important development of Quantum Chaos and quantum graphs~\cite{gnutzmann2006quantum,smilansky2013discrete,haake1991quantum,balasubramanian2022quantum}. This field is not only of fundamental theoretical importance but is also key for assessing possible quantum chaos effects in short term quantum computation.
\subsection{Single systems}

A $d$-dimensional Hilbert space is isomorphic to $\mathbb{C}^d$, the inner product space of $d$-tuples of complex numbers. In what follows, we treat them as the same space for convenience.

Let now $d=2$, making $\mathbb{C}^2$ the relevant space. If we fix an orthonormal basis $\{\Ket{0},\Ket{1}\}$, then some $\Ket{\psi}=\alpha\Ket{0}+\beta\Ket{1}$ becomes the column vector of complex numbers $\Ket{\psi}=(\alpha,\beta)^\top$. Using these basis states is suggestive, and indeed one may associate them with classical bits $0$ and $1$, making $\Ket{\psi}$ a quantum bit, more commonly known as qubit; in this context the basis is also referred to as the computational basis. Such qubits are not simply noisy bits, however, which is best exemplified by considering the density operator. For that we need to know that $\Bra{\psi}=(\alpha^*,\beta^*)$ where for example $\alpha^*$ is the complex conjugate of $\alpha$. The density operator is then
$
\Ket{\psi}\!\Bra{\psi}=
\big(\begin{smallmatrix}
  \alpha\\
 \beta
\end{smallmatrix}\big)
\big(\begin{smallmatrix}
  \alpha^* & \beta^*
\end{smallmatrix}\big)
=\big(\begin{smallmatrix}
  |\alpha|^2 & \alpha\beta^*\\
  \alpha^*\beta & |\beta|^2
\end{smallmatrix}\big)
$
but for a statistical mixture of basis states $\varrho=|\alpha|^2\Ket{0}\!\Bra{0}+|\beta|^2\Ket{1}\!\Bra{1}$ it becomes
$
\varrho=
\big(\begin{smallmatrix}
  |\alpha|^2 & 0\\
  0 & \beta|^2
\end{smallmatrix}\big)
$,
where of course it must be that $\Braket{\psi}{\psi}=|\alpha|^2+|\beta|^2=1$. Due to their important role, the off-diagonal elements are called coherences. The loss of coherences is called decoherence.

An operator with matrix $\mathbf{M}$ is unitary exactly when $\mathbf{M}^\dagger=\mathbf{M}^{-1}$ where $\mathbf{M}^\dagger$ is the conjugate transpose. Examples of unitary operators in the Hilbert space $\mathbb{C}^2$ are the Hadamard gate and the phase gate
\begin{equation}
\mathbf{H}=\frac{1}{\sqrt{2}}\begin{pmatrix}
1 & \ \ 1\\
1 & -1
\end{pmatrix},\quad
\mathbf{P}=\begin{pmatrix}
1 & \ \ 0\\
0 & \ \ \mathrm{i}
\end{pmatrix},
\label{eq:localClifford}
\end{equation}
which act on the vector via standard matrix multiplication. Indeed, a direct calculation shows that for any $\Ket{\psi}$ it holds that $\mathbf{H}^2\Ket{\psi}=\mathbf{H}\mathbf{H}\Ket{\psi}=\Ket{\psi}$, whereas $\mathbf{H}\Ket{0}=(\Ket{0}+\Ket{1})/\sqrt{2}$ and $\mathbf{H}\Ket{1}=(\Ket{0}-\Ket{1})/\sqrt{2}$, which are also denoted by $\Ket{+}$ and $\Ket{-}$, respectively. Recalling that unitary operators are also basis changes, we may immediately deduce that $\{\Ket{+},\Ket{-}\}$ is another orthonormal basis. The phase gate simply adds a complex phase to the coefficient of $\Ket{1}$. Both are widely used for example in quantum computing \cite{nielsen2010quantum} where they and other gates can be used as basic building blocks to, e.g., implement a quantum algorithm.

We may also consider infinite-dimensional systems such as quantum harmonic oscillators which can be constituted, for example, by excitations in optical modes or micro- or nanomechanical oscillators. The relevant Hilbert space has a countable orthonormal basis, the Fock basis $\{\Ket{n}\}_{n=0}^\infty$, and consists of all vectors $\Ket{\psi}=\sum_{n=0}^\infty\Braket{n}{\psi}\Ket{n}$ such that $\sum_{n=0}^\infty|\Braket{n}{\psi}|^2$, or the squared norm, 
is finite. It is convenient to introduce creation and annihilation operators $a^\dagger$ and $a$ defined by
$
a^\dagger\Ket{n}=\sqrt{n+1}\Ket{n+1}$, $ a\Ket{n}=\sqrt{n}\Ket{n-1}
$
because several important unitary and Hermitian operators can be expressed in terms of them. Examples of the former include the displacement operator $D(\alpha)=e^{a^\dagger\alpha-a\alpha^*}$ and the squeezing operator $S(\xi)=e^{(\xi a^{\dagger2}-\xi^*a^2)/2}$ where $\alpha,\xi\in\mathbb{C}$, and of the latter the Hamiltonian of an oscillator with frequency $\omega$ which is $H=\omega(a^\dagger a+1/2)$. From $H$ it is clear that $\{\Ket{n}\}_{n=0}^\infty$ are energy eigenstates since $a^\dagger a\Ket{n}=n\Ket{n}$. Here the ground state $\Ket{0}$ is often called the vacuum. From it one can create the coherent states $\Ket{\alpha}=D(\alpha)\Ket{0}$, the squeezed vacuum states $\Ket{\xi}=S(\xi)\Ket{0}$ and the squeezed coherent states $\Ket{\alpha,\xi}=D(\alpha)S(\xi)\Ket{0}$. These states are widely used in quantum optics pioneered in particular by Roy J. Glauber who later shared a Nobel prize for his crucial contributions \cite{glauber1963photon,glauber1963quantum} to the field in 2005. 

In particular, position and momentum operators may be defined as judicious linear combinations of $a^\dagger$ and $a$. They are Hermitian and so can be measured; in fact, they are continuous variables as the spectrum of both is the entire real line. Importantly, the probability distribution function of either is a Gaussian distribution for any $\Ket{\alpha,\xi}$, fully characterized by just its mean and variance. It turns out that all such pure states of a single oscillator are squeezed coherent states, also called pure Gaussian states. These states can be used as approximations of the eigenstates of position and momentum operators. Informally, an eigenstate $\Ket{0}_q$ of position ($\Ket{0}_p$ of momentum) with eigenvalue $0$ can be approached by $S(\xi)\Ket{0}$ where $|\xi|\gg 1$ and $\mathrm{arg}(\xi)=\pi$ ($\mathrm{arg}(\xi)=0$); states corresponding to different eigenvalues can be achieved by appropriate displacements. At the limit of infinite squeezing one variance vanishes and the other one diverges, informally giving a state of definite position but completely unknown momentum and vice versa. The limit is not in the Hilbert space however, as its squared norm is not finite. The unphysicality of especially $\Ket{0}_p$ has implications for so called continuous variable cluster states, as seen later in Sec.~\ref{sec:properties}.

\subsection{Multiple systems\label{sec:qtheory_correlations}}

Measuring a qubit $\Ket{\psi}=(\alpha,\beta)^\top$ in the computational basis projects it into $\Ket{0}$ with probability $|\alpha|^2$ and to $\Ket{1}$ with probability $|\beta|^2$. What if we measure two qubits? Then we expect the measurements to project the joint system into one of four different combinations which we express as $\{\Ket{00},\Ket{01},\Ket{10},\Ket{11}\}$. They form the basis of $\mathbb{C}^2\otimes\mathbb{C}^2$ where $\otimes$ is the tensor product. More concretely, if the other qubit was $\Ket{\phi}=(\gamma,\delta)^\top$, the product state $\Ket{\psi}\otimes\Ket{\phi}=(\alpha\gamma,\alpha\delta,\beta\gamma,\beta\delta)^\top$ gives the correct outcome probabilities. The local states may be recovered through an operation called the partial trace.

Local gates such as the ones of Eqs.~\eqref{eq:localClifford} can be applied by using the Kronecker product between two matrices; for instance $\mathbf{H}\otimes\mathbf{I}$, where $\mathbf{I}$ is the $2\times2$ identity matrix, applies the Hadamard gate to the first system only. Typically the target is indicated with subindices, in this case by $\mathbf{H}_1$. More generally, any $4\times4$ unitary matrix is a valid operation in $\mathbb{C}^2\otimes\mathbb{C}^2$ but importantly, not all of them can be decomposed into local gates. One example is the CZ or controlled Z gate, determined by $\Ket{A,B}\rightarrow(-1)^{A B}\Ket{A,B}$ where $A,B\in\{0,1\}$. Applications of this gate on multiple qubits initially in the $\Ket{+}$ state can be used to create a so called cluster state, discussed in Sec.~\ref{sec:properties}.

Another example is the CNOT gate, or controlled not gate, which is determined by its action on the basis states via $\Ket{A,B}\rightarrow\Ket{A,A\oplus B}$ where $\oplus$ is addition modulo 2. Here the first system is said to be the control qubit and the second the target qubit. Consider now the states prepared from the four basis states by applying $\mathbf{H}_1$ followed by CNOT. These are, in order, $\Ket{\Phi^+}$, $\Ket{\Psi^+}$, $\Ket{\Phi^-}$, $\Ket{\Psi^-}$ given by
\begin{equation}
\Ket{\Phi^\pm}=(\Ket{00}\pm\Ket{11})/\sqrt{2},\quad
\Ket{\Psi^\pm}=(\Ket{01}\pm\Ket{10})/\sqrt{2}.
\label{eq:Bellstates}
\end{equation}
They are also called the Bell states and as they are formed from an orthonormal basis with a unitary operation they form an alternative orthonormal basis called the Bell basis. Like the gate needed to prepare them, none of the Bell states can be decomposed into a product of two pure states as in $\Ket{\psi}\otimes\Ket{\phi}$. This indicates the presence of correlations. Indeed, if for example the state is $\Ket{\Phi^+}$ and a projective measurement of the first qubit in the computational basis yields the result $\Ket{0}$ then we immediately know that the state of the second one must also be $\Ket{0}$ and vice versa, allowing for example two distant laboratories holding half of the state each to privately share a random bit. Since the measurement outcome for the other qubit is determined completely, Bell states are maximally entangled. In general, bipartite states can be classified into separable and entangled states. A separable state is a product state if it is pure and a statistical mixture of product states otherwise. Entanglement is a multi-faceted and rich phenomenon---here we briefly present only some aspects of it directly relevant to the material reviewed later, such as the quantum communication networks of Sec.~\ref{sec:comms}.

Whereas correlations in all separable states---or any systems obeying classical physics---are amenable to an explanation via local hidden variables, pure entangled states such as Bell states are not. It should be stressed that such non-locality is not the same as entanglement however, since for example mixed entangled states may not exhibit it \cite{werner1989quantum}. Non-locality and its generalization to networks are briefly discussed in Sec.~\ref{sec:comms_entdist}; for a more thorough treatment see, e.g., Sec. 2.6 of \cite{nielsen2010quantum} and Ref.~\cite{tavakoli2022bell} for ordinary and network cases, respectively. 

Entanglement can be applied in teleportation. Consider that laboratory A has a qubit in some unknown state $\Ket{\psi}$ and shares $\Ket{\Phi^+}$ with laboratory B. The joint state reads $\Ket{\psi}_1\Ket{\Phi^+}_{23}$ where qubits 1 and 2 are at A and qubit 3 at B and we have left the tensor product $\otimes$ implicit. But expressing the state of qubits 1 and 2 in the Bell basis, we have
$
\Ket{\psi}_1\Ket{\Phi^+}_{23}=(\Ket{\Phi^+}_{12}\Ket{\psi}_3+\Ket{\Psi^+}_{12}\mathbf{X}_3\Ket{\psi}_3+\Ket{\Phi^-}_{12}\mathbf{Z}_3\Ket{\psi}_3+\Ket{\Psi^-}_{12}\mathbf{X}_3\mathbf{Z}_3\Ket{\psi}_3)/2,
$
or a superposition of states at qubit 3 that are local unitary transformations of $\Ket{\psi}$. Specifically, $\mathbf{X}$, also called NOT gate, is determined by $\Ket{A}\rightarrow\Ket{A\oplus 1}$ and the $\mathbf{Z}$ gate by $\Ket{A}\rightarrow(-1)^A\Ket{A}$; both are their own inverses. If A could project qubits 1 and 2 to one of the Bell states---i.e. perform a Bell state measurement---and communicate the result to B then B could recover the original state by inverting, as necessary, the local gates. The original entangled state $\Ket{\Phi^+}_{23}$ is irreversibly lost however, meaning that A and B need to share a freshly generated Bell state if they wish to teleport another qubit, or if A wants $\Ket{\psi}$ back. Crucially, neither A nor B need to know the state. Otherwise A could just email preparation instructions to B. The state to be teleported can itself be one half of a Bell state; teleporting the entanglement can be used to extend two short hops of shared entanglement into one long hop via local operations and classical communication (LOCC), a process called entanglement swapping.

Given a generic two-qubit entangled state, how many copies of the state on average are needed to facilitate perfect teleportation? This is closely related to the concept of entanglement distillation, where an ensemble of weakly entangled systems is transformed into a smaller ensemble of systems with stronger entanglement. If the initial state is some $\rho^{\otimes n}$ and it is transformed via LOCC into some state $\sigma$ which at the limit of large $n$ approaches
$\Ket{\Psi^+}^{\otimes m_n}$ then the rate is $\lim_{n\rightarrow\infty}m_n/n$, and its supremum over all possible LOCC operations is the distillable entanglement. Its maximum value, 1, is reached by the Bell states and it vanishes for product and separable states whereas entangled pure states have some intermediate value. The case of mixed states is more complicated.

Going beyond entangled qubit pairs, an important example of a state with genuine multipartite entanglement is a so called Greenberger–Horne–Zeilinger (GHZ) state, which can be thought of as a generalization of the Bell state $\Ket{\Phi^+}$ to $M\geq3$ qubits: $\Ket{\mathrm{GHZ}}=(\Ket{0}^{\otimes M}+\Ket{1}^{\otimes M})/\sqrt{2}$. It suffices to say that for suitable multipartite entangled systems a Bell state between given systems may be created by just single qubit operations, exchanging the need to perform Bell state measurements to the need of preparing a more complicated initial entangled state. In the infinite-dimensional case bipartite states can also be classified to product, separable and entangled states and entanglement does not increase under LOCC. Different ways to generalize for example teleportation have been proposed \cite{braunstein2005quantum} but the teleported state might no longer be exactly the same as the original.

\subsection{Infinitely many systems\label{sec:qteory_OQS}}

A quantum system undergoing time evolution can in practice experience phenomena that are unaccounted for by the framework presented so far. This includes irreversibility such as permanent loss of information about the initial state, purity or coherences, suggesting that the dynamics is not unitary in the system's Hilbert space. This is typically the case when the system is open, i.e. coupled to its environment. The environment $E$ of an open quantum system $S$ associated with some Hilbert space $\mathcal{H}_S$ may be defined as a quantum system associated with some Hilbert space $\mathcal{H}_E$ such that the evolution of the total system $SE$ is unitary in the Hilbert space $\mathcal{H}_S\otimes\mathcal{H}_E$. If we are interested in the reduced dynamics of the open system alone we may write the dynamics using the partial trace which strips the environment degrees of freedom, arriving at an exact but formal equation since $E$ could be very large or even infinite, unknown and uncontrollable. Reasonable approximations and assumptions may allow the derivation of tractable equations of motion involving only operators acting in $\mathcal{H}_S$, however. 

In particular, when the initial state is a product state and the initial state of the environment is fixed, we may introduce the dynamical map $\Phi_t$ acting entirely in $\mathcal{H}_S$ such that $\varrho_S(t)=\Phi_t\varrho_S(0)$ and use it to classify the reduced dynamics of the open system \cite{breuer2016colloquium}. In particular, if the open system can only lose information of its initial state and never gain it back it is said that the reduced dynamics is memoryless, or Markovian. Non-Markovianity may be characterized in terms of, e.g., back-flow of information from $E$ to $S$ \cite{breuer2016colloquium}. Results concerning the non-Markovianity of networks of interacting quantum systems are presented in Sec.~\ref{sec:physical}.

A sufficient condition for Markovian dynamics is that the dynamical map has the the semigroup property where $\Phi_{t_1}\Phi_{t_2}=\Phi_{t_1+t_2}$ for any $t_1,t_2\geq 0$. Such dynamics may arise for example if the interaction is weak, the change in environment state is negligible, the intrinsic evolution of $S$ is fast and the environment is a reservoir, meaning that its degrees of freedom form a continuum; for further details see, e.g., Sec.~3.3 of \cite{breuer2002theory}. An important special case is when $E$ is a reservoir in a thermal equilibrium state, i.e. in the stationary state of $H_E$ amenable to a description in terms of just one parameter, its temperature. Such reservoirs are called heat baths. Then under some mild conditions it can be shown that for any $\varrho_S(0)$ the asymptotic state $\varrho_S(t\rightarrow\infty)$ is also a thermal state of the same temperature. In fact, such relaxation to thermal equilibrium is expected at least effectively even when the total system is large but finite, as seen later in Sec.~\ref{sec:physical}.

An example of a dynamical map with the semigroup property arises from a lossy bosonic channel which describes what happens to an optical mode travelling in optical fiber. Ideal fiber is characterized by how losses accumulate with distance and therefore the time the mode is exposed to the environment formed by the fiber. This is typically quantified by $\gamma$ which is in units of dB/km such that $\eta=10^{-\gamma d/10}\in(0,1]$ is the transmissivity of the channel, determining the action of the channel on some Gaussian state as follows. If $x$ is the initial position operator of the mode and the corresponding operator for the vacuum is $x_{\Ket{0}}$ then 
$
x\rightarrow\sqrt{\eta} x+\sqrt{(1-\eta)}x_{\Ket{0}}
$
and similarly for the momentum. The action of this channel for some durations $t_1$ and $t_2$ corresponds to two distances $d_1$ and $d_2$ travelled in the fiber, giving rise to two transmissivities $\eta_1$ and $\eta_2$. Then the total duration $t_1+t_2$ corresponds to applying the above transformation once with $\eta_1$ and again with $\eta_2$, which coincides with applying it once with $\eta_1\eta_2=10^{-\gamma (d_1+d_2)/10}$, leading to the semigroup property. Such channels are considered in Sec.~\ref{sec:comms}. 

Large systems can be studied also outside the open systems framework. Statistical mechanics is a subject in theoretical physics that addresses the many-body properties of systems formed by a large number of classical as well as quantum particles.  
One of the pivotal results of classical statistical mechanics that has been a turning point in physics for the wide acceptance of  the atomistic description of matter, is the Boltzmann distribution. This distribution characterizes the probability that a particle in gas has given energy $\epsilon$ or alternatively the expected occupation  $n_Z(\epsilon)$ of the $\epsilon$ energy level when the gas is in contact with a thermal bath at temperature $T=1/\beta$. The Boltzmann distribution  is given by 
$
n_Z(\epsilon)=e^{-\beta(\epsilon-\mu)},
$
where $\mu$ is the chemical potential of the gas.

Interestingly quantum particles obey different statistical properties than classical particles. Historically, this became evident first by the study of the black-body radiation and then with the formulation of the Fermi-Dirac and Bose-Einstein statistics and the subsequent spin-statistics theorem.
Indeed on top of having a quantized spectrum, quantum particles are also indistinguishable and can be classified according to the values of their spin. Particles with half integer spin are fermions and particles with integer spin are bosons.
Fermions are such that no two particles can occupy the same energy state at once. A property related to their statistics is that fermions have creation and annihilation operators that anti-commute. On the contrary an arbitrary large number of bosons can occupy a single energy state and consequently the creation and annihilation operators for bosons commute.
The Fermi-Dirac $n_F(\epsilon)$ and the Bose-Einstein $n_B(\epsilon)$ statistics determine the occupation numbers of energy states $\epsilon$ in a gas of fermions and boson respectively, and they are given by 
$
n_F=\frac{1}{e^{\beta(\epsilon-\mu)}+1},$ $n_B=\frac{1}{e^{\beta(\epsilon-\mu)}-1} 
$,
where $\beta=1/T$ is the inverse temperature of the gas fixing its average energy and $\mu$ is the chemical potential of the gas fixing its expected number of particles.
Interestingly in the large temperature limit, $T\to \infty$, i..e $\beta\to 0$ both Fermi-Dirac and Bose-Einstein statistics reduce to the Boltzmann statistics.

A key property of the Bose gas is that when the density of states of the particles is such that $g(\epsilon)\to 0$ as $\epsilon\to 0$ (which in a 
non-interacting Bose gas occurs for dimension $d>2$) a notable quantum phase transition can be observed, called the {\em Bose-Einstein condensation} (BEC). In physical systems in which BEC occurs, there is a critical temperature $T_c=1/\beta_c$ such that for $\beta>\beta_c$ the ground state acquires a finite occupation number leading to the macroscopic manifestation of microscopic quantum phenomena such as wavefunction interference. This phase transition, predicted by Einstein in 1026-1927,  has been experimentally detected first in diluted gas of alkali atoms experiments in 1995. Cornell, Wieman and Ketterle shared the 2001 Nobel Prize in Physics for these discoveries.

\subsection{Quantum information}

Formally, information is intimately linked to uncertainty and entropy. Consider a source of quantum information $S_Q$ which generates a pure state $\Ket{\psi_i}$ with probability $p_i$. This defines a random variable associated with a density operator $\varrho=\sum_ip_i\Ket{\psi_i}\!\Bra{\psi_i}$ which in general can have coherences. Schumacher's noiseless channel coding theorem states that the infimum for the number of qubits needed, on average, to describe the use of $S_Q$ over a noiseless channel such as the one achieved via teleportation coincides with the von Neumann entropy
$
S(\varrho)=-\mathrm{Tr}(\varrho\log(\varrho))
$
where the logarithm is base $2$. This number coincides with its classical counterpart, the Shannon entropy, if and only if the states $\Ket{\psi_i}$ are perfectly distinguishable. Otherwise it is in general smaller, but the error vanishes only asymptotically. The distinguishability may be quantified in terms of fidelity \cite{uhlmann1976transition,jozsa1994fidelity}. In the case of pure states it reads $F(\psi,\rho)=|\Braket{\psi}{\rho}|^2$ which can be interpreted as the probability of projecting the state $\Ket{\rho}$ into $\Ket{\psi}$ by performing a projective measurement in an orthonormal basis including $\Ket{\psi}$. Consequently if $F(\psi,\rho)=0$ the states are orthogonal and can be perfectly distinguished by a projective measurement to an orthonormal basis including both $\Ket{\psi}$ and $\Ket{\rho}$. If $0<F(\psi,\rho)<1$ there is no such basis; then $P_{\Ket{\psi}}=\Ket{\psi}\!\Bra{\psi}$ has a non-vanishing chance to project $\Ket{\rho}$ to $\Ket{\psi}$, misidentifying the state. If $F(\psi,\rho)=1$ the states are the same.

Indistinguishability of generic quantum states has several consequences to the nature of quantum information, and in particular rules out some familiar operations used on classical information. In particular, the no-cloning theorem states that there is no unitary operator $U$ that can clone an unknown quantum state---unless it was drawn from a known set of distinguishable states, in which case the state can be identified and cloning becomes trivial. Importantly, this rules out conventional strategies for amplifying the signal in quantum communication networks of Sec.~\ref{sec:comms}.

Entropy of some random variable $X$ can also be thought of as the amount of knowledge we gain if we learn its value, or alternatively as the uncertainty about its value before we learn it. The joint entropy of a quantum system with components $A$ and $B$ is defined in the natural way as $S(A,B)=-\mathrm{Tr}(\varrho_{AB}\log(\varrho_{AB}))$. The mutual information quantifies how much we have learned from one of the variables given that we know the other: it reads $S(A:B)=S(A)+S(B)-S(A,B)$. Importantly, $S(A:B)$ quantifies the total amount of correlations between $A$ and $B$, including both classical and non-classical correlations such as entanglement. It will be seen later in Sec.~\ref{sec:properties} how it can be used to form networks that can reveal nontrivial information about the quantum system.

Finally, the fidelity can also distinguish between quantum states that are classically regular and quantum states that are classically chaotic \cite{chaudhury2009quantum,frahm2004quantum,emerson2002fidelity}. Indeed the classical sensitivity to the initial conditions found in chaotic system corresponds to what is called {\em fidelity decay}, indicating sensibility of the overlap between two wave-functions evolving under Hamiltonian displaying  slight changes of the control parameter.

\section{\label{sec:ntheory}Basics of network theory}

\subsection{Overview of network theory}

Networks are a powerful framework to represent interacting systems as graphs formed by nodes and links. The nodes describe the element of the complex system and the links encode the complex set of their interactions. 
Networks, and in particular lattices, are known to be of fundamental importance for quantum and condensed matter physics.
Indeed, lattices are traditionally used to represent crystal structures and their dimension, together with their spectral decomposition in Fourier modes is pivotal for the study of phonons and electronic structure as well.

When the physical systems under study are 
complex, the underlying architecture of its interactions is captured by a complex network whose topology has a  significant stochastic  element. 
Examples of complex networks are the Internet whose nodes are routers and links are the physical lines connecting them, or the brain whose nodes are neurons and links are synaptic connections between the neurons. Interestingly it has emerged that networks can be also used as mathematical and computational representations of abstract data going beyond the representation of physical interactions. In this regard networks can be seen as a way to encode the complexity of the data structure, which indicates the relevance of developing methods to extract information from networks.

The theory of network science has shown that complex networks are key to embrace complexity and capture the new physics emerging when many (often heterogeneous) elements of a complex system are interacting together. Indeed, it has been shown that seemingly disparate complex systems might be encoded by networks sharing important common properties. These properties are often referred to with the term {\em universalities}. Important universalities include the  small-world networks and the scale-free networks. Relevantly these universalities have been shown to affect the dynamical properties of the networks.

The interest in using networks to represent complex systems goes however also beyond the study of their universalities. Indeed, inference algorithms have been formulated to extract  information from network structure which uniquely characterize single networks. In particular, network measures  allow to identify the specific role of nodes, links and communities of nodes in the particular networks under investigation which might strongly deviate from null models.

Finally, networks are ideal objects to formulate combinatorial and optimization problems. It is not by chance that the birth of graph theory coincides with 1736 date in which Euler solved the famous problem of the seven bridges of K\"onisberg.

For all these reasons, as we will see in the next sections of this review, networks have been key to formulate new research questions in quantum physics, spanning from the study of quantum critical phenomena to quantum communication. In this Section we will review the key elements of network theory. Therefore the expert reader can skip this section. On the other side it  is not our intention to be comprehensive and we refer the interested reader that wants to deepen their understanding of the subject to the relevant monographs~\cite{barabasi,newman2018networks,bianconi2018multilayer,bianconi2021higher}.

\subsection{Graphs and networks}
\subsubsection{Gentle introduction}
A {\em graph} $G=(V,E)$ comprises a set of {\em vertices} or {\em nodes} $V$ and a set of {\em edges} or {\em links} $E$. Strictly speaking, a  {\em network} is a graph $G=(V,E)$ representing the interactions between the elements of a real system. Examples of networks are ubiquitous and include systems as different as crystal lattices, the Internet and the brain. As a matter of fact any pairwise interacting system, being it man-made (like the Internet) or natural (like crystal lattices or the brain), can be represented by a network. In a number of situations however the  distinction between a network and its underlying graph representation has fluid boundaries, therefore  in this review we will use network as a synonym for graph.

A network $G$ can be directed or undirected. An undirected network is a network in which links are bidirectional  and therefore the link $(i,j)$ between node $i$ and node $j$ is not distinct from the link $(j,i)$. An example of undirected link is a chemical bond, or a protein-protein interaction.
A network is directed if its links are directional. Therefore in a directed network we distinguish between the link $(i,j)$ indicating that node $i$ points to node $j$ and the link $(j,i)$ indicating that node $j$ points to node $i$. For instance, in the World-Wide-Web if a webpage $i$ contains a URL link to  a  webpage $j$ we have a directional link $(i,j)$ but we are not guaranteed that the link $(j,i)$ exists.

A network $G$ can also be weighted or unweighted. A network is weighted if we assign to each link a weight given by a  positive real or integer number. For instance, given a quantum spin chain we can construct a network in which every spin is connected to every other spin and the weight of each link is given by the mutual information between the two spins. Therefore the mutual information is the weight associated to the links of this network. In this case, and in every situation in which larger weights are a proxy for stronger interactions, the weights are also called {\em affinity weights}.  Another possibility in spatial networks is to associate to each link a weight indicating the spatial distance between the two connected nodes. In this case the larger is the weight between two nodes the larger is their distance. This latter type of weights are called {\em distance weights}. It is sometimes useful to convert affinity weights to distance weights by inverting the affinity weights, although the distance weights generated in this way will not typically have the properties of metric distances. The one we have mentioned are only specific examples and  one should be reminded that weights can indicate any (non-negative) variable associated to the links, indicating a similarity or dissimilarity measure between the nodes.
An unweighted network is instead a network in which all the links have the same weight, or in which we do not distinguish between different weights of the links, i.e. all interactions are treated on the same footing.
In general, complex quantum networks can be both weighted and unweighted.
In the following paragraph we will introduce several network measures  that are exemplified for simple weighted network in Fig.~\ref{fig:networkbasics}.

\subsubsection{Basic definitions}

All these networks can be simply captured by a matrix, the {\em adjacency matrix} ${\bf A}$ of the network, of size $N\times N$ where $N$ indicates the number of nodes of the network.
For simple networks, i.e. networks that are unweighted and undirected and in which there are no links that start and end on the same node (tadpoles), the adjacency matrix ${\bf A}$ is a symmetric matrix of elements $A_{ij}=1$ if $(i,j)$ is a link of the network, i.e. $(i,j)\in E$, and $A_{ij}=0$ otherwise.
For directed networks, the adjacency matrix has 
essentially the same definition as for undirected networks but since we distinguish between the link $(i,j)$ and the link $(j,i)$ the adjacency matrix is asymmetric.
For weighted networks the adjacency matrix has non-zero elements given by the weights of the links. Therefore $A_{ij}=w_{ij}$ if $(i,j)\in E$ where $w_{ij}>0$ indicates the weight of the link and $A_{ij}=0$ otherwise. 

\begin{figure}[t]
    \centering
   \includegraphics[trim=0cm 13.5cm 18.2cm 0cm,clip=true,width=0.95\textwidth]{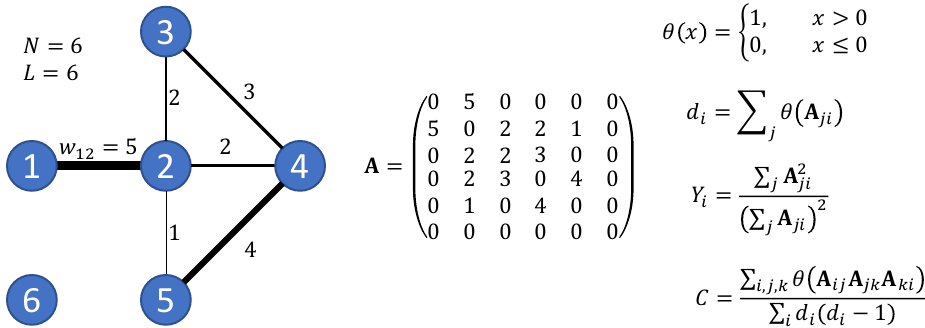}
    \caption{    
A small weighted network and some of the related concepts and measures. The network has $N=6$ nodes and $L=6$ weighted links. Because there is no path connecting node $6$ with any of the other nodes, the network is disconnected. The (weighted) adjacency matrix $\mathbf{A}$ completely determines the network and may be used to calculate various network measures. Both degree $d_i$ and global clustering coefficient $C$ ignore the weights, focusing only on the topology. The weighted degree of node $i$, also sometimes called strength, is simply the $i$-th row sum. In the case at hand clustering is relatively high at $C\approx 0.73$. Disparity $Y_i$ on the other hand takes the weights into account. Node $3$ and node $5$ have both degree $2$ but different disparity. Node $3$ has disparity $Y_3=0.52$ whose inverse is $Y_3^{-1}=1.92$ indicating  the effective number of  links with significant weight.  However node $5$ has disparity $Y_5=0.68$ whose inverse is $Y_5^{-1}=1.47$ indicating that the weight of its links are more unevenly distributed.
    }   
    \label{fig:networkbasics}
\end{figure}

The adjacency matrix of a network captures entirely the structure of a network and plays a fundamental role in determining the dynamics of complex quantum networks. For instance, continuous time quantum walks, that are of pivotal importance in the field of quantum networks, often use the adjacency matrix as their Hamiltonian (see detailed discussion in Section IV and Section VII).
For simple networks (undirected, unweighted networks) the sum of the $i^{th}$ row, or equivalently the sum of the $i^{th}$ column, of the adjacency matrix provides the degree of the node $i$, i.e. the number of links incident to the node (see Fig. \ref{fig:networkbasics} for an extension of this definition to weighted networks). In directed networks we distinguish instead among the in-degree (sum of all incoming links) and the out-degree (sum of all outgoing links) of a node $i$, given respectively by the sum of the $i^{th}$ column and the sum of the $i^{th}$ row of the directed adjacency matrix.
For weighted undirected network the sum of the weights of the links incident to a given node is also called the {\em node strength} or {\em weighted degree}. In order to characterize the heterogeneity of the weights incident to the same node, the {\em disparity}, also called {\em participation ratio }, can be used. The disparity is a quantity between zero and one, that is one if all the strength of a node is concentrated in one link, and zero if every link incident to the same node has the same weight (see Fig. \ref{fig:networkbasics} for an example). The inverse of the disparity can be used to quantify how many links incident to a node have significant weight relative to  its strength.
When weighted networks are fully connected, i.e. a weight is defined for every pair of nodes, one can investigate the similarity between two nodes using the Pearson correlation measured among the vector of all the weights of the links incident to a node and the analogous vector of all the weights of the link incident to the other node. 

An important matrix that captures the structural properties of the network and that is often used as Hamiltonian of continuous time quantum walk instead of the adjacency matrix, is the Laplacian matrix ${\bf L}={\bf D}-{\bf A}$ \cite{chung} where ${\bf D}$ is the diagonal matrix having as diagonal elements the degrees of the nodes. The Laplacian matrix is an operator that classically describes diffusion processes in a network. It is semi positive definite and in a connected network has a single null eigenvalue corresponding to an eigenvector taking the same value over all the nodes of the network. The Laplacian matrix can be normalized in different ways. Very widely used definition of the Laplacian for the classical random walk is $\hat{\bf L}={\bf I}-{\bf D}^{-1}{\bf A}$, where ${\bf I}$ indicates the identity matrix. This matrix is semi-positive definite and has real eigenvalues also if it is not symmetric.
Alternatively, 
the symmetric version of the graph Laplacian $\tilde{\bf L}={\bf D}^{-1/2}{\bf L}{\bf D}^{-1/2}$ is also widely used. This latter definition of the normalized Laplacian has the same spectrum as $\hat{\bf L}$ which is bounded by $2$.

\subsubsection{Network measures}
Network measures are observables that describe a given network structure locally or globally without providing the full information about all the interactions existing in a  network.
For streamlining the presentation, in this section we review only the most relevant network measures for simple networks (unweighted, undirected networks); the reader can refer to more extensive monographs 
on network theory for a full account of all network measures used in network theory.
The most coarse-grained properties of a network are the total number of nodes $N$ and the total number of links $L$. Although in a  network of $N$ nodes there are $N(N-1)/2$ possible connections, in a large variety of real systems the interesting scaling between the number of links and the number of nodes is linear, i.e. $L=O(N)$. These networks are also called {\em sparse networks}. In order to characterize the relation between the number of links and the number of nodes it is possible to use the density of links given by the ratio between the number of links and the number of nodes of the network.

Locally, one of the most important properties of a network are the node degrees that we have already introduced before, indicating how many links are incident to a node. From the full information about the degree of each node of the network, also called degree sequence, it is possible to extract the degree distribution $P(k)$ indicating the probability that a random node has degree $k$ or equivalently the fraction of nodes of degree $k$ in the network.
From the degree distribution one can extract the moments $\avg{k^n}=\mathbb{E}(k^n)$, including most relevantly the average degree $\avg{k}$ of the network and the second moment of the degree distribution $\avg{k^2}$. Note that $\avg{k}N=2L$, therefore in sparse networks the  average degree is asymptotically independent on the network size.
If we want to describe the neighbourhood of a node, not only the number of links incident to a node (the degree of a node) is very important, but also the density of triangles passing through a node is key to express how clustered is the neighbourhood. For instance, in a social network a node of high degree might have many friends that do not know each other or be part of a tight community of friends with high density of closed triangles. 

A very important measure to characterize the density of triangles around a node is the {\em local clustering coefficient} \cite{SW} that, providing that the node has degree greater than one, is given by the fraction among the total number of triangles passing through the node and the maximum possible number of triangles we could observe given the degree of the node.
Therefore the local clustering coefficient is a number between zero and one. The clustering coefficient  is zero if the node is not traversed by any triangle and is one if all the pairs of distinct neighbours of the node are connected by a link.
From the local clustering coefficient of all the nodes one can define  the {\em average clustering coefficient} performed over all the nodes of the network. The average clustering coefficient  provides an important measure to characterize the relevance of triangles in the network. An alternative measure of the density of the triangles in a network is the {\em transitivity} or {\em global clustering coefficient} of the network, given by a suitably normalized expression of the total number of triangles of  the network (see for instance example shown in Fig.~\ref{fig:networkbasics}).

Global network measures often are extracted from information about  the  shortest paths between the nodes of the network.
The paths between two nodes are alternating sequences of nodes and links going from a source node to a target node. The path length in unweighted networks is typically given to be the number of links traversed by the path. This leads to the definition of distance between two nodes as the smallest length of all the paths joining the two nodes. If two nodes are not connected by any path, the distance between them is by definition infinity. Note that although the distance between two nodes is uniquely defined,  there might be multiple shortest paths between two nodes.
Important global properties of a network are the network diameter given by the largest distance between any two nodes of the network, and the average shortest distance, given by the average distance among every distinct pair of nodes in the network. Naturally, the average shortest distance is equal or smaller than the diameter, where the equality holds only for fully connected networks, i.e. networks in which all pair of nodes are linked (at distance 1).
A network can be decomposed into different connected components, which are formed by  sets of connected nodes  such that there is no path connecting pairs of nodes belonging to different connected components. The connected component including a number of nodes of the same order of magnitude of the total number of nodes is called the {\em giant component}. Percolation is a critical phenomenon~\cite{dorogovtsev2008critical} that describes how a network responds to perturbation (damage of nodes or links). The order parameter of this critical phenomenon is  the size of the giant component (the number of nodes belonging to it) and the control parameter is the probability that a node (or a link) is damaged.

An important class of network measures are centrality measures \cite{newman2018networks} that try to quantify how important are nodes for a given network structure. Any centrality measure expresses and quantifies the importance of a node based on some criteria, therefore there is no centrality measure that is better than others in absolute terms, only centrality measures that work better than others for some specific tasks. The most simple centrality measure is the node degree, as nodes with large number of connections might be perceived in some cases to be more relevant (as the number of Facebook friends of a movie star). The  eigenvector centrality ranks the nodes according to the value of the largest eigenvector of the adjacency matrix, and it is based on the assumption that a node is important if many important nodes point to it. This basic idea is also central for the formulation of the Katz and PageRank centrality which however include additional elements. The Katz centrality guarantees that no nodes have zero centrality by assigning a minimal centrality to each node of the network. The PageRank centrality not only assigns a minimal centrality to each node of the network but also takes into account that high central nodes might have many connections, and their contribution to the centrality of the pointed nodes is often  normalized by the node degree. PageRank is among the most important algorithms of network science, and it is the original algorithm that ensured the success of Google with respect to previous search engines. PageRank centrality can be also interpreted as an algorithm that assigns to each node a centrality proportional to the steady state solution of a random walk that can hop from node to node via the links of the network and that sometimes makes a jump to random nodes of the network. Alternative notion of centralities are based on the hypothesis that nodes having small shortest distance with the other nodes of the network are central. This leads to the definition of the closeness centrality given by the inverse of the average shortest distance and the efficiency given by the sum of the inverse of the the shortest distance between each pair of nodes of the network. Finally, the betweenness centrality is high on links that bridge between different highly connected regions of the network.

\subsection{Random graphs}
Physicists have been familiar with lattices since the birth of crystallography. Lattices are regular graphs related to crystallographic symmetry groups. However from the Internet to the brain, complex networks have an important stochastic element.

The groundbreaking idea to consider graphs as the outcome of a stochastic process came  by the famous mathematicians P. Erd{\H o}s and A. R\'enyi which formulated in 1961 the random graph model also called as the Erd{\H o}s-R\'enyi model (ER model) \cite{ER}.
In the canonical version of this model (called $G(N,p)$ model) a random graph between $N$ nodes is generated by drawing each possible link of the graph with probability $p$. The corresponding microcanonical version  (called $G(N,L)$ model) instead consider any random graph of $N$ nodes and $L$ links with equal probability.

The formulation of these models is a very fundamental conceptual step forward in the study of networks, however random graphs are characterized by a very homogeneous properties, for instance in terms of the degree distribution, while as we will discuss in the next paragraph, complex networks are typically characterized by strong heterogeneity.

\subsection{Complex networks}
In network theory the complexity of a network is related to its heterogeneity. For instance a regular square lattice as well as a completely random network where each pair of nodes is connected with the same probability are not complex. Complexity is broadly speaking associated to network topologies that are not regular and therefore include some stochasticity, but they are not completely random either. In other words complex networks live in the wide region of possible topologies between completely regular networks and totally random graphs. Although the possible network topologies that interpolate between these two extremes are exponentially many in the number of nodes of the network, real systems have been shown to display common properties and to follow in {\em  network universality classes}.

{\em Small-world networks} \cite{SW} are networks in which the average shortest (hopping) distance between the nodes, or the diameter (i.e. the largest shortest distance between the nodes), is of the order of magnitude of the logarithm of the network size. In social networks the small world phenomenon is also known as the ``six degrees of separation of  social network", indicating that any two individuals in the world are  only few shaken hands apart in the social network of acquaintances. Interestingly, small-world networks usually combine their small diameter with a high-density of triangles measured by the clustering coefficient of the network. Indeed, the most simple and fundamental model of small world networks, the small-world network model, also known as Watts–Strogatz (WS) model \cite{SW}, rewires random links between the nodes of a 1-dimensional chain with links initially connecting nearest and next nearest nodes on the chain.
Therefore the small-world network model describes topologies that interpolate between randomness and order. Interestingly, while the network retains a significant local structure, adding random links with very low probability $p$ can significantly reduce the diameter of the network making it small-world.

A large variety of real networks display also a significant variability in the node's degree, where the degree of a node indicates the number of links incident it. While the degree $k$ of a node is a local property of the network, the degree distribution $P(k)$, indicating the probability that a random node has degree $k$, is a global property of the network. Therefore the degree distribution is an important property that is key to characterize 
different {network} universality classes. {\em Scale-free networks}~\cite{BA} are networks with degree distribution $P(k)$ decaying as a power-law  with power-law $\gamma\in (2,3]$ for large values of the degree $k$, i.e. $P(k)\simeq Ck^{-\gamma}$ for $k\gg1 $, where $C$ is a constant.
These networks have the important property that the second moment of the degree distribution $\langle k^2\rangle$ diverges as the network size goes to infinity even if the average degree $\langle k\rangle$ remains finite. Consequently even when the average degree is finite, it cannot serve as an internal scale because there are huge variations in the degrees of the nodes.
This phenomenon is due to the highly heterogeneous degree distribution and the significant statistical representation of {\em hub nodes}, i.e. nodes with a degree order of magnitude higher than the average degree. Scale-free networks define a very important universality class and they have been shown to modify significantly the phase diagram of classical critical phenomena including the Ising model, percolation and epidemic spreading \cite{dorogovtsev2008critical}. 

Generative models of scale-free networks can be classified in two class of models: non-equilibrium growing models, and maximum entropy (equilibrium) models. The most fundamental model for generating scale-free networks is the Barab\'asi-Albert network (BA) \cite{BA} which is a non-equilibrium model including just two simple elements: the growth of the network and preferential attachment, determining that new nodes are more likely to link to nodes that have high degree. In particular, the Barab\'asi-Albert model demonstrates that growth and linear preferential attachment (indicating that the probability that a new link connect to an existing node depends linearly on its degree) can generate scale-free models. Therefore the model has an explicative power of the basic mechanism responsible for the emergence of the scale-free distribution.
The maximum entropy models \cite{bianconi2007entropy,bianconi2009,anand2009} of scale-free networks are equilibrium network models. They do not aim at explaining mechanisms for the emergence of the scale-free universality class, rather they are ways to build maximally random networks with scale-free degree distribution that can be used as null models when studying real networks. Maximum entropy models include the configuration model  and the exponential random graphs. The configuration model generates maximum random networks with a given degree sequence determining the degree of each node of the network, this is a specific example of a microcanonical network ensemble \cite{anand2009} that enforces hard constraints. The exponential random graphs, also called canonical network ensembles~\cite{anand2009} generate random networks in which each node has a given expected degree, so from a network realization to another the degree of a given node can change, in this case we say that the model enforces soft constraints. Note that the maximum entropy models we have described can be used to generate network with any given degree distribution or expected degree distribution \cite{anand2009}. Therefore they can also be used to model networks that are not scale-free.

An ubiquitous property of real complex network is also their {\em modular structure} \cite{fortunato2010community}. A network is modular if it can be decomposed in {\em communities} of nodes more densely connected among themselves than with the rest of the network. Although the definition of communities evades mathematical rigour {\em community detection algorithms} are widely used to detect empirically the community structure of networks. Among the most popular community detection algorithms that are able to clusterize efficiently networks of very large network size, are the Louven algorithm \cite{blondel2008fast} based on maximization of the modularity \cite{newman2006modularity} (a measure of how modular or clustered is the network) and the INFOMAP algorithm \cite{rosvall2008maps} that clusterizes the network exploiting the information theory properties of (classical) random walks that are more likely to ``mingle" inside communities. Less computational efficient but very much used due to its  transparent  interpretation,   is the community detection algorithm that finds the hierarchical clustering of the network by iteratively removing links with high betweenness centrality, a network science measure that is higher on links that bridge across different communities. Strong modularity has been found to play a role in for example reservoir computing \cite{nakajima2019boosting,nokkala2021high,ma2023efficient}, a form of machine learning where a classical or a quantum system plays the role of a recurrent neural network.

The models and benchmarks that generate network with communities include the stochastic blockmodels, that partition the nodes in different classes and assigns the probability of links depending on the classes of the two connected nodes. A popular benchmark in this class is the Girvan-Newman~\cite{newman2004finding} benchmark having 4 classes of nodes such that links among nodes of the same class have a given probability and links among nodes of different classes have a smaller probability. Note however that the stochastic blockmodels include also  networks that have more general block structure such as bipartite networks where the link probability among nodes of the same class is zero while the probability of the link among nodes of different classes is different from zero, or networks with a non-trivial core-periphery structure. The stochastic block models have however the limitation that the degree distribution of the network is fairly homogeneous. The Lancichinetti-Radicchi-Fortunato (LRF) model \cite{lancichinetti2008benchmark} is an important benchmark that can instead be used when a non-trivial community structure coexists with very heterogeneous (scale-free) degree distribution of the network.

All these properties are fundamental properties of complex networks. When a new dataset is analysed, an important and useful tool is to characterize the complexity of network by comparing the chosen network observable with a null model~\cite{bianconi2007entropy,bianconi2009}. The most widely used null model is the Erd{\H o}s-R\'enyi (ER) model \cite{ER} of networks (discussed in the previous paragraph) that is constructed by linking every two nodes of the network with the same probability $p$. Clearly this model is not heterogeneous. Indeed,  since the links are placed totally randomly the model does not encode relevant information other than the average number of links. In order to compare a real network to a random ER network the network scientists compare a given observable, being the degree distribution, clustering coefficient, diameter or other network measure with the same observable in a random ER network with the same average number of links of the real network. 
In the relevant case in which the expected total number of links scales linearly with the number of nodes in the network, the degree distribution of the ER networks converges in the large network limit to a Poisson distribution, therefore these networks are also called Poisson networks. When the network is more dense the degree distribution is a binomial distribution.
As expected the  random ER networks have a degree distribution with a very well defined mean and standard deviation, and therefore the degree distribution is fairly homogeneous, which every node having the same expected average degree.
Poisson networks have a diameter that increase proportionally with the logarithm of the network size, i.e. they are small-world but they have a vanishing average clustering coefficient. In particular the expected number of triangles is finite and independent on the network size implying that the networks are locally tree-like.

\begin{figure}[t]
    \centering
    \includegraphics[trim=0cm 0.1cm 0cm 0cm,clip=true,width=0.95\textwidth]{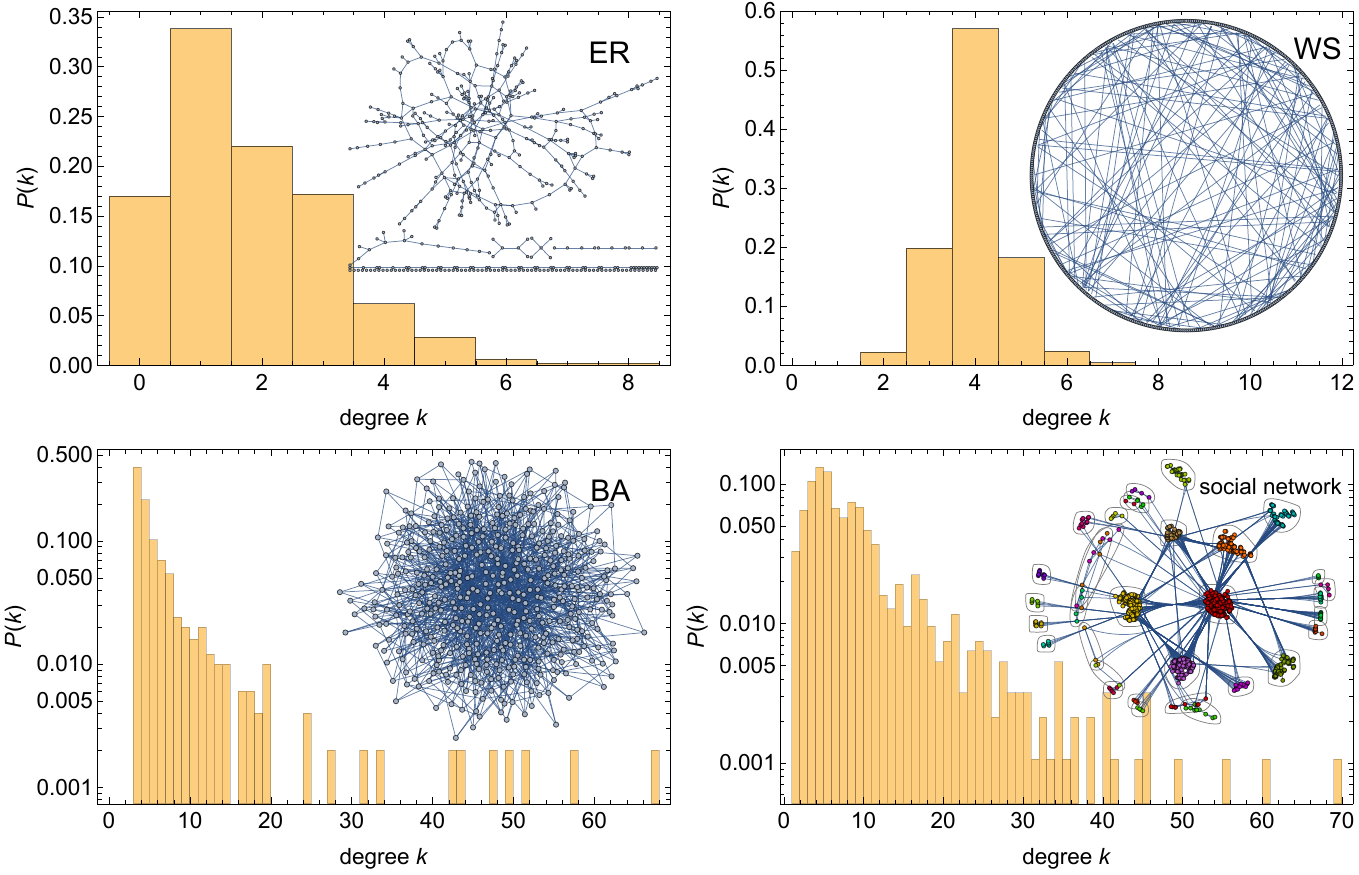}
    \caption{
    The degree distribution is plotted for few  paradigmatic examples of  random network models and for the largest component of the collaboration network in condensed matter from \cite{newman2001structure}. All random networks have $N=500$ nodes. In the Erd{\H o}s-R{\'e}nyi (ER) model each link of a completely connected network is chosen with the same probability, here $p=1.7/N$---equivalently, by percolation with a probability $1-p$. A giant component can be observed. Unlike ER, the Watts–Strogatz (WS) model can generate small-world networks. Here each node in the circular lattice was adjacent to both nearest and next nearest neighbors and the rewiring probability is $p=0.15$---i.e., the probability for each link to change one of its nodes to a random one such that links connect only distinct nodes and there are no duplicate links. Whereas the degree distributions of both ER and WS models are quite unrealistic, the Barab{\'a}si-Albert (BA) model can generate distributions with fat tails and relatively small average degree as might be expected from an empirical network. It lacks a community structure however, which is exemplified by the social network. Nodes are grouped and colored by communities found by a numerical optimization of modularity. 
    }
    \label{fig:randomandcomplexnetworks}
\end{figure}

Some of the previously discussed properties are illustrated in Fig.~\ref{fig:randomandcomplexnetworks}. In particular, the generated ER network has a giant component, the WS network the small-world property, the BA network a power-law degree distribution and the social network a community structure. As will be seen, all of these properties can matter also in various complex quantum networks.

 \subsection{Combinatorial graph theory}
Graph theory is deeply connected with optimization problems \cite{hartmann2006phase}. Graph theory, the mathematical theory of graphs is born with the Euler solution of the famous problem of the seven bridges of K\"onisberg in 1736. This optimization problem  requires to establish whether a network  admits a so called Eulerian cycle that starts from a node and goes back to the same node by traversing each link of the network exactly once. Such graphs are called Eulerian. 
Since then combinatorial graph theory has been a central subject of discrete mathematics. Another notable combinatorial problem is   the determination of whether graphs are Hamiltonian, i.e. they admit an Hamiltonian cycle that starts from a node and end on the same node by traversing each node only once. For instance if you want to place political delegates around a table for an official dinner you might wish to assign the positions around the table  such that neighbour delegates have good political relations (indicating the links of the network).
Interestingly, not only establishing if a network is Eulerian or Hamiltonian is of great interest for combinatorial graph theory but also finding Eulerian and Hamiltonian cycles turns out to be important in a number of combinatorial problems. However the Hamiltonian cycle  problem is NP-complete. 
Among the most important combinatorial problems on graph we mention the matching problem.
The matching problem consists in determining a subset of the links  of the graph (the set of matched links) such that each node is incident to at most one matched link. 
The maximum matching problem is the problem of identifying a matching that minimizes the number of unmatched links. This problem has wide applications in network theory, including most relevantly the recent results relating the maximum matching algorithm to control theory. In particular in Ref. \cite{liu2011controllability} it has been show that the  unmatched nodes of a optimal matching of a network are  the driver nodes of a linear control problem, i.e. they are the nodes to which we can apply external signals that have the ability to  drive the dynamical state of the network to any desired dynamical state. A perfect matching of a network is the matching in which all nodes are incident to exactly one matched link.

So far, Eulerian and Hamiltonian cycles have been found relevant, e.g., when proving certain formal properties of special resource states having a network structure \cite{dahlberg2020transform,dahlberg2020transforming}. Examples related to perfect matching will be given later in Sec.~\ref{sec:properties}.

\subsection{Generalized network structures}

Networks provide a very successful way for extracting information from complex interacting systems. However networks have also intrinsic limitations including the fact that they are not time-varying, the fact that they treat all the interactions on the same footing, and the fact that they only encode pairwise interactions.

In the last decade the network science community has made great progress in overcoming these limitations by  developing new tools and theoretical frameworks for generalized network structures including temporal networks \cite{lambiotte2019networks,holme2012temporal} which change in time, multiplex networks and multilayer network of  networks~\cite{boccaletti2014structure,bianconi2018multilayer}
 that can treat links of different types and higher-order networks  that can encode \cite{battiston2020networks,battiston2021physics,bianconi2021higher}
many-body interactions. 

Multiplex networks \cite{boccaletti2014structure,bianconi2018multilayer}
are a very important framework that allows to capture the multiplicity of types of interaction between a given set of nodes and can be represented by a vector of graphs $\vec{G}=(G^{[1]},G^{[2]}\ldots, G^{[M]})$, each graph describing the network of all the interactions of a given type exiting between the same set of nodes. Any given network $G^{[\alpha]}$ forms a {\em layer } of the multiplex network $\vec{G}$. 
For instance multiplex networks can be constructed by considering different measures of correlation existing between the same set of nodes, or multiplex networks can be used to represent interdependent communications infrastructures. Interestingly, when the layer of a multiplex network are interpreted as the snapshot of a network at a given timestep, the multiplex network (also called in this case multi-slice network) captures temporal networks that evolve in time.
Any multiplex network can be visualized as colored graph in which the same set of nodes is connected by network of different types (color) $G^{[\alpha]}$ with $\alpha$ indicating the color of the interaction, or as a layered structure in which any single node of the multiplex network admits a {\em replica node} in each layer. For instance, Oxford circus bus station and Oxford circus tube station in London are replica nodes of the multiplex (bus/tube) transportation network of London. Replica nodes can be connected to each other by {\em interlinks}.

A multiplex network is not just like a single larger network because it allows us to go beyond the framework of single networks and capture interactions of different types. This aspect of multiplex networks plays a crucial role both in the structure and in the dynamics defined in multiplex networks. The structure of multiplex networks in fact is significantly affected by important correlations, such as the connection of two nodes in more than one layer, called {\em link overlap},  that can be used to extract significant information from the multiplex network data (see for instance \cite{bianconi2013statistical,menichetti2014weighted}). Multiplexity plays also a fundamental role in dynamics as links in different layers and interlinks can be associated to different dynamical processes. In this respect we observe that when defining dynamics on multiplex networks, two main option exists: the first one is to associate a dynamics to each node that is unique,  the second is to associate a different dynamics to each replica node.  Interesting interdependency between the layer of a multiplex network can lead to avalanches of failure events triggering discontinuous percolation phase transitions \cite{buldyrev2010catastrophic}.

Higher-order networks \cite{battiston2020networks,battiston2021physics,bianconi2021higher} are generalized network structures that are fundamental to go beyond pairwise interactions. Higher-order network includes hypergraphs and simplicial complexes. Both types of structures can describe interacting system including higher-order interactions between two ore more nodes. Hypergraphs are formed by nodes and hyperedges with each hyperedge connecting two or more nodes. Simplicial complexes are formed by simplices, that are set of two or more nodes and their faces, where a face of a simplex $\alpha$ is any simplex  formed by a proper subset of the nodes of $\alpha$. The only difference between simplicial complexes and hypegraphs is that simplicial complexes are closed under the inclusion of the faces of their simplices. This comes with the great advantage that the algebraic topology and discrete geometry of simplicial complexes can be studied by algebraic topology \cite{bianconi2021higher} and discrete calculus. Topology is important to characterize the complexity  of the structure of higher-order network and in this respect there are important progress in persistent homology.  Interestingly topology is also of fundamental importance to capture the dynamics of topological signals, i.e. variable associated not only to nodes but also to links or triangles or higher-dimensional simplices. New results are showing that dynamics of topological signals might be key to unlock new higher-order  synchronization phenomena \cite{millan2020explosive,ghorbanchian2021higher} which affect the solenoidal and irrotational component of the dynamics in different ways.

\section{\label{sec:physical}Quantum dynamics in networks}

\subsection{Hamiltonians with a network structure}

Quantum dynamics and critical phenomena are strongly depended on the underlying network structure describing the physical interactions, usually taken to be pairwise. When the former is defined on finite dimensional lattices it is a classic topic on quantum mechanics and in this context it is widely known that quantum critical phenomena are strongly dependent on the lattice dimensionality \cite{sachdev2011quantum}.
Since lattices are nothing else than a special type of networks a very crucial question is whether quantum dynamics displays novel critical behaviour on complex networks strongly departing from lattices. These novel critical phenomena will then reveal a rich interplay between quantum dynamics and complex network topology in line to what happens in the classical domain where anomalous critical behaviour is found for instance for  percolation, Ising model and contact models defined on complex networks \cite{dorogovtsev2008critical}. 
{Even more interestingly in this Section, corresponding  to the network-generalized  block of Fig.~\ref{fig:intro}, we will show that the interplay between quantum dynamics and the underlying network structure can acquire very distinctive and exclusively quantum aspects.} 

Generally speaking, a multipartite quantum system can be dependent from a   graph $G$ describing their physical (pairwise) interactions when  the Hamiltonian $H$ is determined by $G$ and possibly some additional parameters, i.e.
\begin{equation}
    H=H(G,\ldots).
    \label{eq:H(G)}
\end{equation}
There are many examples of quantum systems whose dynamics is dictated by this type of  quantum Hamiltonians. These include networks of nanostructures \cite{mahler1998quantum}, networks of optical fibers \cite{perakis2014small} or waveguides \cite{gaio2019nanophotonic} and even electronic circuits treated in quantum formalism \cite{yurke1984quantum}. Note that  also circuits of quantum gates acting on a registry of qubits are sometimes called quantum networks \cite{deutsch1989quantum}. In this latter case $H$ is not time independent, consisting instead of gates acting on specific qubits at specific times, often involving   also measurements. However this type of quantum complex network can be cast into our classification, considering temporal networks of interacting quantum systems. As explained in Sec.~\ref{sec:properties}, a circuit may be used to prepare a cluster or a graph state \cite{hein2004multiparty} where the links indicate where the gates have acted on the qubits; alternatively, a network description may be assigned to the circuit itself.

Exploring quantum dynamics dictated by a quantum  Hamiltonian $H=H(G,\ldots)$ is fundamental to investigate the interaction between quantum dynamics and the underlying network structure of the interactions and is key to formulate design principles for observing new physics. In this case the network structure is designed and encoded in a Hamiltonian  of the general form of Eq.~\eqref{eq:H(G)}. 
Alternatively the interaction network in the Hamiltonian given by  Eq.~\eqref{eq:H(G)} can also be dictated by physics if such Hamiltonians arise naturally in experimental systems. In this context an important problem is how to infer the network of such interactions using for instance a quantum probe.

The quantumness of  the system  defined by $H=H(G,\ldots)$ depends on the form Hamiltonian and possibly other features such as the quantum states it describes. In this context of greatest interest are usually cases with behavior, properties or applications that go beyond what classical systems can emulate. The complexity of the system  on other hand, is a property of the network $G$. Of particular interest in quantum network context are cases where the latter can be linked to the former, e.g., when a network topology controls some property of interest such as the occurrence of a phase transition \cite{bianconi2012enhancement,bianconi2012superconductor,halu2012phase,bianconi2013superconductor,halu2013phase}, optimal transport \cite{mulken2016complex,maciel2020quantum}, optimal spatial search \cite{chakraborty2016spatial,li2017renormalization,chakraborty2020optimality} or spectral density \cite{nokkala2016complex}.

Often $G$ is taken to be  a weighted undirected network whose nodes are  the subsystems whose links are  the interaction terms, whereas the link weights correspond to the interaction strengths. Such systems are examples of quantum networks formed by interacting quantum systems. Given the Hamiltonian $H$ of such a network, the {topology} of the underlying graph $G$ is  {completely} determined.  
In the case in which one desires to design quantum Hamiltonian by changing the structure of the  networks $G$, clearly  full knowledge of the Hamiltonian $H$ and its parameters should be assumed. {When} $G$ {is partly or fully unknown} inferring its structure can be a challenging problem, however the network aspect can  be important in facilitating certain applications or in controlling the properties of interest, as will be seen in Sec.~\ref{sec:probing}.

Taking the network approach where $G$ and its properties are emphasized, we may ask for example under which condition and design principles  changing  $G$ will significantly change the physics or, alternatively leave the physics unchanged. In this Section we focus on a main research question of establishing which the network structures are particularly  suitable for certain applications or have the ability to exhibit particular  collective or critical behavior. 
Indeed $G$ is a purely classical object unlike $H$, which is why situations where the topology of $G$ controls some key property of the quantum system are of great interest. The network approach can be a powerful tool in such situations especially when $G$ is complex, which can be expected to lead to a nontrivial relationship between its structure and the quantum properties of the system. Ideally, considering a suitable $G$ reveals behavior which is not as readily discernible from $H$ alone. Even if $H$ is a chain as is often the case, there could be a basis change that transforms it into a complex and informative network. An example will be given later where the network approach is used to predict the phase of a spin chain by moving first to the configuration basis \cite{alet2018many}.

In the following we give illustrative examples of the research direction outlined above. The examples are not intended to be exhaustive, but rather useful to  further illustrate the previously presented concepts.

\subsection{Applications and examples}

\subsubsection{Phase transitions and collective phenomena}

Large physical systems can display different states of matter when a parameter is varied. For instance a superconductor can turn into a normal metal if the temperature is raised. In this case one can  observe that a property characteristic of a phase of matter (such as the superconducting gap) vanishes  when an external parameter is varied (in this case when the temperature is above the superconducting critical temperature). More in general such {\em phase transitions} can be controlled by an external parameter such as ambient temperature and pressure, but can also be observed in isolated systems. This latter situation occurs, for instance,  when an internal parameter controlling the system Hamiltonian is varied. In particular, quantum phase transitions take place at absolute zero \cite{vojta2003quantum,carr2010understanding} and consequently pertain to properties of the ground state, whereas in dynamical phase transitions the parameter is time \cite{heyl2018dynamical}. More generally, phase transitions in quantum systems are of great interest as a particular phase might have vanishing electrical resistance, witness unusually long survival of entanglement or control suitability to quantum information processing and machine learning tasks, as will be seen. Here we present some examples with a prominent network aspect. 

\begin{figure}[t]
    \centering
    \includegraphics[trim=0cm 0.15cm 0cm 0cm,clip=true,width=0.5\textwidth]{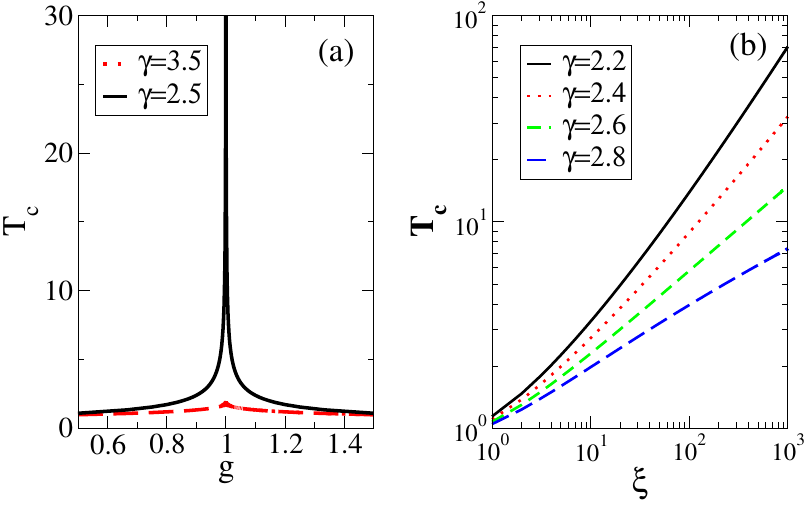}
    \caption{Behavior of the critical temperature $T_c$ as the expected degree distribution of Eq.~\eqref{eq:expecteddegrees} is varied in an annealed random transverse field Ising model with random onsite energies modelling the superconductor-insulator phase transition on a complex network. Left: external parameter $g$ controls the transition to a pure power-law via $\xi\propto |g/g_c-1|^{-1}$, where here we take $g_c=1$. Right: $\xi$ is varied directly. Reprinted figure with permission from \cite{bianconi2012superconductor}. Copyright 2012 by the American Physical Society.}
    \label{fig:highTc}
\end{figure}

A suitable network structure can be a resource for enhancing the critical temperature $T_c$ of the superconducting phase transition in the transverse field Ising model where the spins couple according to some graph $G$. Specifically,  spin systems interacting through a network $G$ whose degree distribution is a power-law with an tunable exponential cut-off  have been investigated in different settings \cite{bianconi2012superconductor,bianconi2012enhancement,bianconi2013superconductor}. In \cite{bianconi2012superconductor,bianconi2012enhancement} the network is generated by a canonical network ensemble (defined in Section III.D) in which each node $i$ has in expectation degree $\theta_i$. The expected degrees $\theta$  of $G$  are taken to be distributed as
\begin{equation}
    p(\theta)=\mathcal{N}\theta^{-\gamma} e^{-\theta/\xi}
    \label{eq:expecteddegrees}
\end{equation}
where $\mathcal{N}$ is a normalization constant and $\xi$ is controlled by external parameter. When the control parameter $\xi$ is infinite, the degree distribution becomes   a pure power-law. In \cite{bianconi2012superconductor} the topology of the network $G$  is dependent on  the  parameter $g$ controlling the transition to a pure scale-free network by modulating the parameter $\xi$ which obeys $\xi\propto |g/g_c-1|^{-1}$. When the   pure scale-free topology is achieved  ($g\to g_c$ and hence $\xi\to \infty$) , the critical temperature $T_c$  determined by the  largest adjacency eigenvalue of the network is maximized as can be seen from Fig.~\ref{fig:highTc}.  Hence this result provides a design principle based on complex networks, to enhance the critical temperature $T_c$ for the superconductor-insulator  phase transition.

A relevant question that arises is whether  Hamiltonian whose network of interactions is scale-free can be realized, and/or designed in specific experimental scenarios. In Ref. \cite{bianconi2013superconductor} it is shown that such scale-free network topologies can be  realized considering  as nodes of the networks $2D$ critical percolation clusters that are joined to each other if their boundary is closer than a threshold distance. This geometry has  important advantages for possible physical realization of these complex quantum networks. Interestingly  such geometry has been also recently adopted to propose new quantum communication algorithms on complex quantum networks in Ref. \cite{chepuri2023complex}. Networks with scale-free underlying $G$ have been analyzed also in the case of Bose-Hubbard \cite{halu2012phase} and Jaynes-Cummings-Hubbard Hamiltonians \cite{halu2013phase} with a Mott insulator or Mott-like phase and superfluid phase, linking in particular the scale-free regime and the maximum eigenvalue of the adjacency matrix to drastic changes in the phase diagram in the thermodynamic limit. For a Bose-Hubbard Hamiltonian, such a $G$ can cause the Mott insulator phase to disappear whereas for a Jaynes-Cummings-Hubbard Hamiltonian it may allow quantum phase transitions even with very weakly interacting optical cavities. Several other quantum critical phenomena have been shown to strongly depend on the complex network topology on which they are defined. Important effect of the interplay between network structure and quantum dynamics have been demonstrated for several other quantum phenomena including Bose-Einstein
condensation in heterogeneous networks \cite{burioni2001bose,burioni2000bose}, and Anderson localization on scale-free networks with increasing
clustering coefficient \cite{sade2005localization,jahnke2008wave}. 

More recently, the first experimental realization of an interdependent network has been reported and demonstrated to lead to novel phenomena \cite{bonamassa2023interdependent}. Generally speaking, such a network is a multilayer network where the layers are in general different networks that depend on each other. Here the layers consist of two disordered superconductors that can be modelled as $2D$ lattices of a type of Jospehson junctions. When uncoupled, the layers experience an independent and typical continuous transition to the superconducting phase as the temperature is lowered. In the interdependent configuration the networks are separated only by an insulating but thermally conducting film, which allows thermal links between the two layers, leading to a regime where the transition becomes abrupt with the critical temperature depending on the properties of both layers. A theoretical model was proposed which reproduced the experimentally observed behavior, suggesting the presence of cascading processes and an abrupt emergence of a giant superconducting component in the network.

In addition to the interest in considering complex networks topologies for the network of physical interactions $G$, important progress has also been recently made in studying fractal architectures \cite{kempkes2019design}.  It is well known that electrons in one dimension form a Luttinger liquid, and in two dimension  exhibit the quantum Hall effect. Exploring the  electron wavefunction on fractal network structures allows to investigate the effects of fractional dimensionality of the underlying lattice. In \cite{kempkes2019design}  the electron wavefunction defined on a artificial array of atoms forming a Sierpisky gasket is shown to inherit the fractional dimension of the fractal lattice. This opens the way for future studies investigating  spin-orbit interactions  and magnetic fields in non-integer dimensions. One open question in this context is whether  this research line could be related to the extensive literature on the non-trivial effect that network topology has on quantum dynamics \cite{pawela2015generalized,de2010bose,almeida2013quantum,xu2008coherent,de2009free,souza2007correlated} defined on (scale-free) 
Apollonian networks \cite{andrade2005apollonian},  which are known to be dual to Sierpinski gaskets.

Recently growing attention is addressed to  synchronization phase transitions and the role of the Kuramoto model \cite{kuramoto1975} and its quantum variations in quantum physics and condensed matter.
Synchronization \cite{arenas2008synchronization} is a  collective phenomena  occurring in network structures. In synchronization multiple oscillators associated to the nodes of the network, and often taken to have different intrinsic frequency,   are coupled to each other through the links of the network. When the coupling of the oscillators is strong enough, the oscillators  assume a common frequency giving rise to a dynamical yet ordered state. The Kuramoto model is the most important classical model displaying this phase transition. The model has been successfully used to describe arrays of coupled Josephson junctions \cite{wiesenfeld1998frequency} and  recently is gaining further attention for study of condensed matter phenomena such as persistent entanglement in isolated quantum systems, exciton delocalization in molecular aggregates, and tunneling of polarons in cuprate superconductors \cite{witthaut2017classical,scholes2020limits,velasco2021evolution}.

At the same time,  the literature is also providing several approaches to capture quantum synchronization dynamics.
In the quantum case few works consider synchronization between expected values of observables such as components of spins or quadratures of optical modes \cite{orth2010dynamics,mari2013measures,galve2017quantum}. Synchronization in quantum networks has mostly focused on networks of interacting quantum harmonic oscillators with a few notable exceptions such as \cite{li2017quantum,
he2019synchronization}. Although, nonlinear oscillators such as the  van der Pol oscillators exhibit richer behavior, the difficulty of solving the dynamics tends to limit the studies to very small systems \cite{lee2013quantum,lee2014entanglement,walter2015quantum}. In harmonic networks synchronization can arise when the network is in contact with a heat bath such that there is a normal mode that decays much more slowly than the others. Then all nodes overlapping with it will assume its frequency for a long transient \cite{manzano2013synchronization}, indicating also the presence of long lasting quantum correlations despite the contact with the bath. In principle, a normal mode can even be completely disconnected from the bath in which case synchronization could last perpetually. The prevalence of such decoherence free normal modes has been studied in \cite{cabot2018unveiling}. In a large network synchronization should also be possible in a small subgraph in the absence of a heat bath if the rest of the network can play the role of a finite environment. This has been confirmed in the minimal case of two oscillators interacting with a large but finite chain \cite{benedetti2016minimal}.

A very impactful, although more  mathematical, research direction has instead lead to formulate the Schr\"oedinger-Lohe synchronization model \cite{lohe2009non,lohe2010quantum} that provides a quantum non-Abelian extension of the classical Kuramoto model \cite{kuramoto1975}. In this model quantum states are distributed among linked  nodes by means of unitary transformations. The distributed states interact with each local state according to a time-dependent interaction Hamiltonian.  The system undergoes a phase transition in which, at sufficiently large coupling,  all qubits become spatially and temporally synchronized as revealed by numerical simulations performed on specific network structures.
Research in the field is growing, aimed at investigating different interesting aspects of the transition also if the model does not have up to now a clear experimental application.

\begin{figure}[t]
    \centering
    \includegraphics[trim=0cm 0cm 0cm 0cm,clip=true,width=0.95\textwidth]{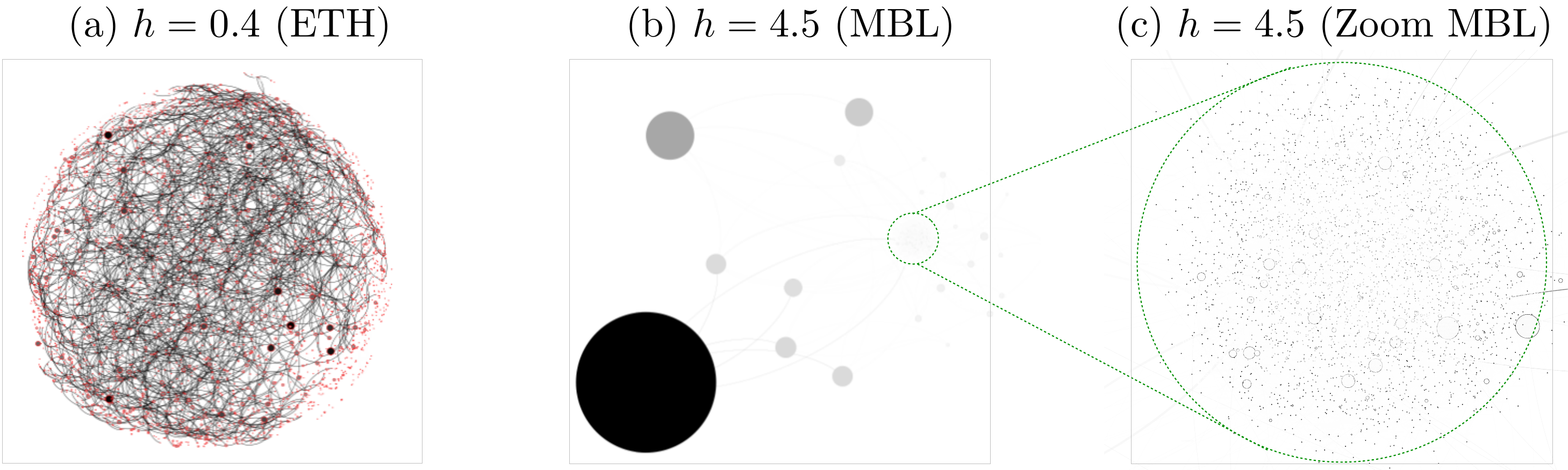}
    \caption{ETH and MBL phases of random-field Heisenberg chain in configuration space, with local fields $h_i$ uniformly distributed in $h_i\in[-h,h]$. The nodes of the network are configurations, weighted by their occupation probabilities (point size) and the links are possible transitions. In ETH phase shown in $(a)$ the probabilities are roughly uniform, as expected. In MBL phase of $(b)$ and $(c)$ the network changes drastically with nearly all probability concentrated on just a few configurations. Figure reproduced from Ref.~\cite{alet2018many}, doi: \text{https://doi.org/10.1016/j.crhy.2018.03.003}, license: \text{https://creativecommons.org/licenses/by-nc-nd/4.0/}}
    \label{fig:MBL}
\end{figure}

Introducing disorder in the form of random local terms in the Hamiltonian can lead to new interesting phenomena. In particular, isolated systems with both disorder and interactions can be in either thermalizing or localized phases \cite{abanin2017recent,alet2018many}, depending on  disorder strength. In a so-called quench experiment such a system is initially prepared into some state $\Ket{\Psi(0)}$ and then allowed to evolve according to its unitary dynamics for a time $t$, reaching the state $\Ket{\Psi(t)}=e^{-i H t}\Ket{\Psi(0)}$. Although the evolution is reversible, according to the eigenstate thermalization hypothesis (ETH) it should hold for any local few-body observable $O$ that $\langle O(t\rightarrow\infty)\rangle\simeq\Bar{O}(E_0)$, where $E_0=\Bra{\Psi(0)}H\Ket{\Psi(0)}$ is the initial energy and $\Bar{O}(E_0)$ the corresponding thermal expectation value. In other words, the local states should become approximately thermal even though $\Ket{\Psi(t)}$ remains pure for any $t$, and this should hold for any $\Ket{\Psi(0)}$. This self-thermalizing phase is characterized also by efficient transport of energy and fast propagation of correlations; intuitively, each local observable $O$ is then able to thermalize by using the rest of the system as a finite environment. In practice, ETH is observed already in spin systems small enough to be amenable to numerical simulations when $E_0$ is sufficiently far from an extremal value. The alternative is many-body localization (MBL) phase, characterized by frozen transport and slow propagation of correlations where typically the limit $\langle O(t\rightarrow\infty)\rangle$ still exists but is sensitive to $\Ket{\Psi(0)}$ and is therefore different from $\Bar{O}(E_0)$. The difference between the phases becomes apparent in the configuration basis where instead of interacting systems one considers a single particle hopping from site to site, in analogy with continuous time quantum walks (see below). In this basis the nodes are configurations and links transitions between them, and the nodes may be weighted by their occupation probabilities. In ETH phase the nodes have similar weights as the system explores all configurations allowed by the global conservation laws; this is why ETH phase is also called the ergodic phase. MBL phase leads to a dramatically different network with the bulk of occupation probabilities concentrated on only a few nodes with the rest of them having negligible weights, as seen in Fig.~\ref{fig:MBL}. To give some examples of the implications, MBL phase has been proposed to be useful for protecting quantum features from decoherence \cite{huse2013localization,bahri2015localization} whereas the ETH phase might be better for unconventional computing \cite{martinez2021dynamical} or quantum annealing \cite{laumann2015quantum}.

The previous example of having to consider a network different from the immediate one to make the network approach useful is not isolated. In fact, modifications of the interaction network that according to classical intuition should be drastic might not change the point where a transition happens at all. This was observed in the case of the transverse field Ising chain at ground state \cite{ostilli2020absence}; adding enough random links to give the new network the small world property was found to have no effect on the transition point. It is however possible to predict phase transitions in the chain with state-of-the-art accuracy by considering networks derived from its ground and thermal states, as will be seen in Sec.~\ref{sec:properties}.

\subsubsection{\label{sec:dynamics_walkers}Walkers and search algorithms}

Classical random walks are stochastic processes where a walker moves in a discrete space. For example for a classical walker in a $d$-dimensional lattice or a graph, the possible moves depend on the current location and their probabilities can vary \cite{masuda2017random}. There is an enormous amount of work concerning their quantum counterparts. Quantum walks~\cite{aharonov2001quantum,kempe2003quantum} are of great interest because they can model both analog systems capable of universal quantum computing \cite{childs2009universal,lovett2010universal,underwood2010universal} and transport of excitations \cite{engel2007evidence,collini2010coherently} or quantum information \cite{bose2003quantum,bose2007quantum} in networks of interacting systems, yet are experimentally convenient as they focus on cases where both the interaction terms and the systems are of the same type. Furthermore, comparing and contrasting classical and quantum walks can deepen our understanding of different facets of quantumness \cite{aharonov1993davidovich,watrous2001quantum,kendon2006random,faccin2013degree} as well as identify situations where there is a possibility for a quantum advantage \cite{childs2003exponential,childs2004spatial}that can provide significant speedups in quantum computation \cite{montanaro2016quantum}. Many excellent in-depth reviews concerning quantum walks are available such as \cite{mulken2011continuous,venegas2012quantum,portugal2013quantum,kadian2021quantum,kempe2003quantum}. Here we highlight a small amount of relevant works from the complex quantum networks perspective, placing them in a wider context. Although many types exist, we focus on so called continuous time quantum walks (CTQW) introduced in the late 1990s \cite{farhi1998quantum} due to the elegant and natural way they generalize to complex networks.

\begin{figure}[t]
    \centering
    \includegraphics[trim=0cm 0cm 0cm 0cm,clip=true,width=0.95\textwidth]{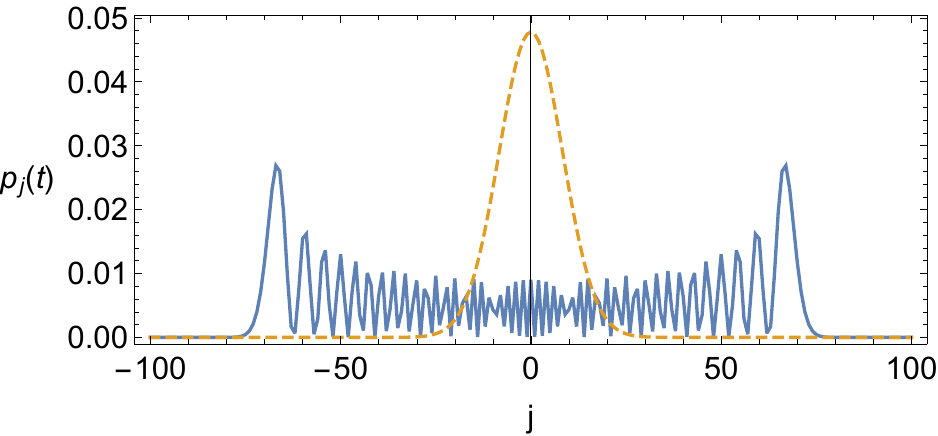}
    \caption{Comparison of classical (dashed line) and quantum (solid line) walks. The walker is localized in the center of the path graph at $t=0$ and at $t=35$ the probabilities are as shown. In the quantum case the evolution obeys Eq.~\eqref{eq:CTQW} where $H=\mathbf{L}$, i.e. the Laplace matrix of the graph. In the classical case $H=-\mathbf{L}$ and the equation is also modified as specified in the main text.}
    \label{fig:CTQW}
\end{figure}

In such walks the network is typically encoded into the Hamiltonian $H$, namely it is taken to be directly proportional to some matrix representation of the network, such as Laplace matrix, adjacency matrix or normalized Laplace matrix \cite{faccin2013degree}. The Hamiltonian acts in an $N$-dimensional Hilbert space, where $N$ is the size of the network. An orthonormal basis is fixed, consisting of states $\Ket{j}$ such that $\sum_j\Ket{j}\!\Bra{j}=\mathbf{I}$, $\langle k\vert j\rangle=\delta_{kj}$. Now a pure state of the walker at time $t\in\mathbb{R}$ reads $\Ket{\psi(t)}=\sum_j q_j(t)\Ket{j}$ where $q_j(t)=\langle j\vert \psi(t)\rangle$ is a complex probability amplitude and $p_j(t)=|q_j(t)|^2\in[0,1]$ is interpreted as the probability that the walker is at network node $j$ at time $t$. The probability amplitudes evolve according to the Schr{\"o}dinger equation as
\begin{equation}
    \mathrm{i}\frac{\mathrm{d}}{\mathrm{d}t}q_j(t)=\sum_kH_{jk}q_k(t),
\label{eq:CTQW}
\end{equation}
where $H\propto\mathbf{M}_G$ and natural units are used such that $\hbar=1$. When $\mathbf{M}_G$ is the Laplace matrix the walker dynamics can be readily compared to continuous time classical random walk by omitting the imaginary unit $\mathrm{i}$ and replacing the probability amplitudes $q_j(t)\in\mathbb{C}$ by probabilities $p_j(t)\in[0,1]$. With these changes the equations of motion describe diffusive spreading over the network \cite{van1992stochastic}. In particular, it can be shown that if the network is connected the long time limit in this case is $p_j(t)=1/N$ for all nodes, independently of the structure of the network. At variance, Eq.~\eqref{eq:CTQW} describes reversible dynamics which rules out a unique long time limit, but one can consider the long-time average distribution. In general, the time-averaged probability distribution gets close to it after a time known as mixing time which has been recently upper bounded for generic graphs when $\mathbf{M}_G$ is the adjacency matrix \cite{chakraborty2020fast,chakraborty2020analog}, revealing that quantum walks typically take longer to mix than classical walks. Furthermore, unlike the probabilities the amplitudes are subject to interference effects which can lead to ballistic instead of diffusive spread \cite{aharonov1993davidovich} as demonstrated in Fig.~\ref{fig:CTQW}. Initially localized in the center of a path graph, the classical walker is likely to be still near the center at a later time unlike the quantum walker.

Fundamental research on CTQW on  complex networks has considered the interplay between transport efficiency and network structure. Sequentially growing networks were considered in \cite{mulken2016complex} where it was found how the mesostructure of these networks affects the global transport efficiency and how changing it can induce the transition to optimal transport. Transport efficiency has been considered also in the case of other types of networks such as scale-free \cite{xu2008coherent,maciel2020quantum}, small-world \cite{quintanilla2007electron,mulken2007quantum,perakis2014small,wang2016quantum} and Apollonian networks \cite{xu2008coherentAp,almeida2013quantum}. Fundamental research has also addressed questions about the difference between classical and quantum random walks \cite{faccin2013degree,gualtieri2020quantum} and provided, among the other results,  a CTQW based method for community detection \cite{faccin2014community} or centrality measure \cite{izaac2017centrality} specifically for quantum networks (see  discussion in Section \ref{sec:propertiesII}). A related research avenue considers mixing continuous time classical and quantum walks and asks what is the optimal ratio and how this depends on the topology; more formally, this amounts to introducing some irreversibility to the dynamics as in quantum stochastic walks introduced in \cite{whitfield2010quantum}. If the initial state of the walker is $\rho$, then
\begin{equation}
    \frac{\mathrm{d}\rho}{\mathrm{d}t}=-(1-p)\mathrm{i}\left[H,\rho\right]+p\sum_{i,j}\left(L_{ij}\rho L_{ij}^\dagger-\frac{1}{2}L_{ij}^\dagger L_{ij}\rho-\frac{1}{2}\rho L_{ij}^\dagger L_{ij}\right)
    \label{eq:stochasticCTQW}
\end{equation}
where one may recognize a convex combination of unitary dynamics given by the commutator---essentially Eq.~\eqref{eq:CTQW} in a different form---and simple Markovian dissipation, as controlled by $p\in[0,1]$. The dissipators $L_{ij}$, accounting for irreversibility, are chosen such that one recovers the classical case at the limit $p=1$. It has been suggested that as a rule of thumb, some classicality can be expected to lead to better transport than the fully quantum case \cite{caruso2014universally}.

Quantum walks can also be viewed as a resource when they are used to implement various algorithms. A prime example is spatial search via CTQW \cite{childs2004spatial}, where the initial state of the walker is typically the equally distributed superposition state $q_j(t)=1/\sqrt{N}$ for all $j$ and the objective is to engineer the dynamics such that $p_w(t)$ for some marked node $w$ rapidly approaches unity, which is taken to indicate that the marked node has been found. To this end the total Hamiltonian is taken to be
\begin{equation}
    H=-\gamma\mathbf{M}_G-\Ket{w}\!\Bra{w}
\end{equation}
where an oracle term $H_w\propto\Ket{w}\!\Bra{w}$ is added to the network Hamiltonian and the uniform link weights are tuned via the real number $\gamma$. The performance of spatial search has been recently investigated in Erd{\H o}s-R\'{e}nyi networks \cite{chakraborty2016spatial} as well as networks characterized by a finite spectral dimension \cite{li2017renormalization}. Steps towards necessary and sufficient conditions for a graph to provide optimal spatial search were taken in \cite{chakraborty2020optimality} and \cite{novo2015systematic,razzoli2022universality} where  the spectral properties of the network and a dimensionality reduction method were leveraged to reach the main conclusions, respectively. Taken together, the results suggest that spatial search and similar algorithms originally proposed for completely connected networks or lattices may continue to work well also in complex networks. CTQW in general and search algorithms in particular are also related to state transfer where both the initial state and the desired final state are localized \cite{kay2011basics,godsil2012state,nikolopoulos2014quantum}. Quantum walks also serve as the basis for several algorithms for network inference, as discussed in more detail in Sec.~\ref{sec:propertiesII}. Here we briefly mention link prediction based on CTQW \cite{moutinho2021quantum} and ranking the nodes of a network based on final occupation probabilities of a quantum stochastic walk \cite{sanchez2012quantum}. 

On a related note, one may consider the complexity of simulating the CTQW itself on a universal quantum computer. It is the case that common algorithms become inefficient in complex networks with hubs \cite{moutinho2022complexity}, however an algorithm for simulating hub sparse networks has been recently proposed \cite{magano2023quantum} as a step towards exploring whether quantum computers can have an advantage in simulating dynamics on complex networks.

There is a large body of research dealing with networks with a predetermined structure in the context of excitation transfer covering notably light harvesting complexes.  Since the networks are typically rather small this line of research is not discussed further here however we suggest to the interested reader Ref.~\cite{plenio2008dephasing} and the articles citing it. Recently larger and more complex networks have appeared in proposals to model quantum dot systems however \cite{cuadra2021modeling,cuadra2021approaching}, where the transport efficiency of such networks is linked to the network structure.

\subsubsection{Structured environments and probing\label{sec:probing}}

As explained in Sec.~\ref{sec:qtheory}, there are fundamental differences between the dynamics of closed and open quantum systems. The dynamics of the former is unitary, which implies reversibility---the information of the initial conditions is always in principle recoverable. Under certain mild conditions the system should also eventually return to a state close to the initial one, although usually this recurrence time is short enough to be of practical relevance only for very small systems \cite{rauer2018recurrences}. For instance,  open systems immersed in a heat bath can undergo irreversible dynamics where quantum information is permanently lost to the environment. The theory of open quantum systems aims to capture the reduced dynamics of the open system in terms of a few relevant quantities describing the environment, which often requires approximations. An alternative is to replace the bath with a finite network, which may allow the study of exactly solvable models mimicking an infinite environment, or facilitate the engineering of highly structured environments leading to interesting phenomena for the open system such as non-Markovianity of its dynamics, i.e. memory effects where some information originally from the system is temporarily recovered. From this starting point one can also investigate what can be deduced of the network from the reduced dynamics, or attempt to control or harness the network via local manipulation of the open system to generate, e.g., entanglement. More generally, open system dynamics can also be harnessed to be a resource for, e.g., quantum computing \cite{ghosh2021realising}.

\begin{figure}[t]
    \centering
    \includegraphics[trim=1cm 0cm 0cm 0cm,clip=true,width=0.75\textwidth]{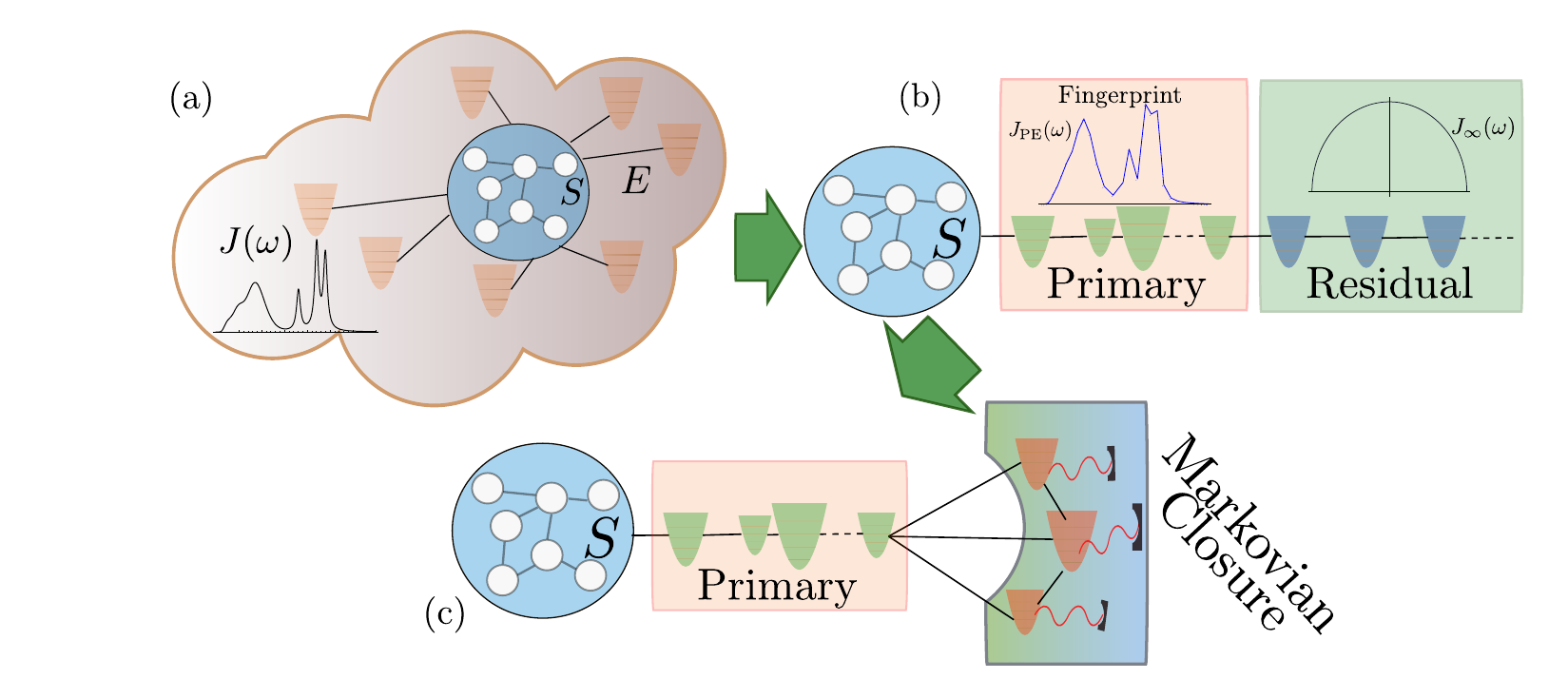}
    \caption{An infinite environment $E$ of an open system $S$ with a continuous spectral density $J(\omega)$ is shown in $(a)$. It can be mapped into a semi-infinite chain tending to a universal tail. Typically one may use the exact coefficients only for the first few oscillators and the asymptotic values for the rest to a good approximation, which in $(b)$ are the primary and residual parts, respectively. Finally, the tail may be replaced with only a finite number of damped oscillators as shown in $(c)$. Reprinted figure with permission from \cite{nusseler2022fingerprint}. Copyright 2022 by the American Physical Society.}
    \label{fig:closure}
\end{figure}

A typical environment is a heat bath consisting of a continuum of unit mass harmonic modes, characterized by its temperature $T$ and the spectral density $J(\omega)$ of environmental couplings, defined as
\begin{equation}
    J(\omega)=\frac{\pi}{2}\sum_i\frac{g_i^2}{\Omega_i}\delta(\omega-\Omega_i)
    \label{eq:spectraldensity}
\end{equation}
where $\delta$ is the Dirac delta function. It encodes the relevant information in environmental modes with frequencies $\Omega_i$, interacting with the open system with coupling strength $g_i$, into a single function of frequency. A given $J(\omega)$ can be discretized to arrive at a finite collection of harmonic modes interacting only with the open system but not with each other, which should mimic the original infinite bath up to some maximum interaction time. Intuitively, the open system cannot resolve the frequencies for sufficiently short times and therefore finite size effects should be negligible in this regime. In \cite{vasile2014spectral}, both discretized and engineered spectral densities arising from finite oscillator chains with tuned nearest neighbor couplings were considered to study the interplay between $J(\omega)$ and non-Markovianity; in the latter case the system was coupled to the first oscillator only. Ref. \cite{guarnieri2016energy} considered the discretized case to study non-Markovianity in strongly interacting systems. To study long time dynamics in this way the network size must be increased which can eventually become a limiting factor. It can be shown that a given $J(\omega)$ can be realized by a tuned chain that ends in a one-way energy and information sink, formally called a Markovian closure, which can be realized by a finite number of damped oscillators with nearest neighbor couplings that undergo relatively simple open system dynamics \cite{nusseler2022fingerprint}, as shown in Fig.~\ref{fig:closure}. Then even complicated dynamics of the open system can be simulated by the system interacting with the first oscillator of a typically short chain which ends in the closure. For any $J(\omega)$ the couplings in the chain tend to a constant value; the chain is truncated and the tail is replaced by the sink.

Going beyond chains, one may consider for example the interplay between $J(\omega)$ and the network structure. A gapped $J(\omega)$ can be created with periodic coupling strengths in a chain. In \cite{nokkala2016complex} it was demonstrated how not only the number of bands could be easily engineered, but also how a similar controllable effect could be achieved by adding to a homogeneous chain just one extra link. The connection between the topology of a random oscillator network and non-Markovianity of the open system dynamics was investigated in \cite{nokkala2017non}, where it was discovered that the latter was affected both by disorder and link density, which increased and decreased non-Markovianity, respectively. The problem of experimental realization of random quantum harmonic oscillator networks has been recently solved in a multimode quantum optics platform \cite{nokkala2018reconfigurable}, where a shaped pulse train pumps an optical parametric process, creating squeezed modes, which can then be measured in a suitable basis to complete the mapping of the network dynamics to that of the optical modes. This has already been used to experimentally realize non-Markovian open system dynamics \cite{renault2021spectral,renault2023experimental}.

Increasing instead the number of open systems facilitates the investigation of using the network as a resource. For example, entanglement generation can be achieved by tuning just the interaction between the systems and a generic oscillator network, as shown in \cite{manzano2013synchronization} where the network itself was also open. Very recently this has been extended to collisions where a series of systems that do not interact with each other collide with a fixed random oscillator network, one by one; by tuning properly the interaction Hamiltonian describing the collisions entanglement can be induced between either consecutive systems or between more distant systems, as shown in \cite{nokkala2023online} where this was called the entangler task. Furthermore, a judicious choice of interaction between a network and several open systems not directly interacting with each other has been shown to be able to realize a universal set of quantum gates on the systems, all the way to complicated gates equivalent to quantum circuits \cite{ghosh2021realising}, potentially leading to very compact quantum computing. The downside for both the entangler and circuit realization is the finding of a suitable interaction Hamiltonian, which appears to be difficult. Closely related to this is the task of realizing quantum computation with the full network but only by local manipulation of a small part of it, which was shown to be possible in the case of a spin chain with nearest neighbor couplings by manipulating its first two spins \cite{burgarth2010scalable}; more broadly, one can investigate controlling the network via such local manipulation, discussed for example in \cite{maruyama2016gateway}. At this point we also mention quantum neural networks \cite{killoran2019continuous,beer2020training,abbas2021power} which are often realized by a circuit acting on a registry of qubits \cite{kwak2021quantum}. As there typically is no graph involved they are not discussed further here.

All previous discussion considers the case where the network Hamiltonian is given; in fact for many of the described applications this knowledge is necessary. In case the Hamiltonian is unknown one may consider various probing schemes. Whereas some are general, others assume specifically that the Hamiltonian has a network structure and exploit this. For instance, this could in practice mean assuming that the network topology is known but the parameters such as the  coupling strengths are not. Alternative this might imply assuming  that the relation between $H$ and $G$ is of a specific nature. 

\begin{figure}[t]
    \centering
    \includegraphics[trim=0cm 0cm 0cm 0cm,clip=true,width=0.95\textwidth]{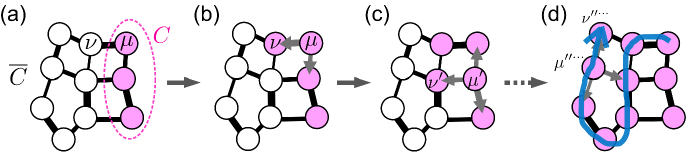}
    \caption{Example of graph infection. Originally the set $C$ is infected and its complement $\overline{C}$ is healthy $(a)$. Because $\nu$ is the only healthy neighbor of $\mu$, it gets infected $(b)$. This makes $\nu'$ the only healhty neighbor of the infected node $\mu'$, so it gets infected as well $(c)$. As eventually the entire graph gets infected $(d)$, the original set $C$ is infecting. Figure reproduced from Ref.~\cite{burgarth2009indirect}, doi: \text{https://doi.org/10.1088/1367-2630/11/10/103019}, license: \text{https://creativecommons.org/licenses/by/4.0/}.}
    \label{fig:infection}
\end{figure}

Multiple works have considered estimating the parameters when the topology is known. For example, the case of spin networks with ferromagnetic interactions in an inhomogeneous magnetic field was considered in \cite{burgarth2009indirect} where it was shown that the coupling and field strengths could be probed via state tomography of any infecting subset of spins. This purely topological condition requires that an "infection" spreads from the set to the entire network by the following rule: an infected node can infect its uninfected neighbor iff it is the only uninfected neighbor, but multiple infection rounds are allowed. As a simple example, either the first or the last node of a chain will suffice, but none of the middle nodes can by themselves infect the chain. Informally speaking, an infecting set is something akin to a surface of the network and can be expected to be small if the interactions are in some sense short range, constraining in particular the node degrees, as shown in the example of Fig.~\ref{fig:infection}. This was later generalized \cite{burgarth2011indirect} to quadratic Hamiltonians of the general form\begin{equation}
    H\propto \alpha^\dagger\mathbf{M}\alpha,
\end{equation}
where the vector $\alpha$ consists of annihiliation and creation operators and the matrix $\mathbf{M}$ is Hermitian. Quantum estimation theory may be used to rigorously compare different schemes and search for the optimal measurement and has been used to analyze different ways to probe the constant coupling strength in linear qubit chains \cite{tamascelli2016characterization} and the constant tunneling amplitude in Hamiltonians describing CTQW when G is known in the case of several common families of graphs \cite{seveso2019walker}. If both the topology and the parameters are known, one can consider probing an unknown network state. This has been done for quadratic oscillator networks both with qubit \cite{tufarelli2012reconstructing} and optomechanical probes \cite{moore2016quantum}. In both cases it suffices to couple the probe to only a single network node, whereas knowledge of the network Hamiltonian is used to find the correct interaction strength profile $g(t)$ between the probe and the node to encode information about the network state into that of the probe.

The case where an unknown structure is probed has been considered both for networks of interacting spins \cite{kato2014structure} and oscillators \cite{nokkala2016complex} and appears to be fairly difficult if $G$ is not constrained. In fact, there are indications that even the easier problem of testing whether two given oscillator networks have the same $G$ up to isomorphism can, in the worst case scenario, be nearly as expensive as probing the entire structure (see, e.g., Sec.~5.3 of \cite{nokkala2018quantum}). Instead of the full structure one may settle for the spectral density $J(\omega)$ of an oscillator network \cite{nokkala2016complex,giorgi2016probing}; while it can be done by coupling the open system acting as the probe to any single network node, each choice has its own corresponding $J(\omega)$. An interesting alternative especially from a quantum networks point of view is the probing of some mesoscopic quantity of $G$ with minimal or limited access to the network, i.e. the ability to couple the probe to only one or few network nodes. Examples include deducing which random graph distribution $G$ belongs to from the behavior of entropy of entanglement of the probe when varying the number of links between it and the network \cite{cardillo2013information}, estimating the degree distribution and constant coupling strength with minimal access by exploiting results from spectral graph theory \cite{nokkala2018local}, and deducing the spectral dimension of the network by probing the frequencies of a subset of the normal modes \cite{nokkala2020probing}.

\subsection{Avenues for further research}

There are multiple ways to go beyond the examples highlighted previously. These include at least generalizing what kind of graph $G$ is or considering novel {encoding rules} of the general form of Eq.~\eqref{eq:H(G)}, taking the presented applications further or pursuing new ones, as well as searching for new opportunities for cross-disciplinary research.

Going beyond undirected simple weighted $G$ has been already considered especially in CTQW, but has recently been done in networks of superconductors as well \cite{bonamassa2023interdependent}. In so called chiral quantum walks some directional bias is introduced by augmenting the link weights with complex phases of the form $e^{\mathrm{i\phi}}$; this is fine as long as $H$ remains Hermitian. Such walks were presented already in 2013 in \cite{zimboras2013quantum} where it was shown how the effect is topology dependent; in some cases transport is unaffected but otherwise effects can range from bias towards a preferred arm in a three-way junction to suppression or enhancement of transport. State transfer in such graphs was considered in \cite{cameron2014universal} the following year. In 2016 some further classification of topologies based on the impact of these phases was carried out in \cite{lu2016chiral} and the case of bipartite graphs specifically was considered in \cite{todtli2016continuous}. 

Chiral quantum walks have seen some renewed interest very recently. Reference~\cite{frigerio2021generalized} proposed that a classical random walk has infinitely many chiral quantum counterparts whereas Ref.~\cite{frigerio2022quantum} considered optimizing the advantage over classical walkers by tuning the phases, which again was found to be topology dependent as in, e.g., even cycles the optimal solution had none, but in some other cases they were found to be beneficial. For example, disordered phases were found to be able to facilitate transport in cases that otherwise would have seen the walker stay near the initial site; a similar result was reported in \cite{frigerio2023swift}. In a related work machine learning was used to study when a chiral walk can beat CTQW and it was found to almost always do so in, e.g., hyper-cube graphs \cite{kryukov2022supervised}. Experimental implementations have been reported both in a special class of quantum circuits \cite{lu2016chiral} and in Floquet systems (experimentally convenient Hamiltonians under periodic driving) \cite{novo2021floquet}. 

Recently CTQW has been considered also in temporal graphs. Ref.~\cite{chakraborty2017optimal} considered the conditions for optimality of spatial search in such a setting. The case where topology is fixed but link weights can randomly alternate between two different values was considered in \cite{benedetti2019continuous}. When also loops, or self-links are present, CTQW in a temporal graph can be used to realize efficient universal quantum computation \cite{herrman2019continuous}; loops affect the evolution of the amplitudes and in particular all isolated nodes were taken to have loops. This framework was revisited and improved in \cite{wong2019isolated} which showed that by allowing also isolated nodes without loops the construction of gates from a universal gate could be further simplified. Yet further possibilities include considering transport in multiplex  \cite{mulken2016enhanced} or fractal graphs \cite{darazs2014transport,xu2021quantum}. Spatial search on fractal graphs has also been considered using a so-called flip-flop quantum walk \cite{patel2012search,tamegai2018spatial,sato2020scaling}.

There is a host of phenomena, such as super- and subradiance \cite{bellomo2017quantum}, that have so far been considered in cases where $G$ is not very interesting from network theory point of view, leaving open the chance to go further. As an example of new applications we mention quantum reservoir computing \cite{mujal2021opportunities} which aims to harness the response of a driven quantum system to solve machine learning tasks; since many works consider completely connected $G$ with random weights the network aspect has not been very prominent so far, although for example the ETH/MBL transition has already been connected to performance \cite{martinez2021dynamical} and there are indications that at least for specific tasks a strong community structure can be beneficial \cite{nakajima2019boosting,nokkala2021high}.

Speaking of ETH/MBL, despite the recent progress the theoretical description of the transition is still lacking. It might be wondered if working in the configuration space instead could pave the way towards it, especially in the light of recent success of applying network theory to correlation networks of spin chains to predict their phase transitions, as explained in more detail in the next Section. A related research avenue could be to consider probing the partial, mesoscopic structure of a network in the ETH phase along the lines of, e.g., \cite{garrison2018does,qi2019determining}.

\section{\label{sec:properties}Network representation of quantum systems}

\subsection{Networks for taming quantum complexity}

Recently it has emerged that networks are a very powerful mathematical and computational  tool to tame quantum complexity. Therefore in this Section we  shift  perspective with respect to the previous Section. Indeed, while in the previous Section networks have been used to encode for the physical interactions of  quantum systems, here  networks are adopted as their  mathematical  and abstract representations. This research line corresponds to the quantum-applied  block of Fig.~\ref{fig:intro}. The works summarized in the Section generally assume that  a quantum system can be represented as a network according to a  suitable rule of the form
\begin{equation}
    G=G(\mathbf{M},\mathbf{H},\mathbf{t},\boldsymbol{\rho}_0),
    \label{eq:G(lots)}
\end{equation}
where $\mathbf{M}$ can indicate a set of measurements or instruments, $\mathbf{H}$ a set of relevant Hamiltonians, $\mathbf{t}$ a set of relevant interaction times and $\boldsymbol{\rho}_0$ a set of initial states. It should be stressed that Eq.~\eqref{eq:G(lots)} is intended to be illustrative in nature, rather than a rigorous definition.  {Whereas complexity of the physical networks in the previous Section is encoded by the network $G$ characterizing the interactions captured by the Hamiltonian, here the graph $G$ is a representation of the quantum system itself whose complexity might arise in a nontrivial and sometimes even surprising manner. Of particular interest are cases where the topology of the network reflects some properties of interest of the underlying physical system. In this case network theory can help tame the complexity of the considered system.}

Due to the general nature of Eq.~\eqref{eq:G(lots)} the networks do not need to indicate physical interactions and might represent more abstract relations. For example, a linear optical setup can be described as a network  by taking the nodes to be states and the links to be optical paths weighted by state transition amplitudes. Hence this approach can naturally lead to a directed network with complex link weights \cite{melo2020directed}. Moreover, one could also adopt a colored network approach where networks describing different experiments  might  be colored according to the used optical setup \cite{krenn2017quantum}, phase shift \cite{gu2019quantumII} or mode number \cite{gu2019quantumIII}. Quantum graphs~\cite{gnutzmann2006quantum,smilansky2013discrete} are a very important mathematical framework to represent these states where links weights indicate distances and on each link the evolution of the state is dictated by a differential equation, typical choices being the Schr\"odinger equations or the Dirac equations.  Quantum graphs can be interpreted as real physical systems and indeed predict the spectrum of free electron in organic molecules and are widely used to  study quantum waveguides. An important breakthrough was Kottos and Smilansky discovery~\cite{kottos1999periodic,kottos2000chaotic} that demonstrated that quantum graphs are a fundamental model for  Quantum Chaos, that can be used alternatively to Random Matrices to represent quantum systems that are classically chaotic. These results have lead to the flourishing at the interface  between mathematics and physics that treats quantum states  defined on metric graphs
\cite{mugnolo2014semigroup,kurasov2001inverse,berkolaiko2017edge}. We refer the interested reader to extensive monographs and review of the subject \cite{berkolaiko2013introduction,kuchment2003quantum,mugnolo2014semigroup}.
Forming instead a network out of pairwise correlation measures to represent some quantum state leads to real but typically nonuniform weights which might reflect both the structure of the Hamiltonian and the phase of the system \cite{valdez2017quantifying,sundar2018complex,bagrov2020detecting}, or give rise to multiplex networks where each layer is associated with a particular measure \cite{garcia2020pairwise}. It should be noted that in such cases the physical network determined by the Hamiltonian need not to be complex. In fact it will be seen that even chains and square lattices of quantum systems can give rise to complexity in the ground state that can be tackled with the help of a suitably defined network. The networks might also be constructed in such a way that certain transformations of the underlying state translate to simple transformation rules of the network, thus demonstrating  the utility of networks to represent quantum dynamics.

The motivations to introduce such networks virtually always revolve around using the networks as a convenient mathematical tool to characterize, explain or simplify the quantum system under consideration. In the following we present examples of such research line to further illustrate the concept. 

\subsection{Applications and examples}

\subsubsection{Network description of states}

A complete description of a generic quantum state $\rho$ depends on the Hilbert space dimension which grows rapidly with the number of systems involved. In recent years networks have been proposed as an alternative ways to describe a quantum state. Most of the approaches use  networks that  encode pairwise correlations in the weights of the links. Such a possibility is attractive not only because the networks scale only quadratically with the number of their nodes, typically taken to be the systems, but also because in the case of pairwise measures tomography of the correlation network is drastically cheaper than full state tomography \cite{garcia2020pairwise}. Typically a transition in the network structure, captured by some appropriate network measure, can be linked to a physical transition. Such a link may then be used to both detect known transitions or discover novel phenomena and characterize them through the network picture. Although information will be lost as the network is not formed using the full state, a plethora of cases have been identified where the network approach is accurate.

\begin{figure}[t]
 \begin{center}
   \begin{minipage}{0.5\linewidth}
      \begin{overpic}[width=1.0 \columnwidth,unit=1mm]{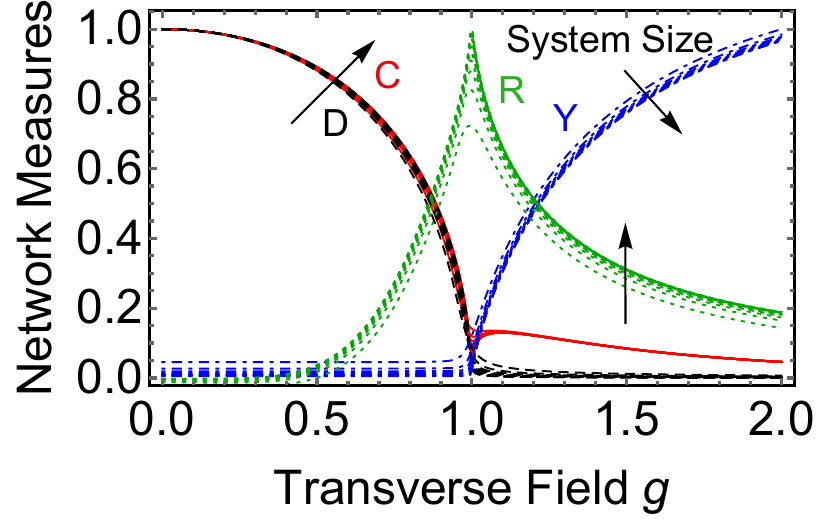}
        \put(2, 2){(a)}
      \end{overpic}
    \end{minipage}\hfill
    \begin{minipage}{0.5\linewidth}
      \begin{overpic}[width=1.0 \columnwidth,unit=1mm]{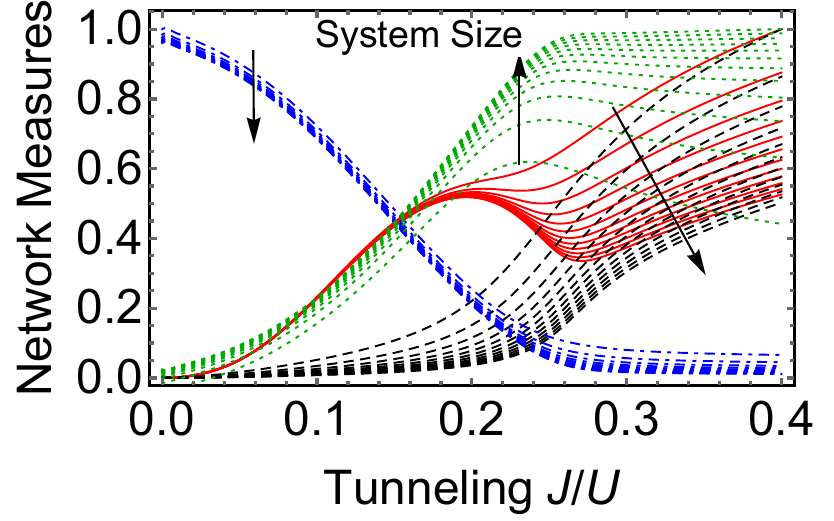}
        \put(2,2){(b)}
      \end{overpic}
    \end{minipage}\hfill
   \end{center}
\caption{Normalized network measures calculated from the mutual information network of the ground state as a function of a parameter in the Hamiltonian for several system sizes. Shown are link density (dashed black line), clustering coefficient (solid red line), Pearson correlation coefficient (dashed green line) and disparity (dot-dashed blue line); arrows indicate the direction in which system size is increased. $(a)$ The transverse-field Ising chain passes from para- to ferro-magnetic phase as field strength increases. All measures display an abrupt change in behavior at the transition which here takes place near $g=1$. $(b)$ Bose-Hubbard chain passes from Mott insulator to superconducting phase as the ratio between particle tunneling $J$ and on-site particle interaction $U$ increases, however the transition is not sharp at the considered system sizes. Nevertheless the measures are clearly useful for identifying the boundary, which should be near $J/U=0.3$. Reprinted figure with permission from \cite{valdez2017quantifying}. Copyright 2017 by the American Physical Society.}
 \label{fig:PTsVSnetmeasures}
\end{figure}

Such use of weighted networks based on (von Neumann) mutual information was introduced in \cite{valdez2017quantifying}, which considered detecting quantum critical points of the transverse field Ising chain from the properties of the correlation network of its ground state. Specifically, the link weight $\mathcal{I}_{ij}$ between some qubits $i$ and $j$ reads
\begin{equation}
\mathcal{I}_{ij}=\frac{1}{2}(S_i+S_j-S_{ij})=\frac{1}{2}(-\mathrm{Tr}(\rho_i\log_2\rho_i)-\mathrm{Tr}(\rho_j\log_2\rho_j)+\mathrm{Tr}(\rho_{ij}\log_2\rho_{ij}))
\label{eq:mutualinfo}
\end{equation}
where $S_i$, $S_j$ and $S_{ij}$ are marginal von Neumann entropies and $\rho_i$, $\rho_j$, $\rho_{ij}$ reduced density matrices of the qubits individually and together, respectively. This work considered as network measures disparity, (global) clustering coefficient, density and similarity between nodes quantified by Pearson correlation coefficient and found that their first and second derivatives or local minima revealed the points with state-of-the-art performance when compared to standard measures for both transverse field Ising and Bose-Hubbard models and three classes of quantum phase transitions. The behavior of the measures against a model parameter is shown in Fig.~\ref{fig:PTsVSnetmeasures}. For example, in the ferromagnetic phase of the Ising model a spin in the chain is correlated with many close spins with similar strength, whereas in the paramagnetic phase a spin is correlated mostly with its nearest neighbors. This transition to short range correlations is captured by the disparity which was shown to exhibit an abrupt change in behavior near the critical point, remaining very small in the ferromagnetic phase but starting to grow with field strength in the paramagnetic phase.

Mutual information networks have been applied also in the case of thermal states. In \cite{bagrov2020detecting} mutual information networks were applied for a Fermi-Hubbard model on a square lattice considering the same measures as in \cite{valdez2017quantifying}. The behavior of the measures were suggested to be connected to the appearance of the pseudogap phase. In \cite{sundar2018complex} a transverse field Ising chain was considered using also other correlation measures such as R{\'e}nyi mutual information, concurrence and negativity and as network measures the ones of \cite{valdez2017quantifying} except node similarity and additionally betweenness centrality, average geodesic distance and diameter. The gradients of these measures were found to exhibit extrema at the transition when temperature and field strength were varied.

\begin{figure}[t]
    \centering
    \includegraphics[trim=0cm 0.cm 0cm 0cm,clip=true,width=0.95\textwidth]{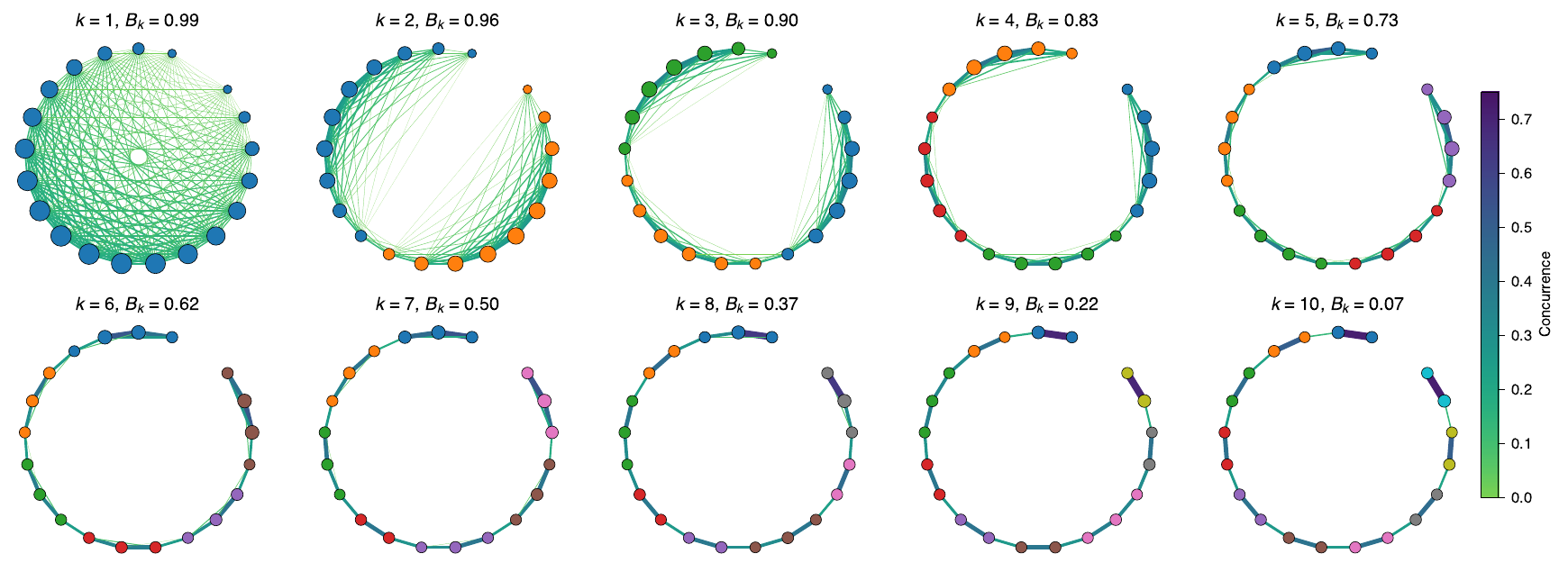}
    \caption{Concurrence networks calculated from the ground state of the XX spin chain for different values of the magnetic field strength $B_k$. Each node is a spin and links indicate concurrence; node size is proportional to the total weighted degree and link width and color to magnitude of concurrence. Nodes are colored according to detected community. The relationship between the community structure and $B_k$ is evident. Used with permission of The Royal Society (U.K.), from \cite{sokolov2022emergent}; permission conveyed through Copyright Clearance Center, Inc.}
    \label{fig:concurrencecommunity}
\end{figure}

Concurrence in particular was applied to study entanglement networks of the ground state of an XX spin chain in \cite{sokolov2022emergent}. Concurrence $C(\rho)$ is a measure of entanglement between two two-level systems with the density matrix $\rho$. It reads
\begin{equation}
C(\rho)=\max(0,\lambda_1-\lambda_2-\lambda_3-\lambda_4)
\label{eq:concurrence}
\end{equation}
where $\lambda_1$, $\lambda_2$, $\lambda_3$ and $\lambda_4$ are the eigenvalues, in decreasing order, of the matrix
\begin{equation}
\mathbf{R}=\sqrt{\sqrt{\rho}(\sigma_y\otimes\sigma_y)\rho^*(\sigma_y\otimes\sigma_y)\sqrt{\rho}}
\end{equation}
where $\sigma_y$ is one of the Pauli spin matrices and $\rho^*$ is the complex conjugate of $\rho$. This work considered both weighted and unweighted variants of degree and local clustering coefficient as well as disparity, and going beyond microscopic structure also the communities as shown in Fig.~\ref{fig:concurrencecommunity}. The network approach was found to reveal new phenomena such as instability of pairwise entanglement with respect to perturbations in the magnetic field strength or community structure in the entanglement network reflecting a global symmetry in the system. Results in a similar vein have been reported also for the quantum critical points of the Kitaev chain in \cite{llodra2022detecting}---whereas the clustering of the mutual information network witnessed the previously known transition from topological to trivial phase, the clustering in the concurrence network revealed previously unknown critical points where entanglement no longer decays with distance.

The different networks arising from multi-qubit states were unified into a single mathematical object in \cite{garcia2020pairwise}, which introduced the concepts of pairwise tomography networks and quantum tomography multiplexes as well as an efficient scheme to construct them with measurements. In a pairwise tomography network nodes are qubits and the links are the associated two qubit reduced density matrices. The presented scheme allows the construction of such a network using a number of measurement settings that scales only logarithmically with the number of qubits. The  pairwise tomography network determines then the quantum tomography multiplex where in each layer the links are weighted by some pairwise quantifier computed from the corresponding two qubit density matrix.

Although the continuous-variables case has received less attention, mutual information networks were considered as early as 2013 for quantum harmonic oscillators in \cite{cardillo2013information} which considered the interplay between the correlation and interaction networks and established how the latter leaves its fingerprint on the former. Although the study focused on ground states the relation was found to be robust to small temperatures.  More recently, the impact of local operations on the structure of a correlation network has been considered \cite{walschaers2023emergent}. The starting point was a Gaussian cluster state with an embedded network structure. If $\mathbf{A}$ is the weighted adjacency matrix of the network, then the corresponding ideal continuous variable cluster state can be made from the product state of $N$ momentum eigenstates with eigenvalue $0$ by acting on some modes $i$ and $j$ with the $C_Z$ gate $\exp{(\mathrm{i}g q_jq_k)}$ if they are connected and where $g$ is the link weight in $\mathbf{A}$, i.e.
\begin{equation}
\Ket{\psi_{\mathbf{A}}}=C_Z[\mathbf{A}]\Ket{0}_p^{\otimes N}=\prod_{j,k}^N\exp{\left(\frac{\mathrm{i}}{2}\mathbf{A}_{jk}q_jq_k\right)}\Ket{0}_p^{\otimes N}.
\label{eq:CVclusterstate} 
\end{equation}
While networks of arbitrary topology can be created deterministically, the states $\Ket{0}_p$ must in practice be approximated. For instance, a way to achieve this is by approximate them by squeezed vacuum states with small but finite variance for $p$. Experimental demonstration of a computational advantage with such states as a resource has been achieved with non-Gaussian measurements \cite{madsen2022quantum}, however with only Gaussian operations efficient classical simulation is always possible \cite{bartlett2002efficient}. Here multi-photon subtractions were considered, de-Gaussifying the state. The state both before and after the operation can also be presented in terms of photon number correlations between the modes; each gate creates such correlations between also the modes adjacent to the target modes---next nearest neighbors---and unlike the cluster state this network has continuous weights between $0$ and $1$. Firstly, the effect of moving from cluster state to correlation network was analyzed. Increase of local clustering coefficient was observed for the Barab\'asi-Albert network, whereas for the Watts–Strogatz network increasing the rewiring probability was found to decrease both degrees and clustering. In general, local multi-photon subtraction was found to increase both the mean degree and the variance and have limited range as biggest effect was on nodes up to two hops away and beyond four hops there was no effect. The impact on higher moments of the degree distribution was sensitive both to the network class, parameter values and the choice of the node however, highlighting also the importance of the topology of the local neighborhood. For example, choosing a low degree node causes a large increase in the correlations in a relatively small subgraph whereas choosing a high degree node modifies a large subgraph but the effect on each individual link is smaller.

We also mention recent results linking the squeezing cost of setting up a Gaussian cluster state to the spectrum of the matrix $\mathbf{A}$ \cite{centrone2023cost}, which are an exact generalization of earlier results of \cite{gu2009quantum} from the large squeezing limit to any squeezing. Importantly, the results imply that co-spectral networks have the same cost and consequently form an equivalence class of cluster states that can be changed into each other applying only passive linear optics. The relationship between cost and topology was also studied, revealing how the scaling with size is strictly topology dependent.

Going beyond correlations, networks have been also constructed based on pairing amplitude in topological superconductors \cite{chou2014network} and electron wave function overlap in quantum dot systems \cite{cuadra2021modeling}. To the best of our knowledge the former work, Ref.~\cite{chou2014network}, was in fact among the earliest to investigate network measures on induced weighted networks as a tool to facilitate understanding, namely to detect topological phase transitions. 
Instead, the  latter work, Ref.~\cite{cuadra2021modeling},  proposes to model quantum dots randomly distributed on a plane as random geometric graphs and considers network measures such as  degree distribution, clustering and average geodesic distance to explain phenomena such as the emergence of transport.

Before concluding we mention the tensor networks formalism, a wholly different approach to network characterization of states. The main idea is to use networks of tensors connected by contractions to efficiently represent physically relevant states~\cite{evenbly2011tensor,evenbly2015tensor,orus2014practical,biamonte2019lectures}, such as low-energy eigenstates of Hamiltonians with local interactions and a finite gap between the ground state energy and first excited state energy. The formalism takes advantage of the limited amount of entanglement in these states, and it may be argued that the overwhelming majority of the inapplicable states are in fact of little practical interest. Such tensor networks lend themselves to diagrammatic manipulation which can be used to reason about the state and find ground states of suitable Hamiltonians when combined with suitable numerical techniques. Although immensely useful, widely applicable and enjoying a strong interest tensor networks are typically lattices and the complex network aspect at least in the research carried out so far is overall weak, making them a borderline case with respect to the classification proposed in Fig.~\ref{fig:intro} which is why they are not considered further here.  

\subsubsection{Network description of experimental data sets}

To apply the results of the previously introduced networks on an unknown state, one needs to perform tomography to estimate with sufficient accurary a suitable bipartite correlation measure that then constitutes the links. In practice one accumulates experimental data by carrying out measurements from a tomographically complete set on a large ensemble of identically prepared systems such that one approaches asymptotically the actual values as the ensemble size increases, and therefore the actual network.

Very recently a more direct approach has been introduced in Ref.~\cite{mendes2023wave} where the network is constructed directly out of the experimental data set based on just a single measurement setting, i.e. projecting a pure many-body state into a fixed basis. The outcomes are called wave function snapshots and they can be probed experimentally as well as numerically; for qubits each is a binary string. The wave function network is constructed out of the snapshots by treating them as nodes, defining a suitable metric (such as the Hamming or the Euclidean distance) and an upper limit $R$ for distance and then creating a metric network where the nodes are connected if their distance is less than $R$ and disconnected otherwise.

Importantly, these choices were shown to generate nontrivial and informative networks. This was exemplified with an Ising model undergoing a quantum phase transition from disordered to ferromagnetic phase where consequently the network degree distribution experiences a transition from a Poisson to scale-free distribution. 

Another example featured experimental data from a Rydberg quantum simulator for spin models, which consist of arrays of trapped atoms whose ground (Rydberg) states play the role of spin down (up) states and interactions can be realized, e.g., via the van der Waals interaction. The network description facilitated the estimation of the Kolmogorov complexity of the simulator output---by definition the minimum length of a computer program written in some fixed language that could generate it---because the degree distribution remained scale-free in which case efficient algorithms can be used on the network, showing a non-monotonic evolution of the complexity. This is remarkable because for generic strings, finding the Kolmogorov complexity is a NP-hard problem. Finally, a cross-certification method based on network similarity was proposed, allowing one to determine whether two devices sample from the same probability distribution by comparing the network degree distributions. The method was demonstrated by comparing the outcomes of two experiments as well as an experiment and a simulation. An interesting research question arises from  the comparisons of these networks with the recently introduced IsingNets networks \cite{sun2023network} constructed from configuration snapshots of  classical Ising models. In particular this comparison will be key to determine the exclusive signature of quantumness in the quantum wave function networks.

\subsubsection{Network description of experimental setups}

Previously, little attention was paid to how the state was or could be created. Shifting focus to this, quite naturally leads to the concept of interpreting experimental setups as networks associated with graphs. This can facilitate intuitive and convenient ways to determine how the state transforms from the initial to the final form, somewhat akin to Feynman diagrams. Moreover, this approach can link experiments to graph theory. In particular, graph theoretical methods may be used to answer experimental questions and experimental methods used to answer graph theoretical questions. Both avenues have been followed particularly in the case of quantum optics.

\begin{figure}[t]
    \centering
    \includegraphics[trim=0cm 0.cm 0cm 0cm,clip=true,width=0.95\textwidth]{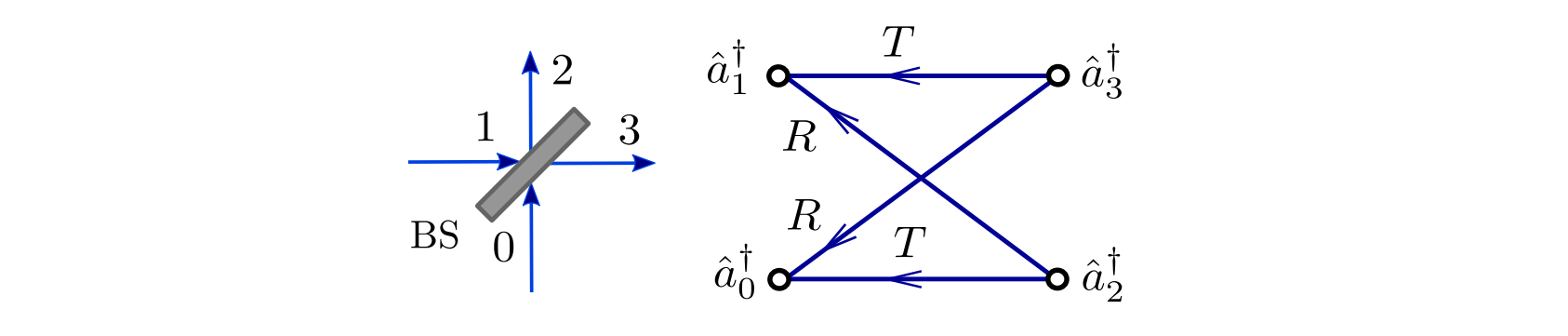}
    \caption{A simple example of the application of the graphical method developed in Refs.~\cite{ataman2014field,ataman2015quantum,ataman2018graphical}. The symmetrical beam splitter on the left has input ports $0$, $1$ and output ports $2$, $3$. The graph depicts the action at the level of creation operators $a^\dagger_i$ indexed by port. The input operators can be either transmitted with weight $T$ or reflected with weight $R$. Here output operators point to input operators, facilitating the computation of the former as functions of the latter: $a_2^\dagger=T a_0^\dagger+R a_1^\dagger$, $a_3^\dagger=R a_0^\dagger+T a_1^\dagger$. Reversing the orientation inverts the functional relationship. Figure reproduced from Ref.~\cite{ataman2018graphical}, doi: \text{https://doi.org/10.1088/2399-6528/aab50f}, license: \text{https://creativecommons.org/licenses/by/4.0/}.}
    \label{fig:atamanBS}
\end{figure}

For an example of the former, Ref. \cite{ataman2014field} has introduced a method to transform a linear lossless device consisting of beam splitters and interferometers into a directed tree connecting input field operators (the roots) to output field operators (the leaves) by optical paths weighted by complex probability amplitudes. Specifically, assuming two input ports and two output ports and monochromatic light in a pure state for simplicity, the input state can be expressed as
\begin{equation}
\Ket{\psi_{in}}=f(a^\dagger_0,a^\dagger_1)\Ket{0}
\end{equation}
where the subscripts stand for input port $0$ and input port $1$. Suppose the output ports are labeled $N$ and $N+1$. If one can find functions such that $a^\dagger_0=g_0(a^\dagger_N,a^\dagger_{N+1})$ and $a^\dagger_1=g_1(a^\dagger_N,a^\dagger_{N+1})$, then one can write the output state as
\begin{equation}
\Ket{\psi_{out}}=f(g_0(a^\dagger_N,a^\dagger_{N+1}),g_1(a^\dagger_N,a^\dagger_{N+1}))\Ket{0}.
\end{equation}
After replacing the optical elements with their corresponding graph elements, an output operator can be computed by simply following every directed path from the roots to it multiplying the amplitudes on a path and summing the products of amplitudes of different paths. Hence, this procedure facilitates the extraction of the sought functions $g_0$ and $g_1$. Inverting the orientation of the graph allows the extraction of their inverses, as shown in Fig.~\ref{fig:atamanBS}. Generalizations to non-monochromatic light and mixed states are discussed in the reference. This method was supplemented with graph elements for nonlinear optical devices performing spontaneous parametric down-conversion in \cite{ataman2015quantum} and demonstrated by explaining three previous experimental results on such setups. It is worth noting that the nonlinear optical elements were presented by directed hyperedges, pointing from two nodes to a third one in the graph. Finally in \cite{ataman2018graphical} optical resonators were treated as directed cycles. Although such a cycle creates infinitely many directed paths to the output the amplitudes are such that the resulting infinite sum converges.

Another closely related approach to interpreting experiments as networks for computational convenience was introduced in \cite{melo2020directed}, with applications to homodyne linear optical setups. Here the nodes are states linked by optical paths, and the links are weighted by state transition amplitudes. Much like previously, the input and output states have a special role as nodes with zero in-degree and out-degree, respectively. Moreover, the overall transition amplitudes from the input to the output are computed by tallying the directed paths. Loops may be present, leading to infinitely many allowed paths which however keeps the computed amplitude finite. Importantly, this work introduces graph simplification rules which can be applied to eliminate intermediate states, parallel paths, loops and the like. In this way the goal is to finally arrive at a graph featuring only the input and output states connected by a single link weighted by the overall amplitude.

The other approach was adopted in \cite{krenn2017quantum} which concerned post-selected states prepared by experiments involving probabilistic photon pair sources and nonlinear down-conversion crystals. Such setups were associated with a graph where the nodes are optical paths and links are the crystals, colored by layer. Notably, the terms in a superposition state created by the setup correspond to perfect matchings of the graph. This means limitations on what kind of terms are possible in a given setup translate then to what kind of matchings are possible in the corresponding graph, facilitating the application of graph theoretic methods to answer such questions. The problem was explored further in \cite{gu2019quantumIII} where the existence and construction of experimental setups for generating different entangled states was solved by finding the graph with the required properties. 

Weighted networks were used in \cite{gu2019quantumII}, concerning post-selected states prepared by experiments involving linear optics elements, nonlinear crystals and probabilistic 2-photon sources, and generalized to $n$-photon sources in \cite{gu2020quantum}. In the former the graph consists of photonic modes linked by photon pair correlations weighted by probability amplitudes for photon pair creation. The weights can be used to account for interference effects. In the latter the graph consists of optical output paths playing the role of nodes, grouped by $n$-photon sources playing the role of hyperedges and weighted by probability amplitudes. In this work in particular the use of experiments to solve graph theoretic problems was proposed, as detection of an $n$-fold coincidence event reveals the existence of a perfect matching in the corresponding hypergraph. Whether a hypergraph admits perfect matchings is in general a difficult problem \cite{keevash2014geometric}.

This line of research was recently generalized beyond the post-selected case in \cite{krenn2021conceptual} and proposed to be used for the design of new quantum optics experiments. The nodes are photonic paths and links correlated photon pairs colored by mode numbers and weighted by complex coefficients. The key novelty over previous works is defining the weights in such a way that they contain the full information of the associated state and the introduction of a function that maps the weights of the network to the corresponding state preparation operator. The presentation was further applied in an algorithm based on optimizing an objective or a loss function depending on the weights, which was found to have superior performance over alternatives in benchmark tasks involving both entangled state enumeration and identifying high-dimensional C-NOT gates. Finally, it was proposed that thanks to the network presentation the algorithm produces human understandable solutions, potentially allowing the user to understand and generalize the concept beyond particular cases.

\subsubsection{Network description of dynamics\label{sec:properties-dynamics}}

Characterizing the properties of a quantum system with a network emerging from its state is appealing in particular when the network is much simpler than the state. But not every state of interest is amenable to a description by a number of parameters quadratic in system size, and therefore not by a network unless the transformation to a network is many-to-one. When it is, all the information is not in the network which makes it unsuited for evolving states since one would have to determine the evolution in the Hilbert space. Indeed, the previously presented Refs.~\cite{valdez2017quantifying,sundar2018complex,bagrov2020detecting,sokolov2022emergent} consider only stationary states. Alternatively a network interpretation can be assigned to an experimental setup for convenience or to benefit from the toolbox of graph theory but the state itself usually remains in conventional form. Sometimes such networks might include special nodes for sources of allowed initial states, as in for example Ref.~\cite{melo2020directed}.

The network description of the dynamics becomes possible when both the state and its transformations can be represented by the network. That is to say given the state and the operations on it one can express the resulting dynamics as a time ordered sequence of networks, or a temporal network.
The final state can then be transformed back to conventional formalism if necessary. This can be achieved by defining a network presentation for the state and a rule to express the operations as network rewrite rules from the set of admissible networks to itself, or creating a graphical calculus. Although both the set of states and the set of operations are typically restricted such approaches have been used both to facilitate classical simulation of the dynamics. For example, this approach can be used  for translating tasks involving entanglement distribution to maximally entangle given systems into network rewiring problems. As a rule of thumb, the more operations one includes the less elegant the rewrite rules become; obviously any transformation not taking us outside the set of admissible states is possible but the way the network transforms might then elude an intuitive interpretation.

A graphical calculus might be created simply for convenience. A prime example of this is the one introduced in \cite{menicucci2011graphical} which accounts for all pure Gaussian states and all unitaries that preserve the Gaussianity as well as quadrature measurements. Whereas most are given as transformations of the adjacency matrix, some admit simple graph rewrite rules. The graphs are undirected and complex weighted, featuring also loops. At the unphysical limit of infinite squeezing the weights become real however, corresponding to ideal Gaussian cluster states of Eq.~\eqref{eq:CVclusterstate}; this is tied to the chief motivation of quantifying, informally speaking, the distance of any physical and therefore approximate cluster state to its ideal limit. Any state that can actually be prepared has finite squeezing and is therefore only an approximation, but such approximations were lacking a network description up until this work was published. Here it was applied to finding suitable Hamiltonians for their adiabatic preparation, supplementing the previously mentioned similar result derived by other means. Its power was also illustrated by finding graphical rules to compute bipartite entanglement for certain states. 

\begin{figure}
    \centering
    \includegraphics[trim=0cm 0.cm 0cm 0cm,clip=true,width=0.75\textwidth]{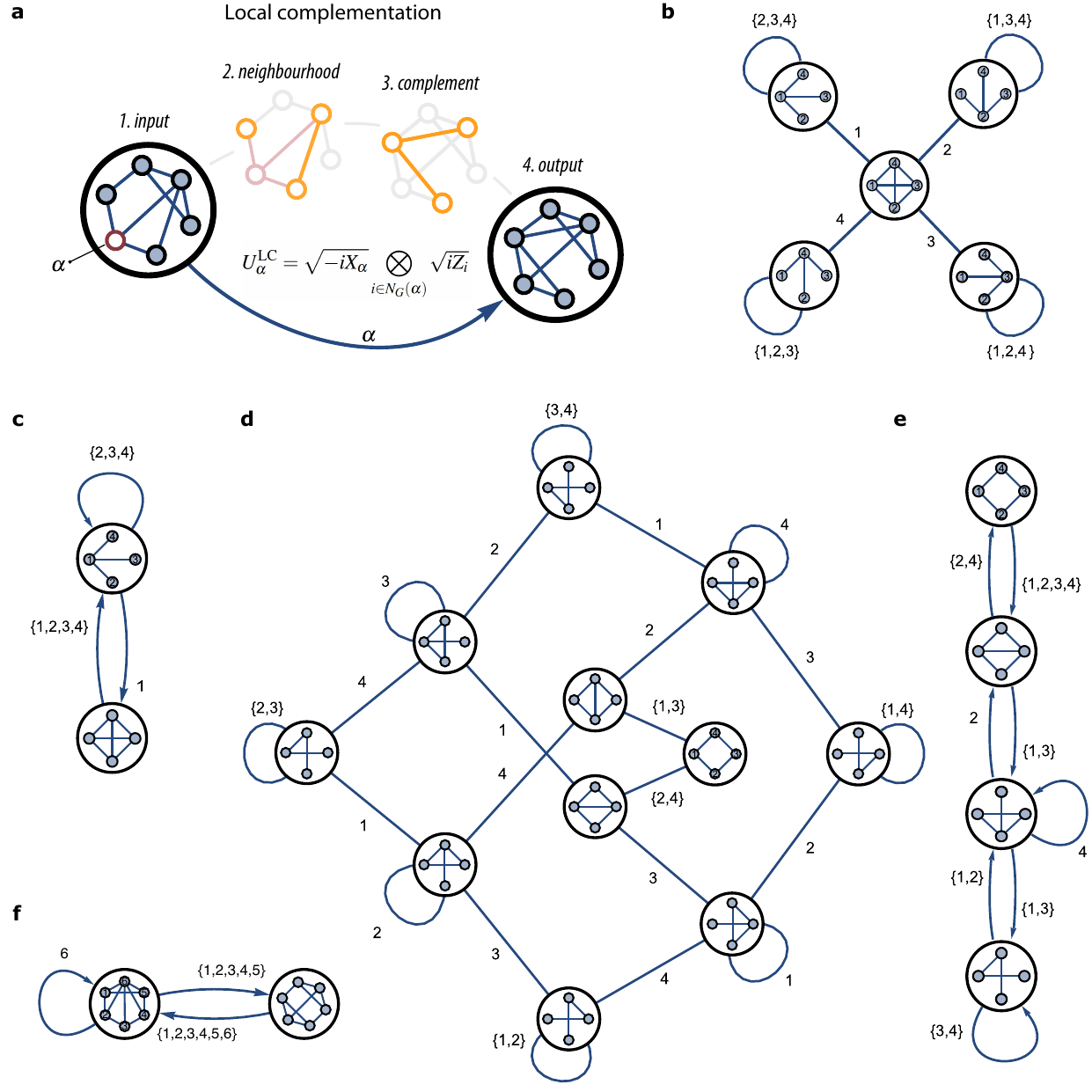}
    \caption{Orbits of qubit cluster states under local complementation. Orbit edges indicate which qubit is chosen. Directed edges indicate that isomorphic graphs are treated as equivalent. $(a)$ Local complementation with respect to some node $\alpha$ can be achieved with local operations on $\alpha$ and its neighborhood, which has implications on its effect on the entanglement in the state as explained in main text. Orbits starting from the completely connected graph of $4$ qubits are shown in $(b)$ and $(c)$. Orbits of the cycle graph of $4$ qubits are shown in $(d)$ and $(e)$. An orbit featuring $6$ qubits is shown in $(f)$. Figure reproduced from Ref.~\cite{adcock2020mapping}, doi: \text{https://doi.org/10.22331/q-2020-08-07-305}, license: \text{https://creativecommons.org/licenses/by/4.0/}.}
    \label{fig:clusterstateorbits}
\end{figure}

In a somewhat similar vein, Ref.~\cite{adcock2020mapping} considered the behavior of qubit cluster states under local complementation.  A qubit cluster state, corresponding to unweighted adjacency matrix $\mathbf{A}$, is
\begin{equation}
\Ket{G}=CZ[\mathbf{A}]\Ket{+}^{\otimes N}=\prod_{j,k}^N\mathbf{A}_{jk}CZ_{jk}\Ket{+}^{\otimes N}
\label{eq:DVclusterstate} 
\end{equation}
where $CZ$ is the controlled $Z$ gate and $\Ket{+}=\frac{1}{\sqrt{2}}(\Ket{0}+\Ket{1})$. Local complementation with respect to some node $\alpha$ toggles every link in the subgraph induced by its neighborhood $n(\alpha)$; if the link was present it is deleted and otherwise it will be inserted. The adjacency matrix $\mathbf{A}$ changes according to
\begin{equation}
\mathbf{A}\mapsto\mathbf{A}\oplus\mathbf{K}_{n(\alpha)}
\label{eq:localcomplementation}     
\end{equation}
where $\oplus$ is addition modulo two and $\mathbf{K}_{n(\alpha)}$ is the adjacency matrix of the complete graph of the nodes adjacent to $\alpha$. Importantly, local complementation of the graph can be achieved by applying local gates on the qubits. Note that, since entanglement cannot increase under local operations, local complementation cannot increase it either. Here it was shown how repeated applications of local complementation creates orbits in the set of qubit cluster states, implying that in fact the entanglement in every state of the orbit must be the same, constituting a graph entanglement class. Several examples are shown in Fig.~\ref{fig:clusterstateorbits}. Other connections between the properties of the orbits and entanglement and also preparation complexity were identified as well, paving the way for follow-up studies where graph theory could perhaps be applied to understand entanglement.

Cluster states in general, both in the continuous and discrete variable case given respectively by Eqs.~\eqref{eq:CVclusterstate} and \eqref{eq:DVclusterstate}, can be viewed as a network presentation of dynamics when used for computation. Specifically, a computation achieved by a quantum circuit featuring also measurements can be emulated by suitable pattern of just local measurements and operations on such a state \cite{raussendorf2003measurement}; whether the overall evolution is deterministic for a given cluster state and measurement pattern depends on a graph property called g-flow \cite{browne2007generalized} or CV-flow \cite{booth2023flow} for discrete and continuous variable cases, respectively. Cluster states themselves are a special case of more general graph states where the requirements for the gates playing the role of links are relaxed somewhat, however in such a way that the state can still be desribed by a simple graph; the discrete variable case is covered in \cite{hein2006entanglement}. Conventions vary, however, and sometimes the terms are used interchangeably.

Using a graphical calculus for simulation purposes has seen use especially for so called (qubit) stabilizer states. These states arise from qubit cluster states of Eq.~\eqref{eq:DVclusterstate} via local Clifford operations and have applications, for example, in quantum error correction and fault tolerant quantum computing. Up to a global phase of the from $e^{\mathrm{i}\phi}$, a local Clifford operator can be generated by the Hadamard gate $\mathbf{H}$ and phase gate $\mathbf{P}$, unitary operators acting on a single qubit given in Eq.~\eqref{eq:localClifford}. The first such software is called GraphSim, introduced in 2006 \cite{anders2006fast}. Despite its respectable age for a piece of software, it still boasts the fastest speed in specific tasks---albeit not in general---when compared to some contemporary alternatives \cite{gidney2021stim}, namely IBM's Qiskit \cite{aleksandrowicz2019qiskit}, Google's Cirq \cite{quantum2020cirq}, and a very recent powerful simulator called Stim \cite{gidney2021stim}. As explained earlier, cluster states are in one-to-one correspondence with graphs. GraphSim uses the corresponding adjacency list and the list of applied local Clifford operators, which may be thought of as node weights or weighted loops although this point of view was not adopted. Any circuit consisting of gates from the Clifford group of operators and local measurements in the computational basis can be simulated. Applying a local gate amounts to just updating the node weight, whereas more complicated operations include also a combination of local complementations and toggling links on or off. The approach has recently been improved by introducing a novel canonical form for stabilizer states as graphs as well as more advanced graph rewrite rules in Ref.~\cite{hu2022improved}, which also features an excellent compact introduction to the stabilizer formalism. Importantly, by introducing both a canonical form and a canonicalization algorithm this work avoids completely the need to test whether two graphs represent the same stabilizer state or not. Another recent development is the generalization to noisy stabilizer states \cite{mor2023noisy}. As a side note, while Stim is both very powerful and not based on graph presentation, it greatly benefits from tallying the action of the circuit essentially in reverse. It might be wondered if the same approach could further benefit the approaches that do use graphs. 

Besides simulation, graph theoretic methods can also be used in conjunction with diagrammatic languages to simplify quantum circuits. Here we mention some relevant examples based on ZX-calculus \cite{coecke2008interacting,coecke2011interacting}. It was applied in Ref.~\cite{duncan2020graph} to circuits acting on a registry of qubits by expressing them as measurements and local operations on a graph state, optimizing, for example the time taken or measurements made, and then returning to circuit formalism. In particular, the simplification took advantage of graph theoretic notions such as local complementation. When applied to Clifford circuits, this approach produced the graph of the cluster state augmented with the local Clifford operators. The difficulty of extracting the circuit from the ZX-diagram limited the approach to measurements in a specific plane of the Bloch sphere only, however this limitation has since been lifted in a follow-up work \cite{backens2021there}. Local complementation was also one of the workhorses of an approach to minimize the number of gates not in the Clifford group of operators in a circuit \cite{kissinger2020reducing}. Very recently a circuit extraction method for the continuous variable case has been introduced, perhaps paving also the way for similar applications \cite{booth2023flow}.

As will be seen in Sec.~\ref{sec:comms}, network description of dynamics has applications in communications. Entanglement percolation and the entanglement distribution primitives of Fig.~\ref{fig:distributionprimitives} can be thought of as relatively simple examples of this. Assuming that the communication network can be initially prepared in a cluster state leads to new opportunities. Starting from an arbitary and possibly complex $G$, Ref.~\cite{sansavini2019continuous} studied how to manipulate the network state to distribute a Bell pair between two given nodes using linear optics operations. On the contrary, in \cite{hahn2019quantum} this was achieved via local complementation achieved by applying local Clifford operations on the nodes, and was observed to lead to fewer measurements than a conventional entanglement distribution protocol. Establishing the initial large-scale cluster state was discussed in \cite{epping2016large} and stabilizer formalism was used to describe both entanglement distribution and error correction. Importantly, the power of the approach was demonstrated by connecting several performance metrics to the topology of the underlying graph and then optimizing them. More recently, noisy stabilizer states have been applied to efficiently simulate very large noisy quantum communication networks \cite{mor2023influence}. Such large-scale applications aside, judicious manipulation of suitable cluster states can be used to realize all-optical entanglement distribution schemes which trade the challenge of requiring powerful quantum memories to the challenge of efficiently preparing and measuring the states \cite{azuma2015all,pant2017rate}.

\subsection{Avenues for further research}

Several authors have suggested applying a network presentation beyond stationary states \cite{sundar2018complex,garcia2020pairwise} to study temporal correlations in the evolving network. Such research would be separate from work focusing on network description of dynamics, as the network at some point of time would not in general determine uniquely the network at later times. It could have applications for example in efficient extraction of nontrivial information of the evolution, providing an alternative to process tomography. Reference~\cite{garcia2020pairwise} in particular suggested studying the topological correlations in multiplex networks formed by associating each layer with a different correlation measure. Aside from generalizing the approach, one can also simply apply it to novel models or classes of states. As for network description of data sets such as wavefunction snapshots, there should be much room for further work as the concept itself is still very new.

Correlation networks in particular can be connected to interaction networks. When such a relation exists it can be applied for example in the preparation of special resource states where both the state and the Hamiltonian that prepares it have a network structure \cite{menicucci2007ultracompact}, or in their adiabatic preparation using Hamiltonians which have them as ground states \cite{aolita2011gapped}. Connections like these might warrant further investigation in the general complex quantum networks context.

Network description of experimental setups such as the one exemplified in Fig.~\ref{fig:atamanBS} have already been further developed by expanding the set of elements covered by it. Similar expansion of such methods in general can be expected to continue. Conversely, one can ask what are the limitations of graphical approaches, especially in terms of convenience. Systematic exploration of such limitations might help characterize the best use cases of these methods, guiding future work and applications.

While a network description of dynamics can be applied for simulation purposes, the alternatives that do not use it featured for example in Ref.~\cite{gidney2021stim} are in general more powerful. Since several theoretical advancements have been made in graphical methods there might very well be room to also introduce simulation software exploiting them. Moreover, the research aiming to shed light on non-classical properties of quantum systems via a network description of their dynamics as in the exploration of cluster state orbits shown in Fig.~\ref{fig:clusterstateorbits} still seems to be quite sparse, perhaps warranting more attention.

\section{\label{sec:emergence}Emergence in network models}

\subsection{Introduction to emergent quantum network models}

In the previous Section we have seen that  networks can be used to encode the information of a quantum system. Here we discuss how classical network models including the Bianconi-Barabasi model \cite{bianconi2001bose,bianconi2001competition}, the growing Cayley tree network \cite{bianconi2002growing} and the {\em Network Geometry with Flavor} (NGF) model \cite{bianconi2017emergent,bianconi2016network,bianconi2021higher} can represent  quantum statistics  and how the Bose-Einstein condensation of a Bose gas can predict  their topological phase transitions. 

In network models quantum statistics can either emerge from a non-equilibrium network dynamics \cite{bianconi2001bose,bianconi2002growing,bianconi2017emergent,bianconi2016network,bianconi2021higher}, or it can be a characteristic property of network ensembles ~\cite{park2004statistical,anand2009entropy,garlaschelli2009generalized} defined following a parallel construction to the statistical mechanics ensembles of quantum particles. 
All these models can be classified as quantum-generalized according to the classification we have proposed in Fig. 1. 

Of special interest are the models in which quantum statistics emerges spontaneously from a 
non-equilibrium network evolution.
These models can be considered as network representations of quantum statistics. In particular a network can be mapped to a Bose gas or to a Fermi gas. Interestingly the   network mapped to the Bose gas can undergo a topological phase transitions, called the  Bose-Einstein condensation in complex networks, \cite{bianconi2001bose} in correspondence to the Bose-Einstein condensation of the Bose gas. 

Emergence is a key property of complex systems and refers to the manifestation of properties that cannot be explained by considering the elements of the complex systems in isolation. Examples of key emergent properties are cognition which cannot be explained by neurons taken in isolation or life itself that cannot be explained by considering separately the constituents of a cell.
In physics, and in particular in quantum gravity it is widely believed that space-time itself should be emergent, and this line of though is nicely summarized by the Roger Penrose quote~\cite{penrose1972nature}: {\em My own view is that ultimately physical laws should find their most natural expression in terms of essentially combinatorial principles, $[\ldots]$ . Thus, in accordance with such a view, should emerge some form of discrete or combinatorial spacetime.}
In this Section it is not our goal to cover the intense activity on quantum gravity approaches at the interface with network science; rather here we would like to discuss the relevance of network science for models in which quantum statistics emerges spontaneously from the network dynamical rules  and briefly cover their relation to questions arising in quantum gravity. 

Interestingly since the beginning of network science, with the formulation of the Bianconi-Barabasi model ~\cite{bianconi2001bose,bianconi2001competition} 
and the growing Cayley tree model~\cite{bianconi2002growing,bianconi2002quantum}
it was realized that non-equilibrium models of networks can display the emergence of quantum statistics. Indeed quantum statistics characterize the statistical properties of the structures of these networks and can determine a topological transition called Bose-Einstein condensation in complex networks. More recently is has been found that models in which quantum statistics are emergent  include not only (pairwise) network models but also higher-order network models (evolving simplicial complexes).
In the simplicial complex models also called NGFs \cite{bianconi2017emergent,bianconi2016network,bianconi2021higher} we observe the remarkable phenomena that different quantum statistics
can describe the statistical property of the same  higher-order network structure. In particular the degrees of the nodes and the generalized degrees of the links and triangles have statistical properties that are captured by different quantum statistics. This implies that  a single higher-order network can represent different quantum statistics at the same time, encoding them in the statistical properties of simplices of different dimension. Interestingly these higher-order networks display also the emergence of hyperbolic geometry \cite{bianconi2017emergent} and the emergence of a non-universal spectral dimension~\cite{mulder2018network} characterizing the diffusion properties of classical and quantum walkers on these structures and their spectral dimension can be inferred using quantum probes \cite{nokkala2020probing}.

This section will provide a guide to all the models representing quantum statistics, emphasizing the research questions that arise in network science as well the relation with fundamental questions in emergent spacetime. We note that however, due to  space limitations, we cannot cover  all the works at the interface between network science and quantum gravity, a field in which research interest is recently growing. This include work on causal sets \cite{surya2019causal,krioukov2012network}, tensor field approaches \cite{jahn2021holographic,sasakura1991tensor,evenbly2011tensor,jahn2022tensor}, combinatorial quantum gravity~\cite{Trugenberger1,Trugenberger2,kelly2019self,trugenberger2023combinatorial} and emergent random graph geometry~\cite{kleftogiannis2022physics,chen2013statistical,akara2021birth,kleftogiannis2022emergent} among other approaches \cite{Astridnet}.
  
\begin{figure}[t]
    \centering
    \includegraphics[width=\textwidth]{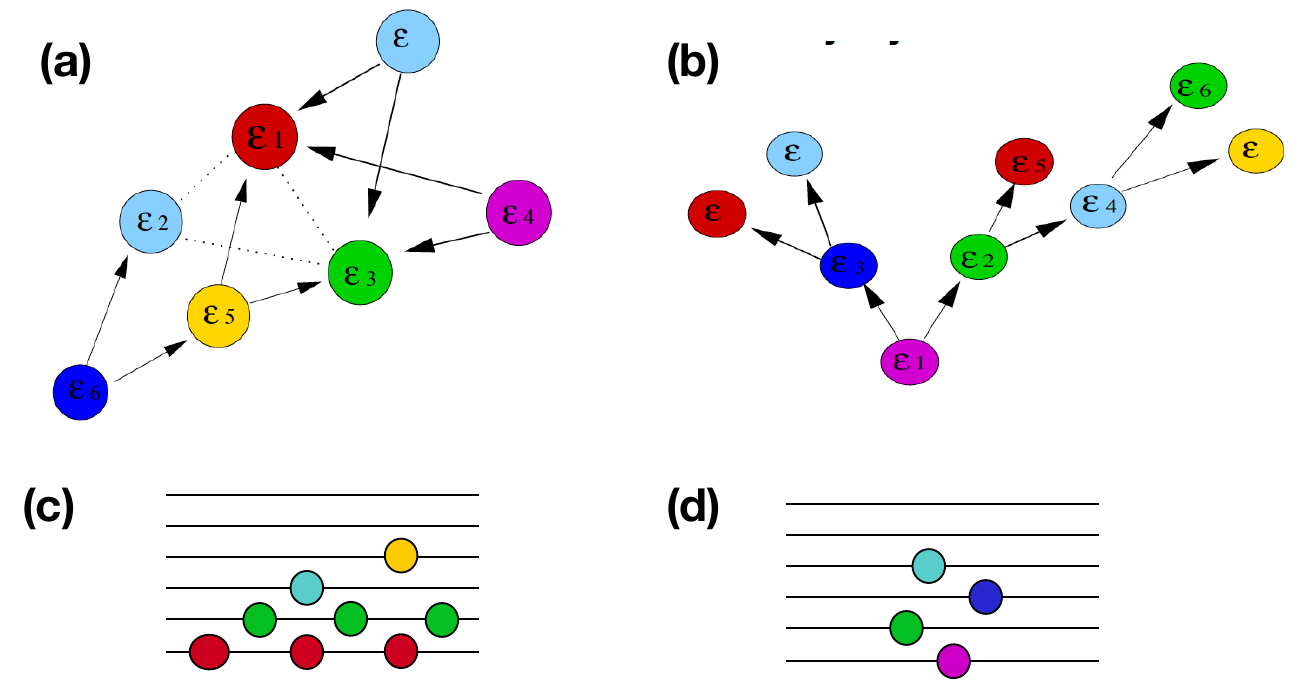}
    \caption{Graphical visualization of two evolving networks representing respectively the Bose (panels (a),(c)) and the Fermi statistics (panels (b),(d)). The network representing the Bose statistics evolves through the Bianconi-Barabasi model~\cite{bianconi2001bose} and grows through the consecutive addition of one node connected to the exiting network by $m'=2$ links, following a generalized preferential rule. The other model representing the Fermi statistics is the Cayley tree defined in Ref.~\cite{bianconi2002growing} that grows by the subsequent selection of nodes giving rise to $m'=2$ offspring nodes. In both models nodes are characterized by intrinsic scalar properties indicating their  {\em energies} $\epsilon$ (here indicates with colours of the nodes). The mapping to the quantum statistics is performed by placing a particle to an energy level $\epsilon$ if in the first model the new node attaches to a node of energy $\epsilon$, and in the second model if a new node of energy $\epsilon$ is selected to give rise to new offsprings. Reprinted figure with permission from \cite{bianconi2002growing}  \copyright Copyright (2002) by the American Physical Society. }
    \label{fig:BE_FD_models}
\end{figure}
\begin{figure}[t]
    \centering
    \includegraphics[width=\textwidth]{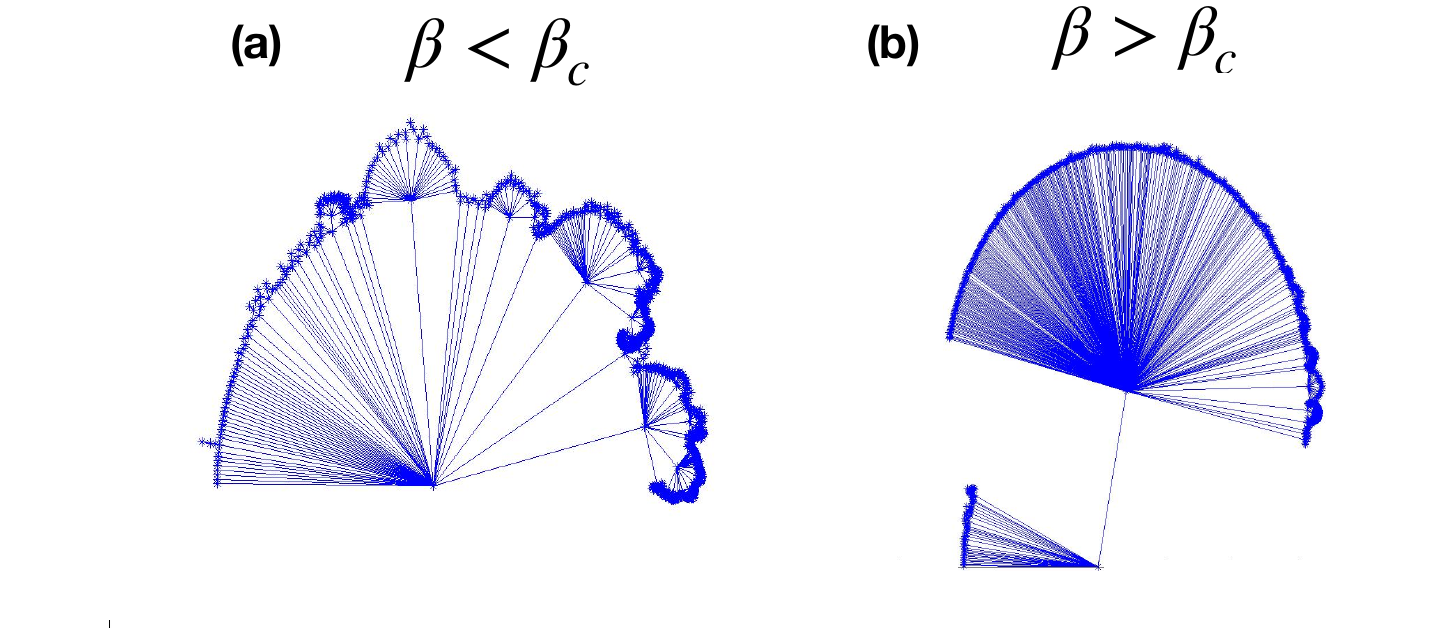}
    \caption{Visualization of the  Bose-Einstein condensation in complex network. The Bianconi-Barabasi model \cite{bianconi2001bose} representing the Bose-Einstein statistics is displayed above (panel (a), $\beta<\beta_c$) and below (panel (b), $\beta>\beta_c$) the critical temperature for the Bose-Einstein condensation occurring at  $\beta=\beta_c$. Reprinted figure with permission from \cite{bianconi2003size}  \copyright Copyright (2003) by the American Physical Society.}
    \label{fig:condensation}
\end{figure}
\subsection{Quantum statistics  and Bose-Einstein condensation in complex networks}

Network Science is a field that  has benefited greatly from  statistical mechanics approaches that have been key to characterize both equilibrium (maximum entropy) and non-equilibrium (growing) network models.
The maximum-entropy models \cite{park2004statistical,anand2009entropy} define ensembles of networks which are maximally random while preserving some network properties. Non-equilibrium network models are instead models of growing networks dictated by simple rules that are able to generate non-trivial complex network topologies including for instance the BA model.
Non-equilibrium models are actually the most promising models for studying emergent properties. Indeed in non-equilibrium models the network evolution, implementing simple combinatorial rules can self-organize leading to  macroscopic network structures with emergent macroscopic properties.
In this context it has been shown that quantum statistics can represent and encode the statistical properties of non-equilibrium growing networks.
In particular the Bianconi-Barabasi~\cite{bianconi2001bose,bianconi2001competition} model of complex networks describes the emergence of network topologies that  represent  quantum Bose-Einstein statistics while the growing Cayley tree with energy (and fitness) of the nodes~\cite{bianconi2002growing}  represents the Fermi-Dirac statistics.
Both models \cite{bianconi2002quantum} describe the growth of a network by the addition of new nodes and links. Moreover both models  are characterized by a dynamical rule that is not only determined by the topological characteristics of the existing network, but is also dependent on an additional feature associated to the nodes called {\em energy}  characterizing the quality of the nodes.
Each node $i$ has an energy $\epsilon_i\geq 0$ drawn from a distribution $g(\epsilon)$ which determine the so called node {\em fitness} given by 
\begin{equation}
\eta_i=e^{-\beta\epsilon_i},
\label{eq:fitness}
\end{equation}
where $\beta>0$ is a model parameter called {\em inverse temperature}.
The definition of the node's fitness implies that nodes with low energy have high fitness. In the considered models nodes with high fitness have either a higher ability to attract new links (in the Bianconi-Barabasi model) or a higher ability to give rise to off-springs (in the growing Cayley tree model).
The dynamics of the Bianconi-Barabasi model includes a preferential attachment of new nodes to nodes with both high-degree and high-fitness. The growing Cayley tree with fitness of the nodes define the growth of a Cayley tree by the subsequent branching of nodes into a constant number of new nodes. In this case the branching nodes are chosen among the nodes that have not yet branched, with a probability proportional on  their fitness.
Interestingly both models \cite{bianconi2002quantum}
 can be shown mathematically to  represent  quantum statistics when nodes are mapped to energy levels and links pointing to old  (for the Bianconi-Barabasi model) or to new   (for the growing Cayley tree model) nodes are mapped to occupation number of energy levels (see Figure $\ref{fig:BE_FD_models}$).

This mapping is not only an interesting mathematical result of these models but is actually a powerful tool to discover an important topological phase transition.
Specifically the Bianconi-Barabasi model displays an important topological phase transition in correspondence to the Bose-Einstein condensation, which is called {\em Bose-Einstein condensation in complex networks}.
Indeed when the inverse temperature of the model $\beta$ exceed the critical temperature $\beta_c$ the network structure is dominated by succession of super-hub nodes that significantly change the topology of the network (see Figure $\ref{fig:condensation}$).
These super-hub nodes are clear leaders of the network acquiring new links linearly in time (albeit eventually with logarithmic corrections) until the emergence of the next leader.
Since the Bianconi-Barabasi model is considered a stylized model which capture salient feature of the evolution of the World-Wide-Web these super-hubs have been usually identified as major players such as Google, Facebook etc. 
These results have been confirmed by numerous studies and mathematical rigorous results
\cite{ergun2002growing,borgs2007first,ferretti2008dynamics,dereich2017nonextensive,iyer2023degree,fountoulakis2022condensation,javarone2013quantum}

At the network level, quantum statistics have been also shown to describe equilibrium network ensembles, such as 
 Exponential Random Graphs~\cite{park2004statistical,anand2009entropy,garlaschelli2009generalized}. In particular the marginal probability of a link takes the form of a Fermi-Dirac occupation number for unweighted networks while takes the form of the Bose-Einstein occupation number for weighted networks. As opposed to the emergence of quantum statistics in the non-equilibrium growing network models discussed before, in equilibrium network ensembles~\cite{park2004statistical,anand2009entropy,garlaschelli2009generalized}  the fundamental reason for the emergence of the quantum statistic is not very surprising. Indeed the network ensembles define statistical mechanic models determining the unweighted and the weighted adjacency matrix entries which can take in one case only values zero or one, and in the other case can take any non-negative integer values and are effectively treated as occupation numbers of energy states. 

\begin{figure}[t]
    \centering
    \includegraphics[width=\textwidth]{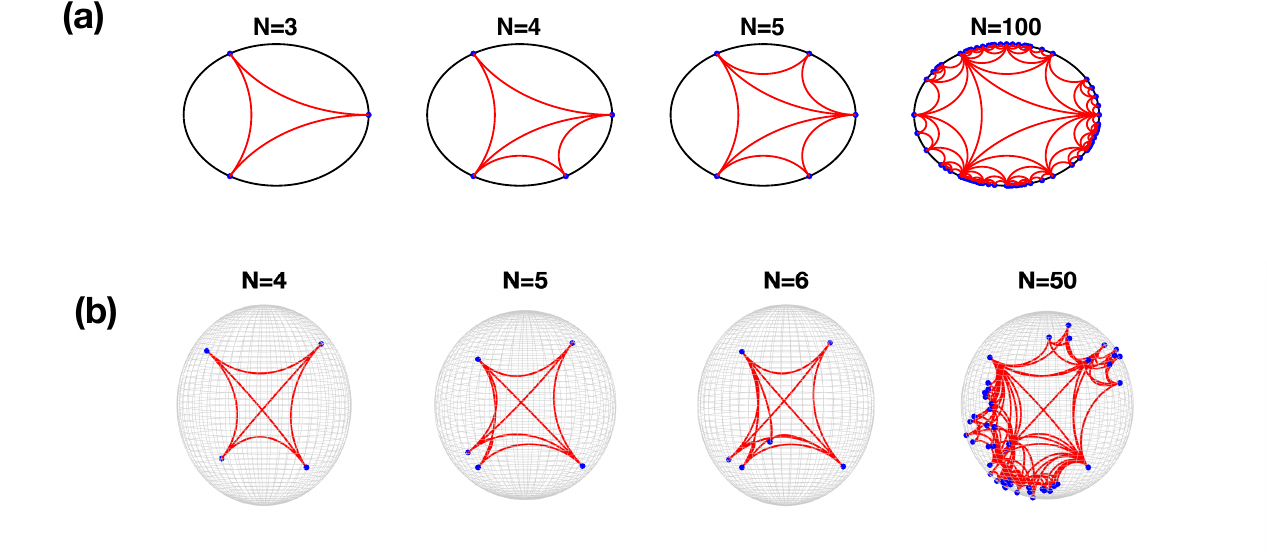}
     \caption{Schematic representation of the evolution of the ``Network Geometry with Flavor" (NGF) ~\cite{bianconi2017emergent,bianconi2016network} with flavor $s=-1$ and inverse temperature $\beta=0$ for dimension $d=2$ (panel (a)) and dimension $d=3$ (panel (b). Reprinted figure from Ref. \cite{bianconi2017emergent}.}
    \label{fig:NGF_model}
\end{figure}
\begin{figure}[t]
    \centering
    \includegraphics[width=\textwidth]{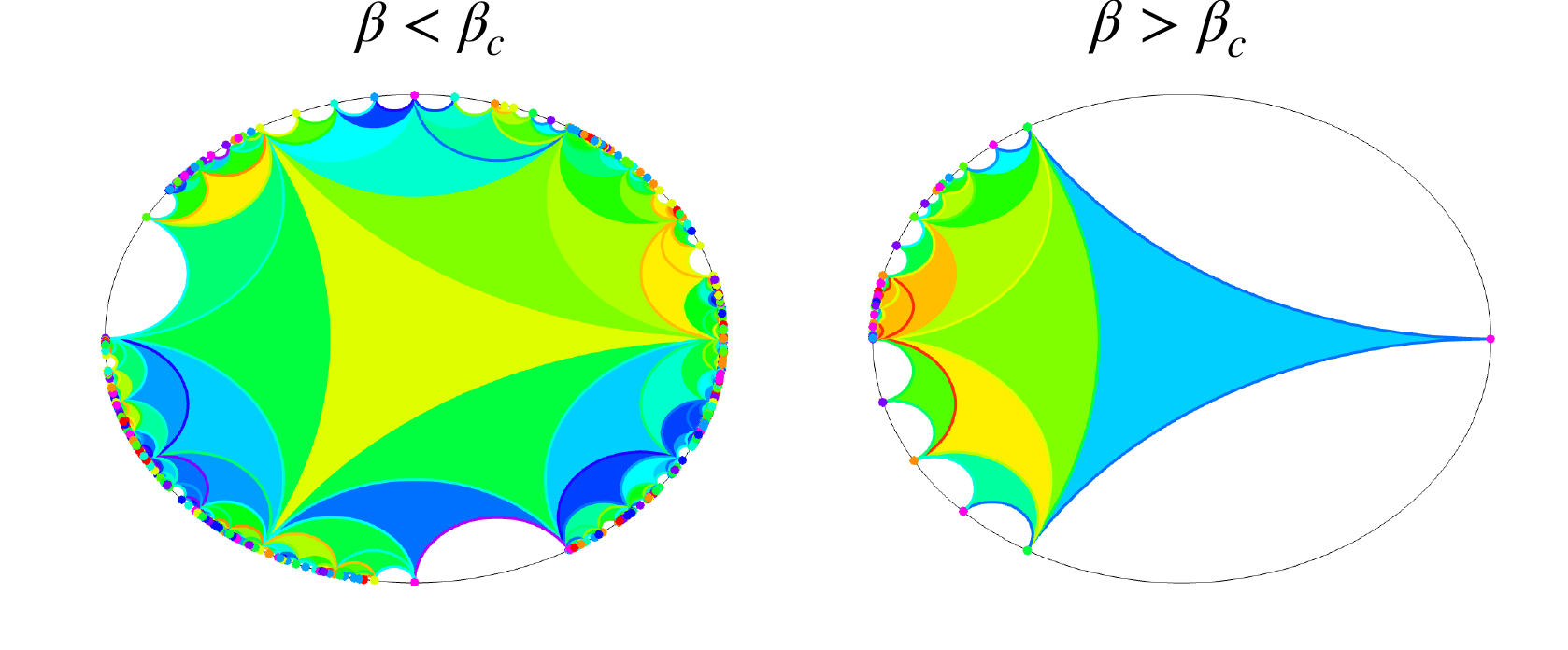}
    \caption{Visualization  ``Network Geometry with Flavor" (NGF) with flavor $s=-1$ model in dimension $d=2$ with intrinsic energies of nodes, links and triangles (visualized with colours) above 
    ($\beta<\beta_c$) and below 
    ($\beta>\beta_c)$ their topological phase transition. The network displays emergent hyperbolic geometry and for $s=-1$ generates a random Farey graph. For $\beta>\beta_c$ the Farey graph develops in a fairly distributed way in every direction of the hyperbolic plane, whereas for $\beta>\beta_c$ the evolution proceed along a spine changing the scaling of the network diameter.}
    \label{fig:NGF_cond}
\end{figure}

\subsection{Emergent quantum statistics in higher-order networks}

Quantum statistics emerge as well in higher-order networks revealing new unexpected interplay with the higher-order network structure.
The  higher-order (simplicial complex) model where quantum statistics emerge are the NGFs\cite{bianconi2017emergent,bianconi2016network,bianconi2021higher}. The NGFs generalize the network models covered in the previous paragraph 
and reduce to the Bianconi-Barabasi model for dimension $d=1$. However they greatly extend this model as they can be formed by gluing triangles $d=2$, tetrahedra $d=3$, etc as well as regular polytopes such as squares, cubes, icosahedra, orthoplexes, etc. \cite{mulder2018network}

NGF are simplicial or cell complexes that grow in time by the subsequent addition of their building blocks (triangles, tetrahedra, etc.) which are attached to the existing simplicial complex by a combinatorial rule that depends on a parameter $s$ called {\em flavor} in addition to the the inverse temperature $\beta$ defining the fitness defined in the previous paragraph.  An interesting property of NGFs is that although their growth obeys a purely combinatorial rule the NGFs display an emergent hyperbolic geometry (obeying Gromov $\delta$-hyperbolicity condition for any value of the flavor $s$ \cite{millan2021local}).

For observing the emergence of the quantum statistics, we need to assign the {\em energies} not only to  each node of the simplicial complex but also to each link, triangle, tetrahedra and so on. This is done by assigning random energies to the nodes and attributing to the link the sum of the energy of its two end nodes, to the triangle the sum of the energies of its three nodes and so on. In this way, using the definition given by Eq.(\ref{eq:fitness}) a fitness value is assigned to each node, link, triangle etc.
Interestingly in this set-up the same higher-order network can represent several  statistics at the same time, each of them characterizing the statistical properties of the $\delta$ dimensional faces of the NGF (see Table \ref{table2}).
For instance in dimension $d=3$ and for flavor $s=-1$ the statistical properties of triangles, links, and nodes are representing respectively the Fermi-Dirac, the Boltzmann and the Bose-Einstein statistics,~\cite{bianconi2016network} whereas
in $d=2$ and flavor $s=-1/m$ with  $m\in \mathbb{N}$, and $m>1$  we have that the statistical properties of  links and nodes represent respectively the Fermi-Dirac and the Bose-Einstein statistics \cite{cinardi2019quantum}.

Interestingly NGF can also undergo a topological phase transition if the temperature is lowered below a threshold $T_c=1/\beta_c$.
In this phase transition the diameter of the network changes scaling \cite{bianconi2015complex} and while for $s=1$ the diameter grows slower than logarithmically with the network size (the network is highly compact) for $s=1,0$ the diameter grows polynomially with the network size for $s=-1$ developing a so called {\em spine} (see Figure \ref{fig:NGF_cond}).

While mathematical rigorous results have confirmed the emergence of quantum statistics in the high-temperature regime~\cite{fountoulakis2022dynamical}, the rigorous mathematical characterization of the topological phase transitions of NGF is quite challenging and many mathematical research questions remain still open  requiring further in depth investigation.

\begin{table}
\center
\caption{\label{table2}  In a $d$-dimensional NGF of flavor $s$,~\cite{bianconi2016network,bianconi2017emergent,cinardi2019quantum} the statistical properties   of the generalized degrees  of $\delta$-dimensional simplices with  energy $\epsilon$   follows either the Fermi-Dirac, the Boltzmann or the Bose-Einstein statistics  depending on the values of the dimensions $d$, $\delta$ and $s$. Reproduced from Ref. \cite{cinardi2019quantum} \copyright IOP Publishing. Reproduced with permission. All rights reserved.” }
\footnotesize

\begin{tabular}{@{}llll}
\hline
flavor &$s=-1$&$s=0$&$s=1$\\
\hline
$\delta=d-1$&Fermi-Dirac &Boltzmann&Bose-Einstein\\
\hline
$\delta=d-2$&Boltzmann&Bose-Einstein& Bose-Einstein\\
\hline
$\delta\leq d-3$&Bose-Einstein&Bose-Einstein& Bose-Einstein\\
\hline
\end{tabular}\quad\quad \quad\quad
\footnotesize
\begin{tabular}{@{}llll}
\hline
flavor &$s=-1/m$\\
\hline
$\delta=d-1$&Fermi-Dirac \\
\hline
$\delta\leq d-2$&Bose-Einstein\\
\hline
\end{tabular}
\end{table}

\subsection{Relation to quantum gravity research questions}

An interesting question is to what extent network models representing quantum statistics relate to quantum gravity approaches. As we discussed in the beginning of this Section, Roger Penrose was the first to postulate a discrete and combinatorial spacetime.
Currently a large variety of quantum gravity approaches describe a  discrete spacetime \cite{CDT,Rovelli,Oriti}. This reflects in some cases  a  fundamental belief that spacetime should be discrete at the Planck scale. Alternatively, a discrete spacetime may be chosen even if the spacetime is assumed to be inherently continuous.
Indeed a discrete spacetime is   mathematically convenient as dealing with discrete structures enforces a cutoff that allows to regularize the theory escaping the dangers of non-renormalizability.
An important scientific problem that arise in this context 
is the identification of the characteristics of discrete spacetimes that correspond to the different quantum gravity approaches. In particular great attention has been addressed in characterizing the spectral properties of these emergent discrete geometries defining the effective spectral dimension ---typically considered to be the measure of dimension--- of the networks. 
This research line has  lead to the development of new concepts and ideas such as the fractal dimensionality of spacetime and the scale-dependent spectral dimension \cite{Dario,benedetti2009spectral}. Interestingly the characterization of the spectral dimension of  discrete spacetimes emerging from different quantum gravity models has been recently considered important to classify quantum gravity approaches and to determine whether  these different theories  define universal predictions valid across different approaches \cite{calcagni2013probing}.
Importantly
the NGFs do not only show the emergence of the quantum statistics and hyperbolic geometry but  they also display the emergence of a (non-universal) spectral dimension \cite{mulder2018network} that   can be inferred by quantum probes \cite{nokkala2020probing}. The emergence of a finite spectral dimension is observed not only  in simplicial but also in  cell complexes, i.e. higher-order networks not only formed by nodes and links triangles but also formed by squares, tetrahedra and so on.
Moreover in  the framework of the NGFs it has also been shown that the notion of the  spectral dimension extends also to the higher-order network level and can be used to characterize the spectrum of higher-order Laplacians \cite{reitz2020higher}.
Interestingly in the NGFs the  spectral dimensions depend on the order of the Laplacian~\cite{reitz2020higher}, the dimension of the simplicial complex~\cite{mulder2018network} and  the nature of  building blocks of the cell-complexes~ \cite{millan2021local}. Therefore while the presence of a finite spectral dimension seems to be a universal property of all the different variants of NGFs, the value of their spectral dimension is highly non-universal~\cite{bianconi2021higher}. It is still an open question whether these results are due to the highly heterogeneous structure of the NGFs.

\subsection{Discussion and future directions}

The Barabasi-Bianconi model has attracted large interest in the network science and in the mathematical community. In these communities most of the attention has been addressed to the characterization of the networks in the Bose-Einstein condensation phase. Indeed while above the critical temperature the mapping to the Bose gas completely captures the  statistical properties of the network, below the critical temperature there are some important differences. In fact the networks in the Bose-Einstein condensed phase are dominated by a succession of super-hub nodes that dominate the network structure, while in the Bose gas it is a single energy level,  the fundamental state that acquires a finite occupation number.
In terms of the application of this model to describe the competition for links in a network what happens in this condensed phase is of major importance. Major questions that arise in this context are: is it always the best (lower energy) node to win? is today's winner destined to be overcome by an even better node arising in the future? Is the fraction of links connected to the winner node really extensive or there are logarithmic corrections to the linear scaling due to this temporal effects?

Although much progress has been made on these very challenging questions many questions remain open. In particular, if the transitions observed on networks displaying emergent quantum statistics is still posing interesting mathematical questions, the transition observed on the NGFs are mostly unexplored so far.

While all these questions are related to the classical consequences of Bose-Einstein condensation for network structures, an important open question  is whether the representation of quantum statistics embodied by these networks can be  harnessed by quantum gravity approaches or even quantum  technologies or applications.

\section{\label{sec:propertiesII}Quantum algorithms for network inference}

\subsection{Quantum concepts useful for complex networks}
\label{sec:algorithms}

In the last decade there has been increasing attention devoted to the formulation of quantum algorithms and observables that can reveal important properties of classical networks \cite{biamonte2017quantum}. 
In this Section we will cover this very innovative research direction which could flourish in the next years with the new generation of quantum computers. This research direction corresponds in our classification to the quantum-enhanced research line (see Fig. 1).

Recently, quantum algorithms have been proposed to capture a wide range of structural properties of classical networks starting from the quantum degree distribution all the way up to quantum community detection and quantum link prediction.
In this class of works the focus is not always to formulate algorithms that are faster than their classical counterpart, rather more often the goal is to capture structural properties that can be neglected or not sufficiently highlighted by classical network measures.

Even more recently with the increasing attention addressed to higher-order networks, it has become clear that quantum  concepts are also key to treat higher-order dynamics. In particular the discrete topological Dirac operator \cite{bianconi2021topological} can be used to formulate a gauge theory \cite{bianconi2023dirac} for topological spinors, which have an explicit geometrical interpretation and capture the dynamics of simple and higher-order networks defined on nodes, links, and even triangles and higher-dimensional simplices of simplicial complexes.  This gauge theory can be used to define an emergent mass of simple and higher-order networks which depends on their geometry and topology \cite{bianconi2023mass}. 
 
The topological Dirac operator can also be used to characterize the dynamics of coupled  (classical) topological signals \cite{battiston2021physics}. In particular the  Dirac operator allows to define a new class of dynamical processes on network and simplicial complexes, revealing new physical phenomena as demonstrated by its application to Dirac synchronization, Dirac Turing patterns and Dirac signal processing~\cite{calmon2022dirac,calmon2023local,giambagli2022diffusion,calmon2023dirac}.

The topological Dirac operator  hence can be quite transformative in our way to treat dynamics on networks, usually only associated to the nodes of the network.

\subsection{Quantum  inference of networks}
\subsubsection{Quantum inference of network structure}
Large scientific attention has been addressed to define algorithms that  probe the network topology using quantum random walks.
For instance in Ref. \cite{faccin2013degree} a quantum definition of the degree of a node of the network is proposed starting from the long-time probability distribution for the location of a  quantum walker on the complex network. In this work it is shown that for low-energy quantum walkers this probability distribution coincides with the one of the classical random walk, and is hence closely related to the degree distribution of the network. However for higher-energy states the classical and the quantum distributions differ, providing a quantum generalization of the concept of degree distribution.
Also based on quantum walks in Ref. \cite{faccin2014community} a quantum algorithm for community detection is proposed. This algorithm is a hierarchical clustering algorithm where the similarity between the nodes are based on quantum transport probability and state fidelity. The proposed algorithm is tested on light-harvesting complex finding
good agreement with the partitioning of nodes used in quantum chemistry.
\begin{figure}[t]
    \centering
    \includegraphics[width=0.5\textwidth]{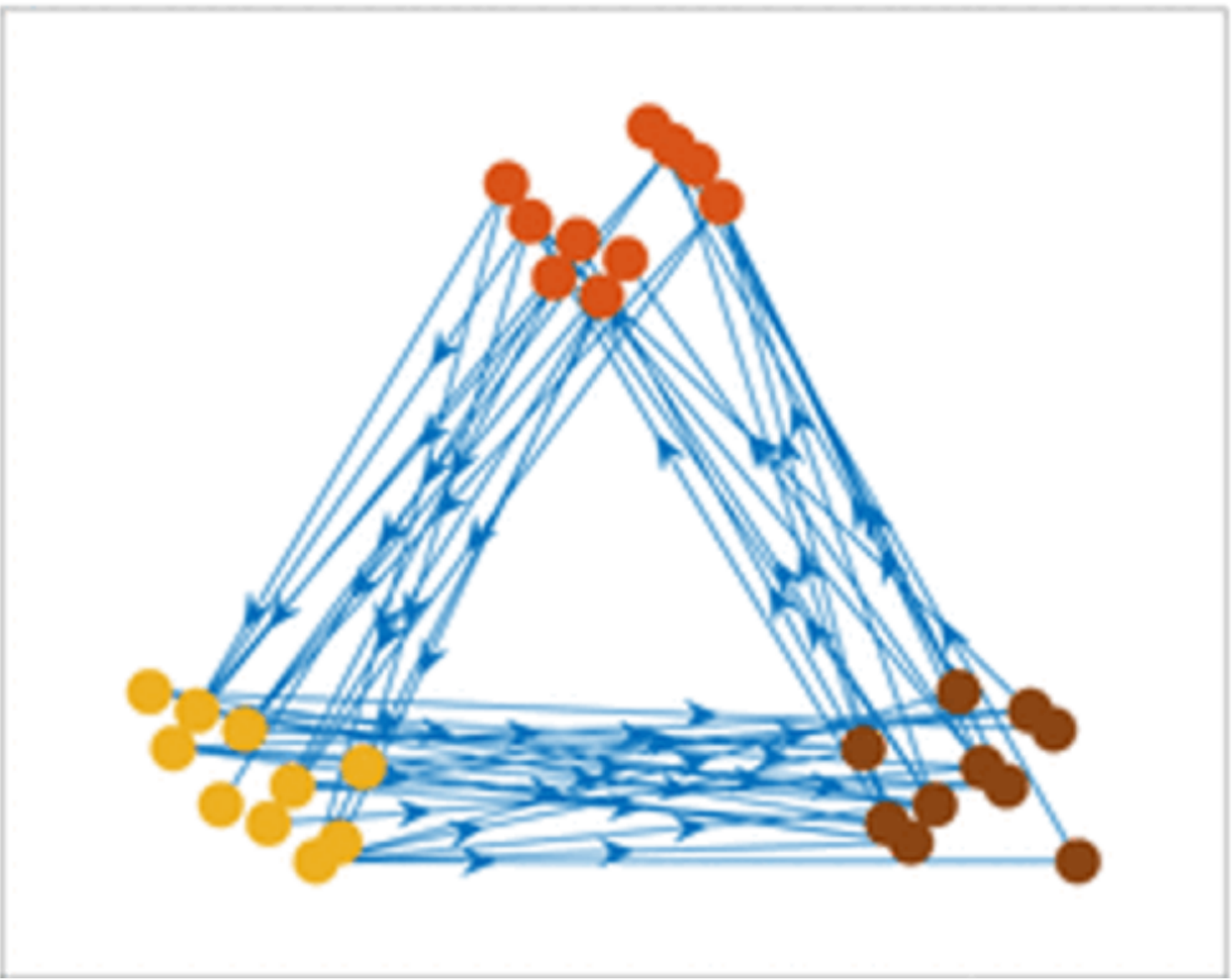}
    \caption{Example of network embedding generated extracting information from the eigenvector of the magnetic Laplacian in which directed edges are associated to a complex angle $2\pi/3$. Reprinted figure from Ref. \cite{gong2021directed}. }
    \label{fig:magneticL}
\end{figure}
An alternative approach for detecting community structure in networks that is becoming increasingly popular uses instead embeddings generated from the eigenvectors of the magnetic Laplacian. The magnetic Laplacian  \cite{shubin1994discrete,smilansky2007quantum,fanuel2017magnetic,fanuel2018magnetic,f2020characterization,gong2021directed} is a complex valued Hermitian variant of the Laplacian that captures non trivial mesoscale network structures. In particular the magnetic Laplacian  \cite{shubin1994discrete,fanuel2017magnetic,fanuel2018magnetic,f2020characterization,gong2021directed} can capture community structure in directed networks and can detect cyclic patterns of connections such as the ones that are formed by three communities of nodes where the nodes of each community link preferentially with the nodes of other communities forming effectively a mesoscale ``triangle" formed by the three communities (see Figure \ref{fig:magneticL}). Having complex matrix elements the eigenvectors of the magnetic Laplacian display a complex phase. For instance in the case of the cyclic community structure described before the phases on the different communities will differ by an angle $2\pi/3$ reflecting the three communities that are present in the network.
The interest on the magnetic Laplacian has recently further fuelled an interesting line of research aimed at exploiting  complex weights in network science. In this framework new  network dynamical processes  have been proposed  including consensus models~\cite{mason} that generalize the Schr\"odinger-Lohe model ~\cite{lohe2010quantum,lohe2009non},  complex weights random walks  \cite{yu} and quantum Hopfield model \cite{torres2023dissipative}.

Link prediction is one of the most challenging and used classical inference algorithm for reconstructing missing or hidden interactions. In Ref. \cite{moutinho2021quantum} a quantum algorithm based on a quantum walk is used to infer missing links extracting information from paths of even and odd path length. Here the emphasis is on the efficiency and speedup of the quantum algorithm with respect to the classical counterparts while showing that the quantum algorithm retains a   performance comparable with classical algorithms.
 
Network symmetries are fundamental properties of lattices and tree structures traditionally studied in condensed matter and in quantum information. Indeed lattices are characterized by well known symmetry groups. Moreover,  symmetric structures formed by two binary trees connected at their leaves have been shown by Farhi and Gutmann~\cite{farhi1998quantum} to allow an exponential speed up of the quantum random walk with respect to the classical random walk. Interestingly also complex networks have non trivial symmetries~\cite{macarthur2008symemtry,sanchez2020exploiting} and detecting isometries is an important NP hard problem of computer science.
Recently  continuous time random walks have been used to detect symmetries or quasi symmetries in network structures and isometries of quasi-isometries among different networks. In Ref. \cite{rossi2013characterizing} the symmetries of a given network are studied. To this end, for every pair of nodes two states localized on them are prepared corresponding to amplitudes either in phase or antiphase. 
The Jensen-Shannon divergence between the density of states corresponding to the two random walks is proved to achieve its maximum value if the two selected nodes are symmetrically placed (note however that is a necessary by not a sufficient condition of ensuring symmetry). Interestingly the  Jensen-Shannon divergence can be used also to detect or infer quasi-symmetries because it is a measure that is robust to the introduction of random perturbations of the symmetries. In Ref. \cite{rossi2012approximate} this research line is extended to compare and reveal symmetries between two different networks. In particular given two unconnected networks, the authors suggest to construct a joined network where links between the nodes of the two networks are inserted so that  each node originally in network 1 is connected to all the nodes originally in network 2 and vice versa.
On this joined networks two continuous random walk walks are studied, which evolve from initial states constructed in such a way that the amplitude corresponding to the states associated to the nodes of the two networks are either in phase or antiphase. The Jensen-Shannon divergence between the two different random walks is then used to detect isometries and more in general to construct kernels between the two networks that can then be used by machine learning approaches to classify networks.

\subsubsection{Quantum centrality measures (Quantum PageRank)}

PageRank \cite{brin1998anatomy} is undoubtedly the most successful Network Science algorithm. It is the original algorithm used by the Google search engines and since then its use has been extended to rank in order of decreasing importance nodes in a variety of networks, including social, technological and biological networks.

The classical algorithm runs polynomially and scales well with the network size, however the PageRank algorithm in practice is run on a continuous basis to rank all the pages on WWW and to address time-dependent changes of the webpage content and the network topology.

From the quantum perspective two major questions arise. The first research question regards  the possible speedup that a quantum algorithm to calculate the classical PageRank can achieve \cite{garnerone2012adiabatic}. The second research question regards the formulation of the Quantum PageRank that can extend the classical definition and retain its good performance while being 
of potential use to rank quantum webpages. Quantum webpages indicate the nodes of a quantum network with quantum capabilities such as reading in/out quantum states. In other words quantum webpages do not require a fully fledged quantum computer and can be realized by quantum storage devices and quantum memories.

Interestingly the answer to these 
questions is based on the definition of the same matrix, the Google matrix $\mathbb{G}$ defined as 
\begin{equation}
\mathbb{G}=\alpha E+(1-\alpha)\frac{{\bf 1}}{N}
\end{equation}
where $\alpha\in (0,1)$ is a parameter of the model analogous of the ``teleportation" parameter of the classical PageRank, ${\bf 1}$ is the matrix of all elements equal to one, $N$ is the network size and $E$ is the transition matrix of a classical random walk where every zero row corresponding to a node with zero out-degree is substituted with a row in which all elements have values $1/N$.
The Google matrix can be shown to be  both irreducible and primitive.

The classical PageRank can be obtained by applying $k$ times the Google matrix to an initial guess for the ranking of the nodes and going in the limit $k\to\infty$.
In Ref. \cite{garnerone2012adiabatic} the Authors have proposed a quantum annealed algorithm to speed up the calculation of the classical PageRank showing that the improvement on the performance of the classical algorithm is more  significant if the out-degree distribution of the network is broad (as it is the case for our current ---classical--- WWW).

\begin{figure}[htb]
    \centering
 \includegraphics[width=1.0\textwidth]{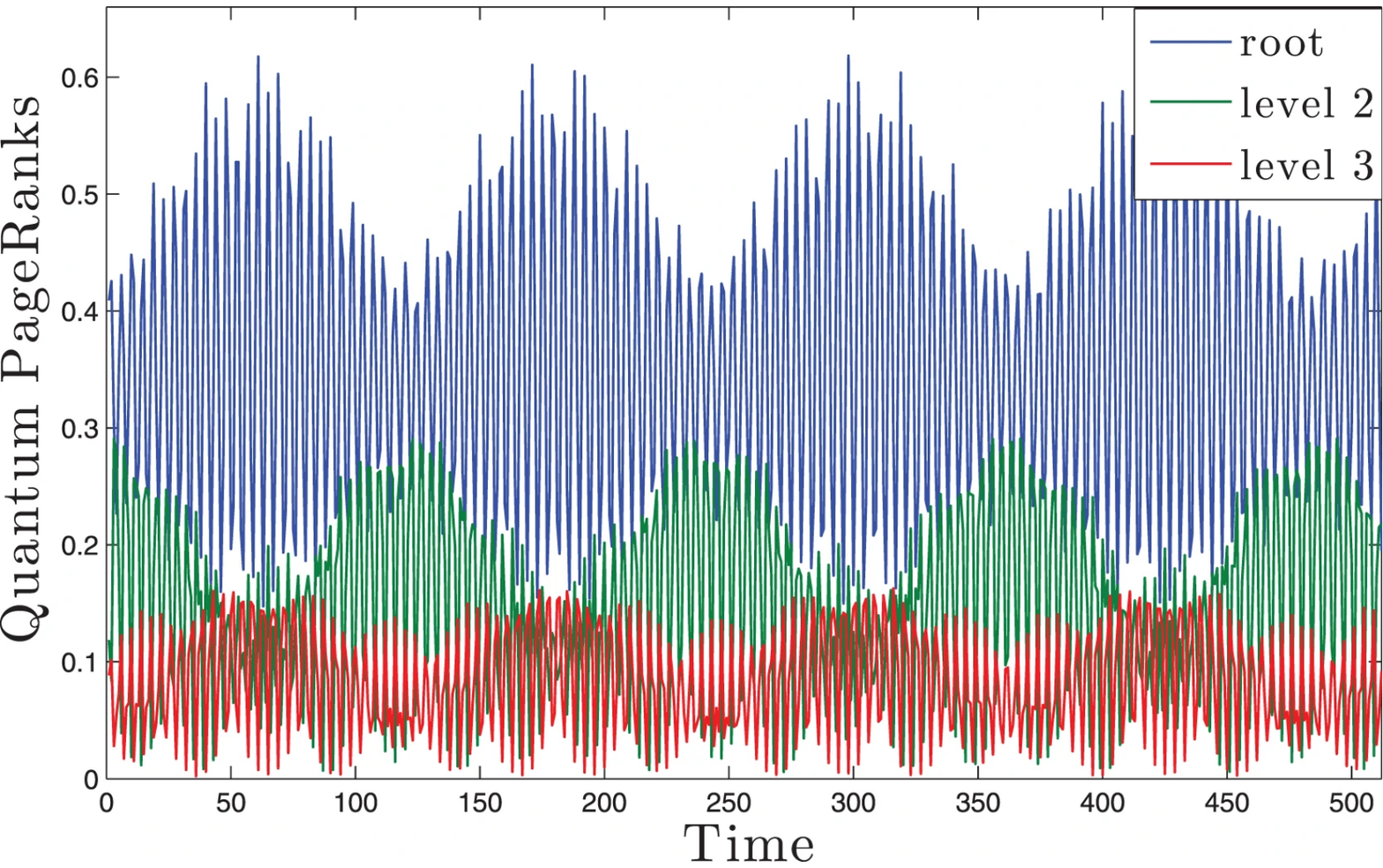}
    \caption{Quantum Page Rank of nodes in the root, in the first level and in the second level of a binary tree plotted as a function of time. Reproduced figure from Ref.~\cite{paparo2012google}. Creative Commons Attribution-NonCommercial-ShareALike 3.0 Unported License \text{http://creativecommons.org/licenses/by-nc-sa/3.0/}}
    \label{fig:QPageRank}
\end{figure}

The quantum PageRank \cite{paparo2012google} provides instead an alternative ranking of the nodes providing a Quantum PageRank class on which the classical PageRank can be embedded and allowing for a classical computation belonging to the complexity class P.
The Quantum PageRank is based on the Google matrix and uses Szegedy procedure to quantize the Markov chain algorithm that provides the classical PageRank.
The resulting quantum Pagerank of the node of the network is a ranking that fluctuates as it depends on the time duration of the quantum evolution (see Figure \ref{fig:QPageRank}). Therefore it is possible to characterize the fluctuating instantaneous ranking given by the quantum PageRank and the ranking provided by the average of the instantaneous quantum PageRank over a suitably large time window.
Several works \cite{paparo2012google,paparo2013quantum,loke2017comparing} have investigated the performance of the quantum PageRank algorithm on real WWW network data and on network models including the ER model, the Barab\'asi-Albert model, and random scale-free networks. The rankings obtained with the quantum PageRank are compared with the ranking of the classical PageRank showing that the classical ranking of the top ranked nodes is always within the range of fluctuation of their corresponding quantum PageRank. However the quantum PageRank in particular for networks with broad degree distribution has the advantage that it can distinguish  better secondary hubs and removes the degeneracy in the nodes with low ranks. Moreover,  comparing the results obtained on different network topologies it has been shown that the quantum PageRank can really capture their differences.  In particular scale free networks are characterized by corresponding random walks that are localized \cite{paparo2013quantum,loke2017comparing,perra2009pagerank} while random ER networks are characterized by random walks that are not localized.

An alternative approach \cite{sanchez2012quantum} uses the theory of quantum open systems to define a Quantum Page Rank algorithm with a single stationary state. The approach allows to interpolate between  purely classical and purely quantum Page Ranks. It is shown that a certain level of quantumness is beneficial to speed up the algorithm.

Pioneering works are using CTQWs 
with a non-Hermitian Hamiltonian given by the directed (asymmetric) adjacency matrix of the network to define quantum centralities of the nodes of the network that generalize well the classical eigenvector centrality. The benefit of these simplified definitions of node centralities is that they can be experimentally implemented via linear optics circuits and single photons \cite{izaac2017centrality,wang2020experimental}.

\subsubsection{Von Neumann entropy of networks}
Statistical mechanics is core for evaluating the information content encoded in networks. The Shannon entropy \cite{anand2009} has been widely used to characterize the information encoded in network ensembles and has been key to classify these networks as microcanonical, canonical and grandcanonical \cite{anand2009,anand2010,bianconi2022grand} according to the corresponding classification of ensembles of particles.\\
The von Neumann entropy allows the investigation of the structure of complex network using tools of quantum mechanics based on the spectral properties of the network. Thus the von Neumann entropy, differently from the Shannon entropy of network ensembles,  can be used to assess the information content encoded in a single network.

Given a definition for the density matrix of a network which is positive semi-definite and normalized to one, the  von Neumann entropy of a network has the usual definition and the quantum Jensen–Shannon divergence to measure the dissimilarity between two networks  can be defined as well.  
Therefore the crucial point for defining the von Neuman entropy is to make a suitable choice for the density matrix. Originally it was proposed \cite{passerini2008neumann} to consider a density matrix given by the  Laplacian, normalized with the sum of degrees i.e.
\begin{equation}
\rho=\frac{L}{\avg{k}N}.
\end{equation}
The resulting von Neumann entropy can be highly affected by the degree distribution if 
the network is scale-free and  it correlates well with other classical entropy measures of the network \cite{anand2009entropy,anand2011shannon}.
This definition of the von Neumann entropy  has been adapted to multiplex networks in \cite{de2013mathematical} and the corresponding Jensen-Shannon divergence has been used to compare and compress/clusterize different layers of real multiplex networks \cite{nicosia}. Other works have also considered the use of the normalized Laplacian   ${\bf\hat{L}}$ instead of the unormalized graph Laplacian {\bf L} \cite{minello2019neumann}. 

\begin{figure}[t]
    \centering
    \includegraphics[width=\textwidth]{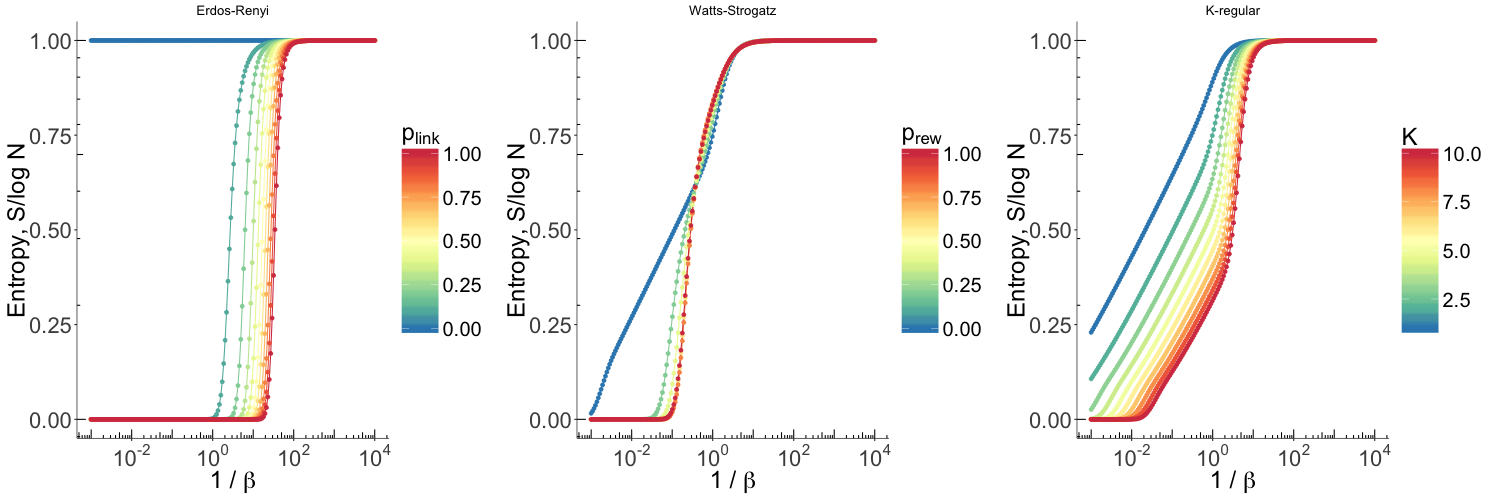}
    \caption{The von Neumann entropy of key network models (from left to right: ER networks, SW networks and $K$-regular networks) is shown as a function of the temperature. Figure reproduced from Ref.~\cite{de2016spectral}, doi: \text{https://doi.org/10.1103/PhysRevX.6.041062}.}
    \label{fig:von_neumann}
\end{figure}

Later it was proposed to use the alternative expression for the density matrix of a network \cite{de2016spectral}
\begin{equation}
\rho=\frac{e^{-\beta L}}{\mbox{Tr}e^{-\beta L}}.
\end{equation}
In Figure $\ref{fig:von_neumann}$ we show the von Neumann entropy of key network models as a function of the inverse temperature $\beta$ as reported in Ref. \cite{de2016spectral}.
This Gibbs like definition of the density matrix is significantly affected by the low eigenvalues of the Laplacian relating to the long time diffusion dynamics on the network.
Consider a classical diffusion dynamics  starting from a localized state on a single node chosen randomly among all the nodes of the network. In expectation, the  probability of return on the seed node after a time $t$ is given by 
\begin{equation}
P(t)=\sum_{\lambda}e^{-\lambda t }.
\end{equation}
Thus the von Neumann entropy   has a classical interpretation as the number of eigenmodes that are important for paths that evolve up to time $\tau=\beta$ \cite{gili}.
The derivative of the von Neumann entropy with respect to $\ln \beta$ has been recently proposed as a measure characterizing the temporal scale at which diffusion processes display significant dynamical transitions or cross-overs \cite{gili,baccini2022weighted}.

Although the von Neumann entropy and its variation is the most popular definition of quantum entropy of a network, alternative definition exist including the one formulated in Ref.~\cite{garnerone2012bipartite} where the Authors first introduce a mapping between the adjacency matrix of a network and pure quantum bipartite states and subsequently show  that the associated entanglement entropy captures important structural properties of the graph.

\subsection{Quantum higher-order networks and the topological Dirac operator}

The research on quantum higher-order networks \cite{bianconi2021higher} constitute a new promising field of research given the increasing interest of the network science community on higher-order interaction networks.
Higher-order interactions are known to be of fundamental importance in quantum physics as revealed by the great scientific interest on the Sachdev-Ye-Kitaev model \cite{chowdhury2022sachdev,maldacena2016remarks}. However the phenomenology of the Sachedev-Ye-Kitaev is not related to network effects as the model is defined on a fully connected network.
 
Early work on higher-order networks include the extension of graph quantum states \cite{hein2004multiparty,hein2006entanglement} to hypergraph quantum states \cite{guhne2014entanglement,rossi2013quantum} prepared from single qubits by performing operations between $k$ connected qubits, with $k\geq 2$. In Ref. \cite{rossi2013quantum} it is  shown that hypergraph states are in one-to-one correspondence with  real-equally-weighted  (REW) states that are essential for quantum algorithms while the graph states in which $k$ is fixed to be equal to two,(i.e. $k=2$) only constitute a subsets of REW states.

Higher-order networks and in particular simplicial complexes are also  interesting because they can shed light in the interplay between network topology on dynamics \cite{bianconi2021higher,battiston2021physics} revealing new physics and phase transitions. Topology is currently gaining significant attention and offers new paradigms for describing classical dynamics inspired by quantum mechanics, which includes among the other applications the characterization of edge currents in biological synthetic biology \cite{tang2021topology} and game theory \cite{yoshida2021chiral}.

Higher-order networks provide a mathematical framework in which the interplay between topology and dynamics is transformative. In particular higher-order networks can sustain topological signals, i.e. dynamical variables not only associated to the nodes of the networks, but also to their links, and in a simplicial complex even triangles, tetrahedra and higher-order simplices. These topological signals can undergo collective phenomena~\cite{millan2020explosive,carletti2023global} which  display very new physics with respect to the associated dynamics defined exclusively on nodes. These topological signals can be studied using Hodge-Laplacian operators \cite{bianconi2021higher} that describe diffusion from $n$-dimensional simplices to $n$-dimensional simplices through either $n-1$ or $n+1$ dimensional simplices. Indeed while the graph Laplacian describes diffusion from nodes to nodes occurring  through links, the $1$-Hodge Laplacian can describe diffusion from link to link occurring through nodes or through triangles. The spectral properties of the Hodge Laplacian encode for important topological features such as the Betti numbers and  allow generalized higher-order diffusion \cite{torres2020simplicial}. The research in the field is rapidly growing and interestingly the Hodge-Laplacians have also been used to define a higher-order von Neumann entropy that encodes relevant higher-order network properties \cite{baccini2022weighted}. 

A very active and very promising research direction for treating the dynamics of coupled topological signals involves the discrete topological Dirac operator. The discrete topological Dirac operator \cite{bianconi2021topological} of networks and simplicial complexes is a topological operator, rooted in quantum physics, that has the ability to couple topological signals of different dimension.
The discrete topological Dirac operator has first been proposed in lattices by Kogut and Susskind \cite{kogut1975hamiltonian} and is fundamental to define the staggered fermions. Successively, it has been used in non-commutative geometry \cite{davies1993analysis} and then its spectral properties have been further investigated  in the framework of quantum graphs \cite{post2009first,hinz2013dirac}.

\begin{figure}[t]
    \centering
    \includegraphics[width=0.9\textwidth]{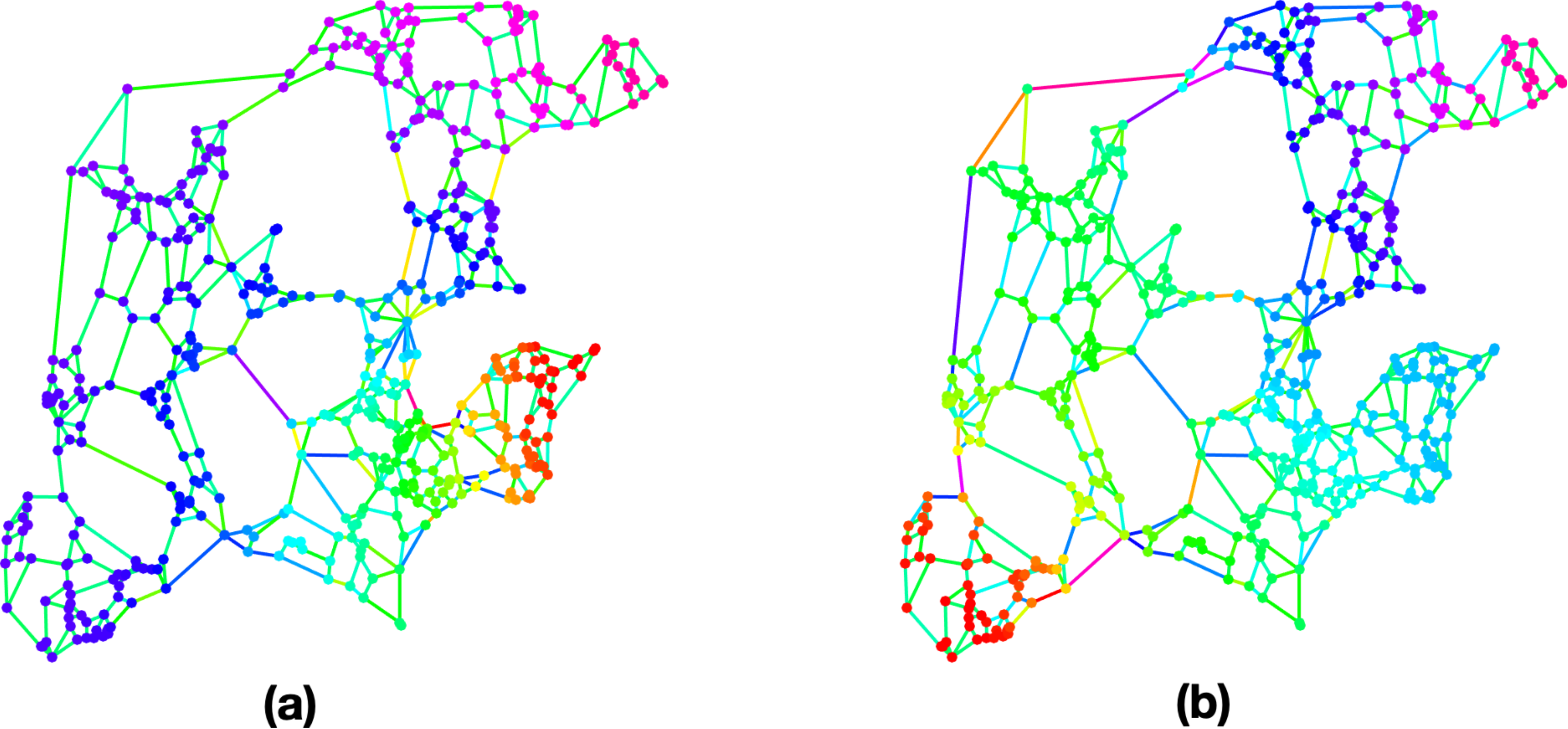}
    \caption{ Visualization of two eigenstates of the topological Dirac equation defined on the nodes and links of  a real  network. Figure reproduced from Ref.~\cite{bianconi2021topological}. The real network used for this figure is published in Ref. \cite{lee2017mesoscale}.}
    \label{fig:Dirac}
\end{figure}

The topological Dirac operator can be used to formulate a topological discrete Dirac equation \cite{bianconi2021topological} in which the spinor acquires a geometrical interpretation and is defined on each node and link of the network. Interestingly on general networks the harmonic model break the charge conjugation symmetry, leading to different matter-antimatter sectors\cite{bianconi2021topological}.
Moreover the  topological Dirac equation can be also generalized to  simplicial complexes by considering topological spinors defined on nodes, links, triangles, tetrahedra and so on. The topological spinors are determined by dynamical variables (cochains) defined on each simplex of the simplicial complex. In particular on a network the topological  Dirac operator acts on the topological spinor by coupling the dynamics on the links to the dynamics of the nodes. As the continuous Dirac operator, the discrete topological Dirac operator can be considered as a square-root of the Laplacian operator and admits both positive and negative eigenvalues. Interestingly for each positive eigenvalue there is a corresponding negative eigenvalue corresponding to the matter-antimatter symmetry. The corresponding eigenvectors are related by chirality. However an interesting aspect of the discrete topological Dirac equation is that in general the harmonic eigenvectors of the topological Dirac operator do not display the matter-antimatter symmetry \cite{bianconi2021topological}. 
 
The topological Dirac equation can be adapted to capture different directions of regular lattices, and can be used to study the interplay between topology and quantum dynamics on multiplex networks and simplicial complexes. Interestingly the topological Dirac operator allows to define a lattice gauge theory in which the fermion fields are taking values on both nodes and directed links which play the role of the fiber bundle of the network \cite{bianconi2023dirac}.

The discrete Dirac operator is currently gaining increasing attention in non-commutative geometry \cite{majid2023dirac} and in quantum graph literature \cite{anne2015gauss,athmouni2021magnetic,requardt2002dirac,parra2017spectral,miranda2023continuum}.

Recently the Dirac operator has been proposed to define classical higher-order dynamical models displaying relevant new physics.
In particular in Refs. \cite{calmon2022dirac,calmon2023local} Dirac synchronization  is proposed. Dirac synchronization is a  higher-order Kuramoto model with  phases associated to  nodes and links of a network, coupled to each other by the Dirac operator which display a discontinuous synchronization and an emergent rhythmic phase.
Moreover in Ref. \cite{giambagli2022diffusion} the Dirac operator is used to reveal novel mechanisms for the emergence of Turing patterns on both  nodes and links.

The topological Dirac operator is also useful to define inference algorithms. 
In  \cite{lloyd2016quantum}  the topological Dirac operator is used to formulate a quantum algorithm  that  calculates the homology of simplicial complexes. This algorithm is based on a representation of the simplicial complex as a quantum state  having as basis the set of the simplices of the simplicial complex. The  proposed quantum algorithm is shown to display an exponential speed-up over the best known classical algorithms for calculating homology. Recently in Refs. \cite{vasileios,xia} this approach has been extended also to propose a quantum algorithm for the calculation of persistent homology of simplicial complexes.

Another inference algorithm using the Dirac operator is  Dirac signal processing \cite{calmon2023dirac} that allows to jointly process signals defined on simplices of different dimensions.

\subsection{Discussion and future directions}

Quantum algorithms to infer relevant information from networks are only in their infancy however the field has already obtained important results that provide good foundations for further development.

One of the important aspects of quantum algorithms for complex networks is their strong connection to the network spectral properties. Indeed the most important operators and observables that have been defined, from the magnetic Laplacian, and the quantum Google matrix, to the Dirac operator and the von Neumann entropy, are based on the spectral properties of the discrete network structures under investigation.

However the algorithms differ significantly from their classical counterparts. The magnetic Laplacian introduces complex valued weights of the links capturing for their direction. The quantum Google matrix obeys dynamics that is not dissipative like the classical PageRank algorithm. Finally  the Dirac operator provides a fundamental change of the understanding of dynamical processed on networks because it describes a topological coupling between dynamical variables associated to nodes, links, and higher-dimensional simplices of  higher-order network structures.

These different ways to use quantum concepts to model, understand and extract information from networks have already shown their clear advantage as demonstrated in particular by the large success of von Neumann entropy measures of networks and the important role of the Dirac operator to describe new physics in networks and simplicial complexes.

One open problem that emerges in this context are whether quantum algorithms can further transform the landscape of combinatorial algorithms on networks, providing progress for instance in the  graph isomorphism problem. 

Additionally one important question is whether new mathematics is needed to treat networks and simplicial complexes.
In particular in the continuous Dirac equation, the Dirac operator is coupled to the algebra of gamma matrices.
An interesting question is whether also to analyse networks coupling the Dirac operator to a group could be key to capture the full geometry and topology of the data and to model coupled topological signals.

\section{\label{sec:comms}Quantum communication networks}

\subsection{\label{sec:comms_intro}Fundamentally different communication}

In quantum communication networks, photons are exchanged between distant nodes to facilitate distribution of cryptographic keys and entanglement as well as transmission of quantum information \cite{bassoli2021quantum}. Such networks have applications beyond those of their classical counterparts \cite{wehner2018quantum}, making this very vibrant research area fall into the quantum enhanced class in Fig.~\ref{fig:intro}, although certain aspects can also be considered to be network-generalized. They may be roughly divided into within reach and theoretical networks, which now in the early 2020s still correspond to quantum key distribution (QKD) and quantum information (QI) networks, although not as strongly as for example just in mid 2010s. Both are subject to the same fundamental limitations arising from the properties of quantum information, particularly to communication rate limits that decrease rapidly with distance travelled in optical fiber, making them metric networks. Indeed, much of the research has focused on designing architectures that tolerate the limitations \cite{briegel1998quantum,acin2007entanglement} or on improving the basic building blocks \cite{muralidharan2016optimal,heshami2016quantum,awschalom2021development}. There are many excellent contemporary treatises on the topic \cite{wehner2018quantum,razavi2018introduction,rohde2021quantum,wei2022towards}, however here a more concise account of the field is provided with a unique point of view emphasizing the network aspect. In particular, we focus on research where the architecture is fixed and it is asked what kind of topology it tends to lead to \cite{brito2020statistical,brito2021satellite} or how a given topology controls its performance \cite{perseguers2008entanglement,cuquet2009entanglement,harney2022analytical,harney2022end}. To this end we will also briefly introduce the relevant architectures.

\begin{figure}
    \centering
    \includegraphics[trim=0cm 0cm 0cm 0cm,clip=true,width=0.9\textwidth]{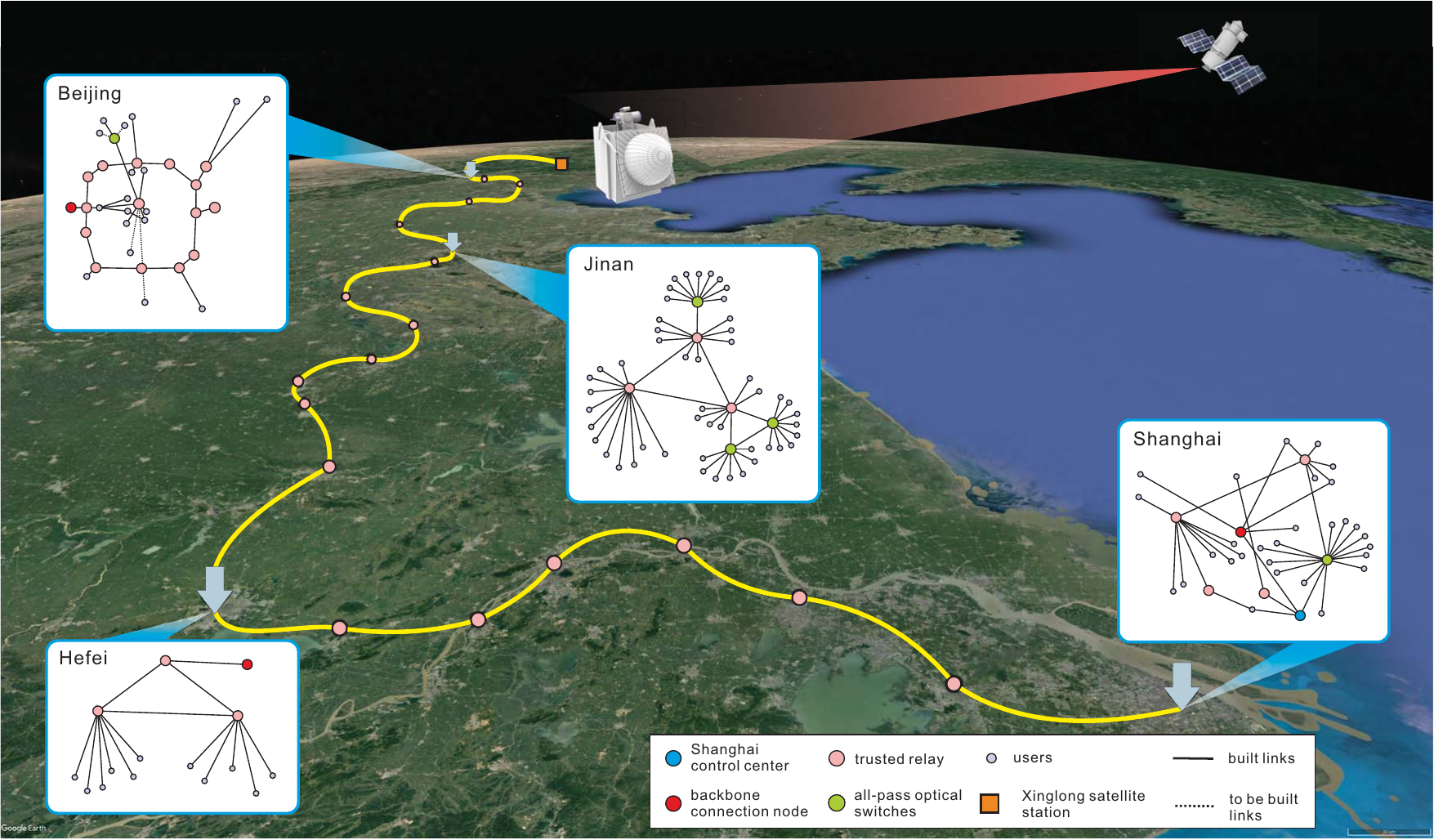}
    \caption{The Beijing-Shanghai backbone network connects four metropolitan area QKD networks and also features a long distance quantum satellite link covering 2600 km between Xinlong and Nanshan (not shown). Reprinted figure with permission from \cite{xu2020secure}. Copyright 2020 by the American Physical Society.}
    \label{fig:chinaQKDnetwork}
\end{figure}

By QKD networks we mean specifically the case where the transmission of photonic quantum states is limited by the total distance between participants and inbound photons can be only measured or forwarded. Then the actual messages will be classical and transmission of the states merely facilitates its encryption with quantum secure keys; consequently such networks have also been called semi-classical \cite{zhuang2021quantum} or partially quantum \cite{epping2016large}. Examples of past and present QKD networks include the DARPA network in Boston \cite{elliott2005current}, SECOQC network in Vienna \cite{peev2009secoqc}, the Tokyo quantum network \cite{sasaki2011field}, the Hefei quantum network \cite{chen2009field,chen2010metropolitan}, the London Quantum-Secured Metro Network \cite{lord2023london} and the Beijing-Shanghai backbone quantum network \cite{qiu2014quantum}. This latter network, shown in Fig.~\ref{fig:chinaQKDnetwork}, has since been used to connect four metropolitan area networks in Shanghai, Hefei, Jinan and Beijing and has been supplemented by a satellite link connecting two ground stations around 2600 km apart \cite{chen2021integrated}. In contrast, direct fiber links achieving reasonable rates cover distances of around 100 km  \cite{bedington2017progress,xu2020secure}. There is an interest in the integration with existing optical fibres both to save costs and because the communication protocols virtually always require classical communication as well \cite{orieux2016recent,razavi2018integrating}; the feasibility of such co-existence of both quantum and classical layers in the same fiber network has been demonstrated in particular in the Madrid Quantum Communication Infrastructure \cite{lopez2021madrid}. Although significant as testbeds for research and development which might later benefit QI networks as well, it is the case that fiber based QKD networks are either limited to a small service area or lack end-to-end security, inhibiting their growth. Recent theoretical results \cite{brito2021satellite,pirandola2021satellite,goswami2023satellite} for satellite based networks are quite encouraging, however, as are novel fiber schemes connecting next nearest neighbors \cite{lucamarini2018overcoming,minder2019experimental} which have been reported to achieve reasonable experimental rates beyond 400 km \cite{wang2022twin}. The boundaries are moving.

In QI networks also quantum information can be transmitted over long distances. While they form a broad class with many subcategories, the holy grail is the quantum Internet \cite{kimble2008quantum}: a global public commercial network capable of storing, processing and transmitting quantum information and entanglement. Photonic qubits can be transformed to stationary qubits stored in quantum memories and back as necessary---a process called quantum transduction \cite{kurizki2015quantum} achieved by quantum interconnects---and the entire network can be prepared into a nonclassical state. In conventional proposals the quantum layer is divided into physical and virtual layers, where the former is used only to distribute entanglement. All quantum information is transmitted over the virtual layer consisting of teleportation channels which consume the distributed entanglement as fuel. The network must then be able to at least generate and distribute, but ideally also store and accumulate entanglement. As networks, they would be not only metric but also multiplex and the virtual layer would be temporal. Capable of much more than just QKD, such networks could revolutionize the world much like classical Internet did. For now, they face daunting technological challenges related in particular to sufficiently powerful quantum memories.

Proposals to achieve QI networks in the near-term future feature prominently all-optical schemes \cite{azuma2015all,pant2017rate,hasegawa2019experimental,li2019experimental,goswami2023satellite} as this could eliminate the need for such memories, albeit at the cost of some applications. At the other end of the spectrum are all solid state schemes, which however are envisioned to for example facilitate state transfer on a chip with minimal control and less sources of errors. This is achieved by the coherent dynamics of interacting but stationary carriers of information, much like in CTQW described in Sec.~\ref{sec:dynamics_walkers}, as opposed to physically moving, i.e. shuttling, the systems. Originally proposed in the early 2000s to facilitate communication between nearby  quantum processors \cite{bose2003quantum} it has since been studied in both spin \cite{christandl2004perfect,christandl2005perfect,burgarth2005conclusive,wojcik2007multiuser,paganelli2013routing} and oscillator systems \cite{plenio2004dynamics,plenio2005high,chudzicki2010parallel,portes2013perfect,nicacio2016coupled} with or without some limited control or additional operations. More recently a scheme for transferring logical qubits in quantum dot arrays has been proposed and found to be favorable over shuttling in terms of energy cost \cite{lewis2023low}. Results concerning complex networks are scarce, however, and consequently this otherwise very important and highly active area of research will not be discussed further here; we recommend instead Refs.~\cite{kay2011basics,godsil2012state,nikolopoulos2014quantum}.

In the following we first focus on networks within reach of current technology. Although quite different in terms of applications, as networks they share for example the natural weights with theoretical networks and provide an opportunity to introduce them in a simpler setting. We then present two common approaches for achieving entanglement distribution, considering mostly ideal conditions for the sake of simplicity, and briefly review network-generalized nonlocality. Finally, we consider the quantum Internet and its applications at various stages of development and along the way present results concerning what could be called noisy intermediate scale quantum (NISQ) networks, covering some of the vast and still somewhat unexplored landscape between within reach and ideal networks. As we focus on the network aspect we refer the reader to: \cite{rohde2021quantum} for motivation and applications and to \cite{alleaume2014using,diamanti2016practical,mehic2020quantum,pirandola2020advances} for QKD in particular, \cite{razavi2018introduction,wei2022towards,ramakrishnan2022quantum} for quantum repeaters, quantum memories and their candidate platforms, \cite{awschalom2021development} for quantum interconnects, \cite{roffe2019quantum} for quantum error correction, \cite{illiano2022quantum} for quantum Internet protocol stack, \cite{bassoli2021quantum} for classical simulation of the networks and \cite{bedington2017progress,pirandola2021satellite,sidhu2021advances,kaltenbaek2021quantum,de2023satellite} for the use of quantum satellites with their simulation discussed in particular in \cite{de2023satellite}.

For readers interested mostly in networks within the reach of current technology or almost, just the QKD references could be adequate. Otherwise, the books \cite{rohde2021quantum,bassoli2021quantum} are the most comprehensive ones. Reference \cite{illiano2022quantum} can be expected to be particularly interesting for readers with a communications engineering background and takes steps to cover compactly the salient parts of quantum theory. The references concerning repeaters, memories and interconnects are more hardware than network oriented. The rest are fairly self-explanatory.

\subsection{\label{sec:comms_long}Metric quantum networks}

\subsubsection{\label{sec:comms_QKD}Within reach: quantum secure networks}

Considering only technology mature enough to be deployed in the field now, the basic building blocks are nodes capable of generating, detecting or possibly forwarding quantum states, connected by quantum channels constituted by optical fiber or free space. The ideal carriers for the transmitted quantum states are photons. Besides moving at the speed of light and being highly resilient to decoherence, non-classical photonic states can nowadays be routinely generated, transmitted and measured. As anticipated, such networks already suffice for QKD. The working principle is to use the states to share a random string of classical data, upper bound the amount of leaked information by taking advantage of certain fundamental properties of quantum information, and finally to distill a secret key from the data that can then be used to encrypt the actual messages.  QKD is attractive because its security is based on the laws of Nature, implying for example that it is future technology proof as long as for example experimental imperfections can be accounted for. In contrast, conventional cryptography protocols are based on plausible assumptions about the difficulty of inverting certain functions and are secure only if the computational power of the adversary is limited and the assumptions indeed hold. 

\begin{figure}[t]
    \centering
    \includegraphics[trim=0cm 7.5cm 6.5cm 0cm,clip=true,width=0.97\textwidth]{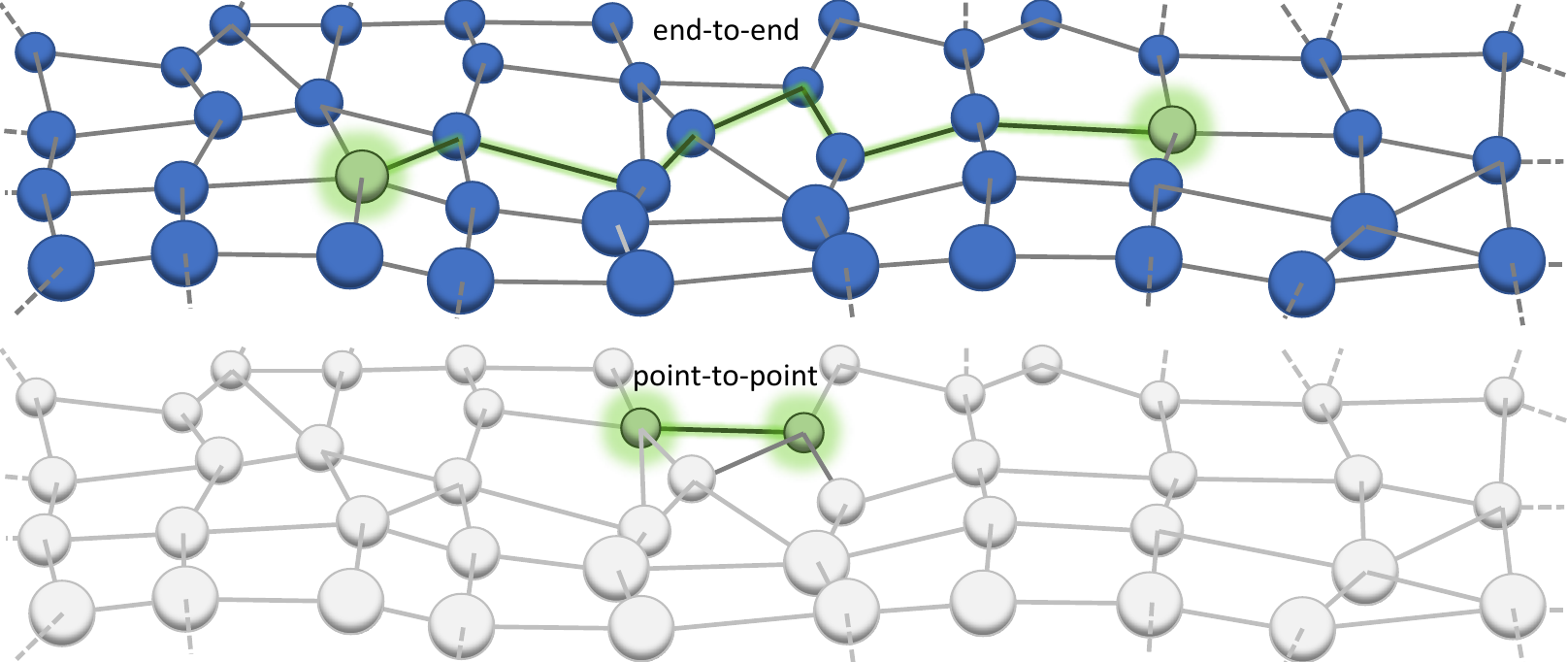}
    \caption{End-to-end transmission concerns two nodes communicating over a network using, for example, single or multiple paths. Point-to-point transmission takes place when two adjacent nodes communicate directly and the rest of the network plays no role. Upper bounds for end-to-end capacities may be found by considering the point-to-point capacities where the the PLOB bound of Eq.~\eqref{eq:PLOB} applies directly.}
    \label{fig:e2eVSp2p}
\end{figure}

Unlike conventional protocols, QKD is greatly limited by distance however. Specifically, the secret key bit rate achieved via point-to-point transmission depicted in Fig.~\ref{fig:e2eVSp2p} of quantum systems over lossy bosonic channels (see Sec.~\ref{sec:qteory_OQS}) is limited by the channel capacity, or in network terms the maximum flow. It is in units of the average number of (already distilled) secret key bits transmitted per channel use, which in turn can in principle reach but never exceed a fundamental, protocol-independent limit known as the Pirandola–Laurenza–Ottaviani–Banchi (PLOB) bound \cite{pirandola2017fundamental}. For such channels this ultimate capacity is 
\begin{equation}
\mathcal{C}(\eta)=-\log_2(1-\eta)
\label{eq:PLOB} 
\end{equation}
but approximately $1.44\eta$ for $\eta\ll 1$, where transmissivity $\eta$---the fraction of photons that survive the transmission---drops exponentially with distance. Assuming state-of-the-art optical fiber of length $d$, one would typically use $\eta=10^{-\gamma d/10}$ where $\gamma=0.2$ dB/km quantifies losses per distance. In free space one would include for example a factor accounting for the geometric position of the source with respect to the receiver \cite{pirandola2021satellite}. In principle, the rate merely decreases rapidly with distance but in practice collapses abruptly to zero due to detector noise washing out the quantum signal \cite{boaron2018secure}, making distance a hard limit. This is further exacerbated by high consumption rates: it is natural to pair QKD with encryption providing the highest security, where each key must be as long as the message and used only once. From now on this is assumed unless stated otherwise.

This on the one hand provides the links with their natural weights, i.e. the physical distance and the resulting values of $\eta$ and $\mathcal{C}(\eta)$, and on the other hand strongly limits the networks. The ultimate capacity can be generalized both to chains and arbitary networks \cite{pirandola2017fundamental,pirandola2019end}; conveniently, these results are also applicable to QI networks as seen later. In the following we present examples with a particular protocol, however it must be underscored that the fact that a protocol independent upper bound coincides with the rates achieved is highly nontrivial.

Under the considered limitations Eq.~\eqref{eq:PLOB} can be applied to a transmission over some path $P$ by simply considering the total transmissivity  
\begin{equation}
\eta_P=\prod_{e\in P}\eta_e,
\label{eq:etapath}     
\end{equation}
where $\eta_e$ is the transmissivity of link $e$. For constant $\gamma$, it coincides with that of a direct optical link. Therefore only nodes connected by a short enough path can directly share a secret key.

Covering larger areas can be achieved with trusted nodes or relays. Such a trusted chain $P_t$ could operate for example as follows. First every link, i.e. adjacent pair of nodes, generates locally stored secret key bits for $n$ rounds. In the large $n$ limit the number of bits in some link $e$ tends to $n \mathcal{C}(\eta_e)$. Next, secret bits are transmitted end-to-end (see Fig.~\ref{fig:e2eVSp2p}) by encrypting and decrypting them using locally stored bits. Since strongest encryption is assumed each transmitted bit consumes a local bit from a link, and once the link with the smallest $\mathcal{C}(\eta_e)$ runs out we are done. Therefore the rate per use of the chain tends to $\mathcal{C}(\eta_{P_t})$ where   
\begin{equation}
\eta_{P_t}=\min_{e\in P_t}\eta_e,
\label{eq:etatrustedpath}     
\end{equation}
making the rate limited by the bottleneck rather than the total distance. $P_t$ can be thought of as a chain of classical repeaters---conversely, Eq.~\eqref{eq:etapath} is also known as the repeaterless PLOB bound. A conventional repeater receives, amplifies and repeats a signal to extend its range; here each intermediate node increases the bit rate between end nodes in a scalable manner but deals with classical information. In the example $P_t$ follows a continuous generation protocol where links constantly accumulate resources, allowing operation near the theoretical achievable rate as any fluctuations caused by the probabilistic loss of some of the photons average out. Indeed, this is the standard operation mode of QKD links in experimental networks since DARPA and SECOQC networks \cite{mehic2020quantum}.

Generalization to QKD networks of arbitrary topology is straightforward, although the resulting capacity is of limited practical interest as discussed in Ref.~\cite{mehic2020quantum}. Like before, each link accumulates secret bits which are then consumed by the transmission. When a link is out it may be discarded; when the considered nodes are disconnected we are done. Any set of links $C$ that disconnects the network is called a cut(-set), which here is understood to specifically disconnect the sender from the receiver. The minimum cut
\begin{equation}
C_{\mathrm{min}}=\argmin_C\sum_{e\in C}\mathcal{C}(\eta_e)
\label{eq:floodingcapacity}
\end{equation}
formalizes the bottleneck, and the  capacity is given by the corresponding minimum sum. This is of course just a network flow problem, with maximum flow of secret bits achieved by flooding the network such that every unique path from the source to the sink is utilized at the highest possible capacity.

The enormous drawback of trusted networks is that every link can know the secret bits it transmitted. As any of the nodes involved could in principle leak them they must be assumed to be isolated from any unauthorized parties, which is the trusted node hypothesis. Although there are ways to mitigate this somewhat \cite{mehic2020quantum}, the lack of end-to-end security makes real-world QKD networks strongly gravitate towards private and non-commercial, inhibiting their growth---conversely, this is why having few or no trusted nodes is significant. The properties of complex QKD networks are best analyzed using suitable random network models. In particular, it turns out that under reasonable additional assumptions and a fixed protocol a fiber network leads to a Poisson degree distribution but a satellite network to a log-normal distribution. We also present results showing how satellites are a game changer in long haul communications, making them a strong candidate for a backbone that connects smaller networks together \cite{liorni2021quantum,harney2022end}.

The case of a fiber network was recently considered in \cite{brito2020statistical}. The nodes are embedded in a disk and the fibers are distributed according to the Waxman model \cite{waxman1988routing}. Then the probability of a link decreases exponentially with distance but is adjusted by parameters controlling the maximum distance, typical distance and average degree, chosen here to mimic the U.S. fiber-optics network. Next each fiber connecting some nodes $i,j$ is assigned a probability
\begin{equation}
p_{i,j}=1-(1-q_{i,j}(d_{i,j}))^{n_p}
\label{eq:britofiberlink}
\end{equation}
of a successful photonic link where its length $d_{i,j}$ controls the transmissivity $q_{i,j}(d_{i,j})=10^{-\gamma d_{i,j}/10}$ with $\gamma=0.2$ dB/km and where $n_p$ is a free parameter controlling the number of attempts made. Although a fixed value $n_p=1000$ was used for main results it was reported that the properties were not sensitive to the value of $n_p$. The degrees were found to follow a Poisson distribution controlled by the density of nodes with a giant component appearing at relatively low densities. The model was found to exhibit large clustering, but perhaps unsurprisingly not the small-world property as far away nodes required many intermediate nodes to reach one another.

This model was compared in \cite{brito2021satellite} to a network where a satellite shares Bell pairs to ground stations uniformly distributed in a disk, playing the role of the nodes. This kind of architecture where a central node merely generates and distributes Bell pairs is known as entanglement access network.  Remarkably, the central node can be untrusted as the secret bit is created when the halves are measured, not when the state is prepared. Here the cost is requiring a simultaneous line of sight which provides a hard limit to the size of the disk; in experiments, such satellite links have achieved 1200 km \cite{yin2017satellite}. The satellite is assumed to be stationary at $h_{\mathrm{sat}}=500$ km above the disk's center, which could correspond to a sun-synchronous orbit---daily transmission bursts can be imagined. The probability that an entangled photon is received by some ground station $i$ is $p_i(d_i)\in(0,\eta_0]$ which decreases exponentially with distance $d_i$ to the satellite and where $\eta_0\approx0.1$ is an empirical value accounting for various imperfections. Two nodes $i$, $j$ are taken to be connected if after $n_p$ trials, at least once both nodes receive their half. The probability is
\begin{equation}
\Pi_{i,j}=1-(1-p_i(d_i)p_j(d_j))^{n_p}.
\end{equation}
Crucially, here each node has its own distance. The smaller the distance $d_i$ is for some node $i$ the higher the probability $\Pi_{i,j}$ for any $j$, making nodes near the center more attractive than nodes in the periphery. This bias was found to lead to the appearance of hubs as well as the the small-world property, and the degree distribution was found to be closely approximated by a log-normal distribution. The satellite network was also found to cover large areas with less nodes for a fixed number of trials $n_p$ whereas the hubs increased robustness to random failures but decreased it against targeted attacks.

Fiber based local or metropolitan area entanglement access networks have been envisioned. Many have also been built \cite{chang2016experimental,wengerowsky2018entanglement,joshi2020trusted}, however scaling such networks to a large number of users is challenging as discussed at length in the cited works. This and the limited reach have been proposed to be alleviated by a hybrid architecture where many such networks are connected by a single shared trusted user \cite{ottaviani2019modular}. Importantly, this would still leave all the other nodes untrusted. 

\begin{figure}[t]
    \centering
    \includegraphics[trim=0cm 0cm 0cm 0cm,clip=true,width=0.5\textwidth]{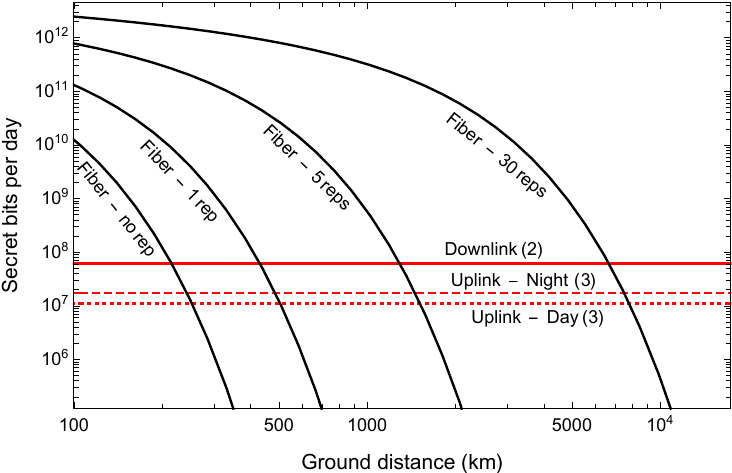}
    \caption{Comparison of daily secret key rate achievable with a repeater chain (black solid lines) and a satellite (red lines). The capacity in the former case is essentially based on Eq.~\eqref{eq:etatrustedpath} and increases with separation between repeaters, whereas the satellite operates as a classical trusted repeater. As it passes over the stations once a day the daily rate is distance independent. A clock speed of $10$ MHz is assumed. Figure reproduced from Ref.~\cite{pirandola2021satellite}, doi: \text{https://doi.org/10.1103/PhysRevX.6.041062}, license: \text{https://creativecommons.org/licenses/by/4.0/}.}
    \label{fig:satelliteVSfiber}
\end{figure}

A thorough approach to satellites was taken in \cite{pirandola2021satellite} which derived practically achievable daily secret key rates between two distant ground stations connected by a single sun-synchronous satellite. Importantly, although the rate is still limited by the PLOB bound modified by the effect of the geometric position, the rate-distance scaling is more favorable \cite{bonato2009feasibility}. As one round always takes a day the rate is distance independent, however as there is no  simultaneous line of sight the satellite must be trusted. It should be stressed however that it is remarkable how a global distance can be covered just by a single untrusted node which is hard for unauthorized parties to directly access as it is in orbit. These rates have been benchmarked against two ideal fiber based alternatives: a chain \cite{pirandola2021satellite,harney2022analytical} and a lattice like network \cite{harney2022analytical} utilizing ideal quantum repeaters. As anticipated, the chain achieves $\min_{e\in P}\mathcal{C}(\eta_e)$ and the network $\sum_{e\in C_{\mathrm{min}}}\mathcal{C}(\eta_e)$ \textit{with} end-to-end security. For any fixed number of links $L$ in the chain there is a total distance beyond which the satellite is superior \cite{pirandola2021satellite} as seen in Fig.~\ref{fig:satelliteVSfiber}, whereas to reach a superior distance independent rate the maximum link length should be around $200$ km or less \cite{harney2022analytical}. The network was taken to be degree regular with restrictions on neighbor-sharing properties of adjacent nodes to facilitate analytical treatment. Distance independent rate requires that the minimum cut $C_{\mathrm{min}}$ is distance independent, which in this case can be connected to both maximum link length and nodal density, and critical values to beat the satellite may be derived for different unit cells. All in all it was found that for long distances, a single trusted satellite can already achieve rates that would be very costly to beat even with highly idealized fiber networks. 

Before concluding we highlight two exciting and potentially disruptive avenues to push networks within reach further: trusted node free QKD between next nearest neighbors \cite{lucamarini2018overcoming} and long distance transmission of quantum states with a chain of co-moving untrusted satellites equipped with reflecting telescopes \cite{goswami2023satellite}. Remarkably, the former scheme can already break the PLOB bound of Eq.~\eqref{eq:PLOB}, achieving a secret bit key rate that scales with $\sqrt{\eta}$. Although the node is not a repeater, meaning that the rate cannot be boosted further by introducing more such nodes, the scheme can be realized with existing technology and recent experimental results are very promising \cite{wang2022twin}, achieving a record distance of 830 km in fiber. The satellite train on the other hand could receive a photonic state from a ground station and reflect it from satellite to satellite, bending with the surface of the Earth, finally reflecting it to the receiving ground station. Simulations are encouraging, predicting acceptable losses over global distances. Together with other presented results this underscores the indispensability of satellites for achieving such coverage in the near-term future.

\subsubsection{\label{sec:comms_entdist}Entanglement distribution: prerequisite for quantum information networks}

Moving from secret bits to qubits prevents the use of classical trusted repeaters. A QI network utilizing quantum repeaters can be imagined, but such a network is then subject to the no-cloning of quantum information which rules out signal amplification and also prevents making back-up copies: the transmission of a single unknown qubit can only ever be attempted once. Under these circumstances the network would need a perfect quantum channel which is noiseless, always succeeds and can cover as much distance as classical channels. Teleportation can achieve this; given pre-shared entanglement it can be consumed to swap the qubit to the receiver via local operations and classical communication (LOCC). This requires entanglement distribution, namely preparing entanglement between two marked nodes in a network. Due to non-increase of entanglement under LOCC \cite{chitambar2014everything}, this unavoidably involves transmitting entanglement bits, or halves of a Bell state. Importantly, there is a crucial difference between unknown qubits and entanglement: we are free to prepare as many Bell states as we like and use them only as fuel for the virtual teleportation channels that will handle the actual communication of quantum information.

\begin{figure}[t]
    \centering
    \includegraphics[trim=0cm 10cm 8cm 0cm,clip=true,width=0.97\textwidth]{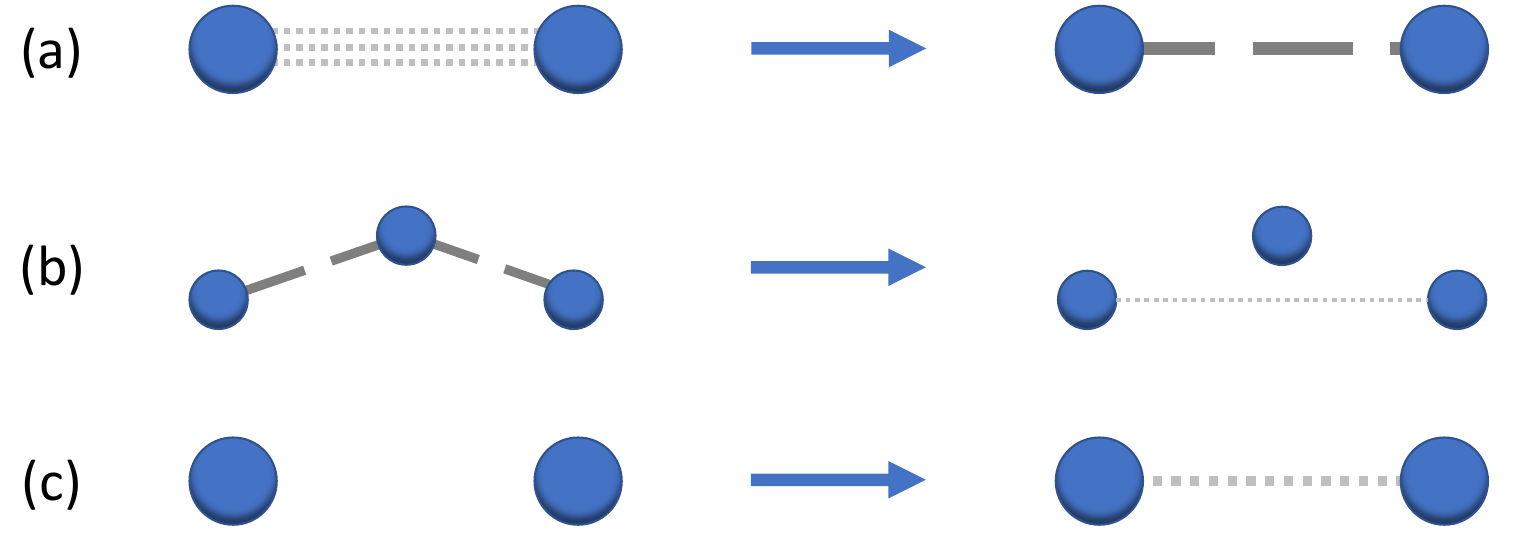}
    \caption{Common entanglement distribution primitives. Dashed lines depict bipartite entangled states shared by the nodes. $(a)$ Entanglement distillation (or purification) turns several states with relatively weak entanglement and low purity into a state with stronger entanglement and higher purity. $(b)$ Entanglement swapping turns two incident links into a single, longer link connecting the end points, in general at the cost of purity. $(c)$ Elementary link creation, also called remote entanglement generation, creates the initial short distance links with the help of, e.g, optical fiber and distillation. The rate cannot exceed the repeaterless PLOB bound.}
    \label{fig:distributionprimitives}
\end{figure}

The pioneering work \cite{pirandola2009direct} in the study of fundamental rate limits of quantum channels  introduced a lower bound applicable in particular to pure loss bosonic channels  considered here. Finding an upper bound coinciding with it was later achieved in \cite{pirandola2017fundamental}, the PLOB paper. Therefore while rate limits have been quantified in many ways \cite{takeoka2014fundamental,azuma2016fundamental,rigovacca2018versatile,azuma2021tools}, here we still focus on the PLOB bound. It turns out that for lossy bosonic channels the ultimate capacities for secret bit, qubit and entanglement bits all coincide. Indeed, a shared Bell state can either be converted into a secret bit or a qubit. Importantly, these rates correspond to exact Bell states which can be expected to require entanglement distillation where many sufficiently entangled noisy states can be probabilistically converted into less states with stronger entanglement and higher purity via LOCC, not increasing it on average. The PLOB bound is closely related to ultimate entanglement distillation rates, which in particular require an unlimited mean photon number to be achieved; this is why $C(\eta)\xrightarrow[\eta\to 1]{}\infty$. Remarkably, an explicit distillation protocol achieving these limits has very recently been introduced \cite{winnel2022achieving}. Initial links are created by transmitting entangled photons and is known as remote entanglement generation. Once distilled, short entanglement links can be converted into longer ones with entanglement swapping, which replaces two incident links by a longer link, effectively "detaching" from the shared node. At this point nodes adjacent in the entanglement layer no longer need to be adjacent in the channel layer. These common entanglement distribution primitives are depicted in Fig.~\ref{fig:distributionprimitives}. Some more recent proposals consider quantum error correction which might reduce the classical communication overhead \cite{muralidharan2016optimal}, but the same ultimate capacities still hold.

Some form of quantum memory is typically assumed to facilitate the repeated use of the primitives. For simplicity we assume that the memories can store an arbitrary number of qubits and have infinite coherence time, any local operations can be carried out, and there can be unlimited classical communication. Now we are in a position to relatively easily introduce quantum repeaters. In fact under such strongly ideal conditions they can operate analogously to the classical trusted repeaters with shared random string links replaced by entanglement links, secret key distillation replaced by entanglement distillation and secret bit swapping---transmission of secret bits by consuming local secret bits---by entanglement swapping. Such networks can achieve the capacities of Eqs.~\eqref{eq:etatrustedpath} and \eqref{eq:floodingcapacity} for example by operating in continuous generation mode with the crucial difference that the capacities now concern also entanglement bits and qubits and the repeaters can remain untrusted---consequently the networks could be public and commercial, fostering growth. As pointed out previously, it is highly nontrivial that the capacities cannot be exceeded; this was proven in full generality in Ref.~\cite{pirandola2019end} which considered also other types of channels.

One may ask what kind of capacity distributions can be expected for the links and the nodes; the latter is just the total capacity of incident links, or the weighted degree. Considering expected end-to-end capacity, it can be argued why both unusually high and unusually low capacity links might be absent. For former any capacity in excess of the bottleneck will be wasted, whereas for latter the link is a bottleneck at worst and not particularly useful at best. The capacity distribution can then be expected to be relatively narrow around the mean value, as in for example a Poisson distribution. The expected node capacity is arguably the simplest upper bound for the expected end-to-end capacity since the bottleneck cannot be larger. Similar arguments apply also here. If link capacities are indeed all rather similar then it follows that not only node capacities but also (unweighted) degrees will be distributed close to the mean value. These speculations are in line with recent results comparing Waxman and scale-free networks \cite{zhuang2021quantum}, where for the latter the probability of a new link was $p_{i,j}\propto k_i/d_{i,j}$ where $j$ is a node to be added and $k_i$ the current degree of an old node $i$. Each new node is connected to $m$ old ones. This results in hubs, which however were found to inhibit the expected end-to-end capacity since they attract links from great distances which leads to an abundance of low capacity links and nodes. This is exacerbated by limiting the number of links to $m$ per node which means that every low capacity link is one less decent to high capacity link. Indeed, for scale-free networks the capacity was found to saturate as node density was increased, but for the Waxman networks it increased linearly. For both the expected capacity was found to abruptly start increasing after a critical node density which importantly was higher than the density required for the giant component. One may also consider the robustness of such networks to different imperfections such as loss of nodes or links. This was done in \cite{zhang2021quantum} where it was found that while the capacity decreased linearly under random breakdowns for both, the scale-free network was very vulnerable to targeted attacks as the loss of only a relatively few hubs in terms of either capacity or degree significantly decreased the average end-to-end capacity. 
The results hold as is also for the ultimate secret bit capacities of fiber based trusted repeater networks.

First proposed in 1998 \cite{briegel1998quantum}, the original and later repeater protocols made various assumptions about imperfections but not about memory until recently. Unfortunately, an imperfect memory is both unavoidable in realistic models and arguably the Achilles' heel of repeater networks as they have been designed assuming scalable accumulation of resources to facilitate a repeat-until-success approach for every subtask, as will be elaborated on in the next Section. For now, we introduce entanglement percolation, proposed in 2007 \cite{acin2007entanglement} as an alternative for repeater networks designed specifically to operate entirely on-demand to ease the memory requirements. In short, starting from a given initial state it makes a single attempt at distributing the entanglement such that there is a phase transition in the success probability where it abruptly becomes distance independent at a critical value of initial entanglement. If it fails the protocol must start from scratch.

Assuming an initial state for the network where each link shares an identical pure but non-maximally entangled state, entanglement percolation focuses on singlet conversion probability (SCP), or the probability to reach from a given initial state a Bell state shared by given nodes---including adjacent nodes as a special case---using distillation and swapping. The links have some $\mathrm{SCP}=p<1$; if conversion fails the link is lost. Swapping preserves SCP but not purity \cite{acin2007entanglement,cuquet2011limited,perseguers2013distribution}, and swapping the resulting mixed state again is not done as this would decrease SCP \cite{cuquet2011limited}. The goal is then to use probabilistic conversion permitted for any link and deterministic swapping permitted for pure state links to form at least one path of maximally entangled states between the given nodes. The nodes may then be directly connected using swappings. The central question concerns the sufficient amount of preshared short range entanglement, as quantified by $p$, for entanglement distribution to beat the exponential scaling of direct transmission.

\begin{figure}[t]
    \centering
    \includegraphics[trim=0cm 0cm 0cm 0cm,clip=true,width=0.5\textwidth]{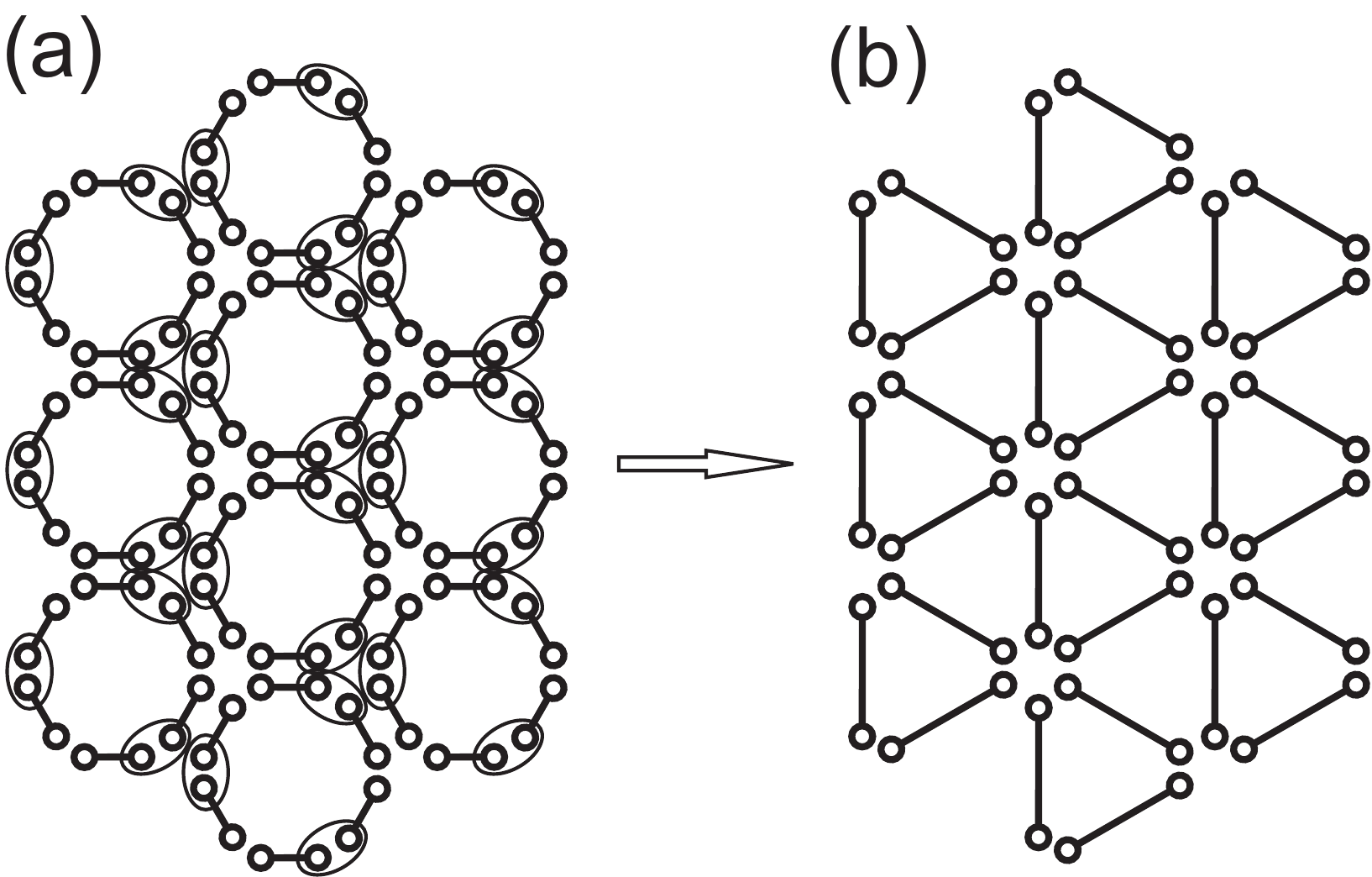}
    \caption{An example of quantum entanglement percolation. The lattice in panel $(a)$ is transformed into the triangular lattice of panel $(b)$ by swapping the entanglement at the circled nodes, lowering the percolation threshold. Reprinted figure with permission from \cite{perseguers2008entanglement}. Copyright 2008 by the American Physical Society.}
    \label{fig:QEP}
\end{figure}

In a strategy called classical entanglement percolation (CEP), first simultaneous conversion of all links is attempted, which divides the network into connected components where links are now maximally entangled. CEP succeeds if the target nodes are in the same component and otherwise fails. The anticipated phase transition occurs at $p\geq p_{\mathrm{th}}$ where $p_{\mathrm{th}}$ is the network percolation threshold, since for $p\geq p_{\mathrm{th}}$ a giant component appears and $\mathrm{SCP}=\theta(p)^2$ where $\theta(p)$ is the probability that a node is in the giant component. For $p< p_{\mathrm{th}}$ SCP decreases exponentially with distance, making CEP useless. Remarkably, often the percolation threshold $p_{\mathrm{th}}$ may be lowered by first reshaping the network with swapping, facilitating entanglement distribution even when $p$ is not enough for CEP; this strategy is called quantum entanglement percolation (QEP). Typically when QEP is used each link is assumed to be a product state of two identical states to facilitate reshaping the network, whereas in the reshaped network links have only one state. 

It was shown in the seminal work \cite{acin2007entanglement} that in open chains CEP is not optimal but gives the correct asymptotic scaling which is exponential for all $p<1$; a single failed conversion is fatal. In 2D lattices the possibility of QEP was demonstrated with a honeycomb lattice which was reshaped into a triangular lattice as shown in Fig.~\ref{fig:QEP}. Optimizing SCP in lattices was considered in \cite{perseguers2008entanglement}. QEP was successfully generalized beyond lattices in \cite{cuquet2009entanglement} where it was shown that reshaping could be done based on local information only and moreover the advantage of QEP over CEP can be significantly larger in random networks. Vulnerability of such networks to attacks was considered in \cite{wu2020structural}. The framework has also been generalized to $n$-partite maximally entangled states and generalized swapping called $n$-fusion. The case $n=3$ was shown to lead to advantages over $n=2$ (Bell states) in lattices in \cite{perseguers2010multipartite}, and recently it has been shown that for $n\geq3$, distance independence of the success probability remains possible even if the $n$-fusion is probabilistic and sometimes fails \cite{patil2022entanglement}. Going beyond two states per link, a strong advantage of QEP may be achieved even in chains but at the cost of more LOCC operations per node \cite{siomau2016quantum}. Finally, the related problem of when various subgraphs can appear as $p$ changes was considered in \cite{Perseguers-random}, where a quantum strategy was introduced such that all possible subgraphs appear at the same threshold value.

Entanglement percolation has recently been reviewed in great detail in Ref.~\cite{meng2023percolation} which also compares it to a novel approach called concurrence percolation \cite{meng2021concurrence,malik2022concurrence,meng2023deterministic}. Switching from SCP to concurrence, a measure of bipartite entanglement, serves as a basis for a new type of percolation that still uses essentially swapping and distillation---conversion of series and parallel links to single links---but in general no longer attempts to convert any of the states to Bell states. Informally speaking, this leads to a more economical use of the available resources. Indeed, concurrence percolation has been found to achieve a lower critical threshold for success than other approaches in many lattices  \cite{meng2021concurrence}, whereas in random networks the advantage is supported by numerical evidence  \cite{malik2022concurrence}. Furthermore, a non-trivial saturation point can appear where a non-maximal amount of initial entanglement can suffice for entanglement distribution to both succeed with certainty and lead to a Bell state between distant nodes. In contrast, CEP/QEP have only the trivial saturation point at $p=1$. As noted in the review, there are still open questions and work continues.

Like repeater networks, percolation networks are not ready for deployment. Whereas early repeater protocols took perfect memories for granted, early work on CEP/QEP took a pure initial resource state for granted. A more realistic initial state would be mixed but as will be seen in the next Section this leads to problems. Furthermore, conventional proposals require a high percolation threshold whereas compensation with more states per link might require some accumulation as the creation of each initial entanglement link must still respect the PLOB bound of Eq.~\eqref{eq:PLOB}. For the same reason the physical link length is still limited for CEP/QEP to work. Concurrence percolation is promising and its generalization to mixed states is an important open research direction.

We conclude by pivoting from entanglement distribution to network-generalized nonlocality. Consider two nodes receiving a Bell state from an untrusted source. They can measure it in either, say, basis $\{\Ket{0},\Ket{1}\}$ or $\{\Ket{+},\Ket{-}\}$. If the two nodes happened by chance to choose the same basis they have shared a secret bit because there can be no local hidden variable involved in the preparation of the state that, if known, would allow the prediction of the measurement outcomes before the measurements have been carried out. This is in fact a consequence of the Bell state violating the celebrated Clauser–Horne–Shimony–Holt Bell inequality \cite{clauser1969proposed} which must be obeyed by all bipartite correlations with binary outcomes and two possible measurement settings amenable to a local hidden variable model. In general, Bell inequalities separate local and nonlocal correlations and under certain mild assumptions are amenable to a geometric interpretation as hyperplanes that define all local correlations as their convex hull \cite{fine1982hidden}, the so called local polytope. The set of quantum correlations contains the local polytope as a proper subset, meaning that some of them violate a general Bell inequality.

In the past decade it has been recognized that in a more general network where links are Bell pairs generated by independent sources between each adjacent pair of nodes, qualitatively new inequalities arise that separate local and nonlocal correlations at the network level, as recently reviewed in great detail in \cite{tavakoli2022bell}. Importantly, the set of network local correlations is contained inside the local polytope as a proper subset, meaning that there are correlations which violate a network Bell inequality without violating any of the ordinary Bell inequalities---if correlations are all assumed to be local then this means that the assumed network structure must be false. Furthermore, the set of network local correlations is not convex, complicating its characterization. One may consider even more general scenarios if multipartite entanglement is introduced. As pointed out in \cite{tavakoli2022bell}, the field is still facing many open questions.

Very recently also quantum steering has been generalized to networks \cite{jones2021network}. Unlike previously, here some of the nodes are trusted and one considers the conditional states that the untrusted nodes can prepare for the trusted nodes by performing local measurements. In the absence of any correlations the states of the trusted nodes are of course independent of any local operations the untrusted nodes perform, whereas in the case of a shared Bell state the untrusted node can choose to project it to an arbitary basis simply by measuring in that basis. Between these two extremes one may consider whether the effect can be explained with local hidden variables, and for quantum correlations in particular this is not always the case. When it is not, it is said that the shared state is steerable. In networks and under certain conditions, it was found that the set of steerable and network steerable states are not necessarily the same, however several no-go results where also introduced forbidding network steering in many scenarios.

These research avenues are closely related to the study of the relationship between the structures of the fiber and entanglement layers. This in turn naturally depends very strongly on the assumptions one makes about the initial short range correlations and the allowed operations. On the one hand they may lead to forbidden correlation structures that the underlying fiber network simply cannot create as in, e.g., \cite{hansenne2022symmetries} and as mentioned above. On the other hand one may conclude that in fact even a physical chain suffices for the creation of a variety of entanglement networks such as lattices, random networks and small-world networks \cite{siomau2016structural}.

\subsubsection{\label{sec:comms_qinternet}Road to quantum Internet: public commercial quantum information network}

The most important takeaway of a recently proposed roadmap \cite{wehner2018quantum} towards a full blown quantum Internet is that we may reap benefits not only at the end but also continuously along the way, with each new stage unlocking previously unavailable applications. This is great news because the road is long and rocky and our maps unreliable. Some of the different activity sectors benefiting from the developing quantum Internet were identified in \cite{de2023satellite} as industry, critical infrastructures, finance, administrations and operational as well as fundamental science. Importantly, different sectors were proposed to have different requirements; for example, whereas industry and science might tolerate relatively high latencies and low entanglement distribution rates, administrations would not. Returning to the roadmap, it envisions three stages of networks capable of QKD and some related protocols not discussed here, followed by another three for QI networks.

The former achieve QKD with trusted nodes, without trusted nodes and with device independence. Although not exactly corresponding to the proposed architecture the first two have already been reached by fiber based trusted repeater and entanglement access networks, respectively. If the end nodes of the latter could also carry out deterministically any single qubit measurements they could switch to device independent protocols which both relax certain conventional assumptions and close some loopholes related to experimental imperfections or vulnerabilities as presented, e.g., in \cite{xu2020secure}. We are certainly at least at stage one but trusted node QKD cannot be expected to be valuable enough for the networks to grow to their theoretical continental \cite{brito2020statistical} or even global service area \cite{pirandola2021satellite}. It is arguable whether we are past it already for example because the diameter of an entanglement access network is limited by the repeaterless PLOB bound, which in fiber translates to roughly 100 km, and the number of users by some technical difficulties. Some recent developments and proposals discussed in the end of Sec.~\ref{sec:comms_QKD} might push the limit of trusted node free networks much farther than this in the near-term future however, which might be interpreted as reaching stage two or three.

The QI networks might be described as teleportation, distributed quantum computing and quantum computing networks. Reliable teleportation of unknown qubits can be achieved if the network is equipped with quantum memories and is capable of arbitrary local unitary operations. While the network diameter and service area could be limited by imperfect memories and lossy operations it could provide for example secure cloud quantum computing where clients with limited capabilities outsource demanding computations to an untrusted service provider. Using so called homomorphic encryption \cite{broadbent2015quantum} the data provided by the client remains private; using blind quantum computing \cite{broadbent2009universal} even the algorithm remains private, a feat which cannot be achieved for all algorithms in classical computing. Other proposed applications include improved clock synchronization \cite{komar2014quantum} and extending the baseline of telescopes \cite{gottesman2012longer}. The final stages introduce fault tolerant quantum computing to all end nodes, performing classically intractable computations either at the network level with distributed quantum computing or also at the single end node level. In the latter case the network could perform tasks related to facilitating efficient co-operation of local computers with an advantage over their classical counterparts \cite{ben2005fast,tani2005exact}. 

Aside from the development of its abilities, one may also consider how the quantum Internet could develop as a network. Distance is crucial as even in ideal conditions it controls the overhead. Tentatively three different regimes may be identified: short, intermediate and long. Point-to-point optical links are feasible only for the first, whereas for intermediate distances the overhead and cost of using entanglement distribution in fiber would still be acceptable. Long distances require solutions where even the overhead is (almost) distance independent. This could be achieved by powerful quantum memories moving on board a satellite or as freight \cite{devitt2016high,rohde2021quantum}; the latter case is known as the quantum sneakernet. Crucially, provided that local entanglement stores are maintained the bottleneck would be the time it takes to carry out a teleportation protocol, i.e. it would be limited mainly by the classical communication rate. As a side note, the classical capacities are still very rarely taken into account although they affect both use cases \cite{de2023satellite} as well as efficiency of QKD \cite{mehic2017analysis} and presumably other applications. Both sneakernet \cite{devitt2016high,rohde2021quantum} and satellite links \cite{liorni2021quantum,harney2022end} have been proposed as the backbone for the quantum Internet. The backbone would stitch together networks where distances are in the short or intermediate regime.

\begin{figure}
    \centering
    \includegraphics[trim=0cm 5.5cm 1cm 0cm,clip=true,width=0.95\textwidth]{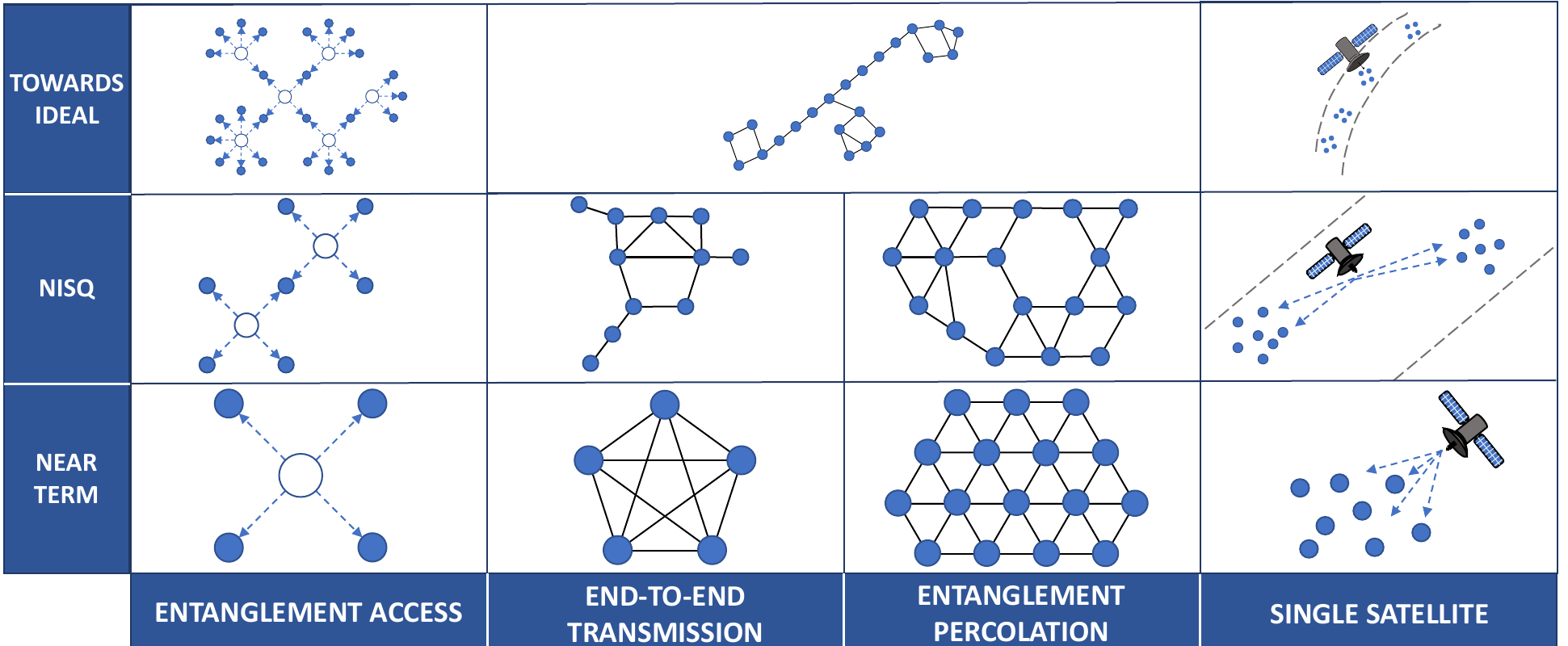}
    \caption{Examination of how trusted node free QI networks could evolve for certain generic architectures as elementary link creation probability, memory coherence times and fidelities increase. See text for details.}
    \label{fig:limiteddiameternetworks}
\end{figure}

For the remainder of this Section we very briefly examine possible evolution of fiber networks based on entanglement access, end-to-end transmission and entanglement percolation as well as satellite based entanglement access networks, shown in Fig.~\ref{fig:limiteddiameternetworks}. Specifically, an entanglement percolation network first attempts to create a large entanglement cluster which is then manipulated to distribute entanglement where possible and needed, whereas in end-to-end transmission the nodes that would like to communicate are predetermined and the network might opt for the single highest probability path connecting them. For simplicity, only a single satellite is considered; using many offers new possibilities as shown in, e.g., \cite{gundougan2021proposal,khatri2021spooky,wallnofer2022simulating,goswami2023satellite}.

In the near term rates are still restricted by the repeaterless bound which then strongly restricts the diameter. The appeal of entanglement access networks is that the end nodes do not need to be able to generate the initial short distance entanglement whereas in end-to-end case they do. Using entanglement percolation might increase the range especially if the physical links are short enough to keep the states almost pure, at the cost of requirements for the percolation threshold which should translate to requirements for the link and node density. Satellite links have more lenient rate-distance scaling but without memory an untrusted satellite requires a simultaneous line of sight. This leads to perhaps the largest but still limited area of service. The case of networks approaching the ultimate limits is not too difficult to envision as we may now assume strongly ideal conditions similar to those in the previous Section. For entanglement access networks a modular structure where several such networks are connected together by a shared user may be envisioned, as already proposed for networks running QKD and related protocols \cite{ottaviani2019modular}. The line between end-to-end transmission and percolation networks blurs as both could switch to flooding; fiber backbones could appear, limited mostly by cost-effectiveness since now a satellite could connect any two nodes near its ground track. 

The NISQ case introduces both soft and hard diameter restrictions arising from lossy memories, noisy probabilistic operations and mixed states of limited fidelity and energy. It must be stressed that it both covers an enormous leap in technology---even assuming that one day we will have a fully developed quantum Internet, it might be expected that QI networks will stay in the NISQ regime much, much longer than in the near term regime---and has features which are missing from both near term and nearly ideal networks. At best, they limit how much the service area can grow from near term regime; for both fiber and satellite based entanglement access networks at least a second end node cluster should become possible, in the latter without a simultaneous line of sight.

Considering percolation, for simplicity we may imagine a kind of 0th order approximation for imperfect memories: the entangled resource states are maintained perfectly up to some maximum time and then vanish, forcing us to minimize temporal overhead by relying on parallelism. But considering mixed states in general leads to a $\mathrm{SCP}=0$---although in special cases it may survive if the number of resource states is increased and adapted strategies are used \cite{broadfoot2009entanglement,broadfoot2010singlet,broadfoot2010long}---makes QEP impossible and the final fidelity path length dependent \cite{cuquet2011limited,lapeyre2011distribution}.  We will not consider imperfect repeaters with perfect memory, which are covered quite well in Ref.~\cite{muralidharan2016optimal}. When states in memory decohere, both maximum  \cite{laurenza2022rate} and minimum \cite{pant2017rate} distances where the repeaters can beat the repeaterless PLOB bound may appear. On-demand generation is commonly assumed, leading to complicated waiting time distributions \cite{azuma2021tools} as then stochasticity dominates, which in particular complicates the prospects of a neat network description. Decoherence introduces a maximum waiting time \cite{hartmann2007role} after which resources can no longer be distilled and the entire protocol might have to be restarted. When also the operations such as swapping are probabilistic the physical overhead quantified for example in terms of memory qubits grows rapidly \cite{razavi2009quantum,pant2017rate}.

These challenges have in part prompted the introduction of hybrid strategies. One approach is to replace solid state memories used in conventional repeaters with highly entangled resource states called repeater graph states \cite{azuma2015all}; proof-of-principle experiments have already been carried out \cite{hasegawa2019experimental,li2019experimental}. Like in percolation, temporal overhead is reduced by massive parallelism and losses are managed by introducing many alternative paths. Also like in percolation the challenge is shifted to the generation and manipulation of the resource states, although there has been some recent progress regarding the former \cite{zhan2020deterministic}. As noted in Ref.~\cite{pant2017rate}, this approach in particular suffers also from a poorly scaling physical overhead and a minimum distance to beat the repeaterless bound. A somewhat related example includes taking advantage of a specific two-dimensional lattice structure to connect at least one pair of nodes belonging to its opposite sides without requiring memories or complicated resource states \cite{das2018robust}. Combining entanglement percolation with lossy repeaters was considered in \cite{khatri2019practical} where it was found that allowing for some repeated attempts at the subtasks lowered the critical probability of initial link creation for the emergence of a giant entanglement component.

At this stage, the growing networks create a need for concrete and practical protocols for distributing the entanglement. Focusing on end-to-end case, conventional path finding and routing algorithms cannot be used directly but with suitable modifications may achieve good \cite{van2013path} or even optimal performance \cite{chakraborty2020entanglement}, however in general optimality may require specific properties from the network \cite{caleffi2017optimal}. Efficient algorithms for finding the shortest path are possible even under quite general conditions \cite{santos2023shortest}. If there are many overlapping requests the network might need to operate on-demand to decrease the average latency \cite{chakraborty2019distributed}. If demand is low or higher latency is acceptable the requests may be handled one at a time, in which case multipath routing may more easily allow to beat the repeaterless rate \cite{pant2019routing}, or in batches, in which case computationally efficient and optimal routing algorithms have been proposed for specific architectures \cite{chakraborty2020entanglement}. Routing multipartite entanglement may also be considered. Reference~\cite{bugalho2023distributing} introduced multiple such routing algorithms and in particular one which simultaneously optimizes both the rate and final fidelity for GHZ states; here it was also demonstrated how focusing only on the bottleneck in a non-ideal network can lead to drastically worse results, underscoring the importance of taking into account the imperfections in any practical routing algorithm. Multipath routing can be advantageous also for multipartite entanglement, as recently shown in Ref.~\cite{sutcliffe2023multiuser}. Optimizing the repeater scheme itself has also been considered \cite{goodenough2021optimizing} and practical figures of merit proposed, such as average connection time and largest entanglement cluster size \cite{khatri2019practical}.

Robustness of routing with mixed states was considered in \cite{coutinho2022robustness} where a finite amount of mixed initial resource states was considered. Since the cost in resource states to satisfy a given target fidelity is path length dependent, transitivity in who can reach who may then be broken: if Alice can achieve a non-vanishing rate at target fidelity with Bob and Bob with Charlie, it does not ensure that also Alice can achieve one with Charlie. Indeed it was found that under these circumstances networks can experience an abrupt transition to overlapping connected components in terms of such a rate as a function of both the amount of initial resources and the probability of random link failures. Assuming identical initial resources in all links, critical router efficiencies were derived to suppress such transitions for various topologies. Although scale free networks were found to be the most promising, the ones considered here would in practice require links with both long distance and relatively high capacity; adding satellite links to a fiber network was tentatively proposed.

We briefly mention also the importance of developing practical methods to certify successful resource state generation. Device independent methods robust to noise have been developed and experimentally tested \cite{agresti2021experimental}, but their computational complexity scales unfavorably with the complexity of the network. A physics aware machine learning approach directly applicable to noisy raw data can be used for such cases \cite{d2023machine}.

To conclude, it can tentatively be said that with improving efficiency the possible variation in
local node density and complexity could increase rapidly. What kind of complexity should be expected would depend on the growth principles of these networks, which in turn should depend on the one hand on the fundamental limitations and on the other hand on the incentives such as service requirements and interest.

\subsection{Avenues for further research}

Scalability requires managing the impact of losses and operation errors and in particular either the development of sufficiently powerful quantum memories or significant improvements in memoryless alternatives. Proof-of-principle experiments in the near term should demonstrate beating the repeaterless PLOB bound up to a few lossy repeaters, trusted node free QKD networks with a diameter in this regime and the use of a quantum satellite equipped with memory. Percolation and related approaches have experienced somewhat of a renaissance in the form of hybrid approaches \cite{azuma2015all,khatri2019practical} and concurrence percolation \cite{meng2021concurrence}. This very promising theoretical framework opens new important research questions for the years to come. For instance it would be crucial to incorporate in this framework the preparation of the initial entanglement links, which quite naturally makes the network metric, and to investigate how well the main results tolerate mixedness. Meanwhile, standardisation of mainly trusted node QKD networks but also others is pursued by several organizations (see Ref.~\cite{de2023satellite} for a recent summary) and in particular the interplay between the network structure, key management policy and degree of practical security might benefit from further applications of network theory. Good examples of the latter include trading some rate to increased security using multiple \cite{mehic2020quantum} or single paths \cite{cirigliano2023optimal}. In particular, Ref.~\cite{cirigliano2023optimal} introduces the concept of quantum efficiency to describe the tradeoff and maximum efficiency networks that optimize it, as well as an algorithm to find them for a given relative importance between rate and security. Alternatively, one may consider the case where only some of the nodes are trusted and ask about the connected components for a given maximum number of hops between trusted nodes. This can be tackled with the recently developed extended-range percolation framework \cite{cirigliano2023extended} applicable to both random and complex networks.

One can also never quite rule out disruptive novel ideas such as the recently introduced QKD protocol able to break the repeaterless PLOB bound with a single memoryless intermediate node \cite{lucamarini2018overcoming,ma2018phase,lin2018simple,minder2019experimental}; although the advantage is limited to that of a single repeater, this is already substantial and ranges of 830
km have been reported \cite{wang2022twin}. We also mention in passing emerging new directions such as going beyond definite causal order \cite{ebler2018enhanced,rubino2021experimental,miguel2021genuine}, 
generalizing to entanglement-assisted LOCC by introducing short distance entangled catalyst states to the network \cite{santra2021quantum}, quantum network coding \cite{lu2019experimental} and pursuing hybrid technology applying, e.g, both quantum theory describing qubits and continuous variable states of light \cite{pirandola2016physics,guccione2020connecting}.

So far a great deal of attention has been given to restrictions concerning both near term and ideal networks, however there is still much room for further work. For example, whereas entanglement distribution and simulation of intermediate stage networks has been considered, random network models incorporating their limitations and objectives such that they could be used without deep understanding of the microscopic theory are still missing. A good example of what would be needed to build such models are ideal capacity weighted networks: they can be readily applied with just a superficial understanding of the physics and the engineering, have a clear interpretation as networks of ideal quantum repeaters connected by pure loss channels and provide meaningful benchmarks. Introducing something similar for intermediate stage networks is of course a great challenge for the quantum networks community as, e.g, deriving the waiting time distributions or the final fidelity can become involved already in chains \cite{brand2020efficient}. There has been some recent progress regarding this. For example, Ref.~\cite{sadhu2023practical} introduces networks weighted by judiciously chosen functions of elementary link creation probability and uses graph theoretical methods to, e.g., find critical values for quantities such as storage time and link length for successful sharing of resource states. Reference \cite{mor2023influence} considers a unified model for imperfections in preparation, memories and measurements all treated in noisy stabilizer formalism (See Sec.~\ref{sec:properties-dynamics}) to efficiently simulate very large noisy networks---an excellent example of a beneficial application of one kind of quantum network to another.

Interesting connections between these and induced quantum networks introduced in Sec.~\ref{sec:properties} could also be further explored in the context of resource states for all-optical repeaters, new simulation tools and analysis or improvement of entanglement management policies. Alternatively, one might consider distribution of graph states in large scale quantum networks as in \cite{epping2016large,pirker2018modular,meignant2019distributing}, which might facilitate new applications. Furthermore, more research on growth principles for networks at all stages is needed. Such principles can be expected to consist of both limitations and incentives that together govern the evolution of future quantum networks, and work on especially the incentives is scarce, with some notable exceptions such as Part VIII of Ref.~\cite{rohde2021quantum}.

\section{\label{sec:discus}Discussion and future directions}

Whereas complexity is what empirical networks seem to naturally gravitate towards, quantumness is coy and needs to be cajoled to manifest by isolation of the systems. The two meet in the following research lines: network-generalized quantum problems, quantum-applied network theory,   quantum-generalized approaches for complex networks and quantum-enhanced communications, which are currently pursued by scientists and researchers from a variety of backgrounds. In this review we have introduced the four main research lines, unified them under the broader context of complex quantum networks and provided a comprehensive overview of the field. 

There are multiple promising directions for further development of the field. From the quantum systems perspective, new ways to generalize quantum phenomena to a network scenario can be envisioned. We highlight chiral quantum walks as an example where the underlying graph is no longer undirected and which can have advantages over both classical and sometimes also conventional quantum walks \cite{frigerio2022quantum,kryukov2022supervised,frigerio2023swift}. Regarding quantum correlations in a network scenario, most work still focuses on Bell nonlocality, leaving room for other types: steering \cite{jones2021network}, entanglement and discord. Speaking of applications of network theory to the quantum case, a network description of a stationary state has already been explored as a cheaper alternative to state tomography \cite{valdez2017quantifying,sundar2018complex,bagrov2020detecting,garcia2020pairwise} but generalization to evolving states and process tomography has only been suggested. Using a network description to not only detect but to discover previously unknown phenomena \cite{sokolov2022emergent,llodra2022detecting} remains in its infancy. Pivoting to quantum enhanced communication networks, both within reach trusted node and ideal quantum repeater networks can be modeled compactly as just a network of channels weighted by capacities. This description can be applied without a deep understanding of the relevant physics and the results have a clear interpretation. Introducing similar models for the NISQ (noisy intermediate-scale quantum) stage covering the large gap between near term and nearly ideal is undoubtedly challenging but if it could be done the field could have contributions from researchers from a much wider variety of backgrounds and specializations.

From the network science perspective the field of  complex quantum networks could  also be transformative. 
For instance the new generation of quantum computers could lead to the flourishing of quantum algorithms for classical network inference leading to further understanding of their complexity. Moreover new quantum technologies could be key to design complex quantum networks in experimental set-up leading to novel quantum phenomena displaying a rich interplay between topology and dynamics. From the dynamical point of view  directions that are particularly promising and that  lie at the classical/quantum interface are progress on quantum synchronization \cite{lohe2010quantum} and on network dynamics dictated by the topological Dirac operator \cite{bianconi2021topological}.
Finally the full potential of networks as a powerful tool to understand quantum matter  is not yet fully explored and provides a very promising direction for unsupervised detection of quantum phase transitions.   

As we have seen, the field has already produced important contributions in each of its research lines which have so far progressed and evolved mostly independently with some notable exceptions. For example, Hamiltonians derived from a graph can be both simulated with cluster or graph states \cite{nokkala2018reconfigurable,renault2021spectral} or be used for their adiabatic preparation \cite{aolita2011gapped}. The states in turn could be used to replace conventional quantum memories \cite{azuma2015all} or for error correction in quantum communication networks \cite{raussendorf2007topological}, but for very short distances one may consider state transfer or transport with suitable Hamiltonians again \cite{nikolopoulos2014quantum,mulken2016complex}. If the communication network could be prepared into a continuous variable cluster state entanglement could then be distributed using the protocol of \cite{centrone2023cost}, found to be efficient in particular in sparse complex networks. Finally, networks with comparable complexity to the classical Internet can be constructed even from the ground state of a two dimensional spin lattice by taking as nodes not individual spins but regions of spins of varying sizes and as links entangled clusters of spins shared by exactly two nodes, constituting communication channels \cite{chepuri2023complex}. However we believe that there remains much more potential in the ways the lines could further couple together. In the light of the above the interaction between the different research lines and the interdisciplinary collaboration between physicists and network scientists will be key to foster new discoveries in the field and to address  the new challenges 
that the next generation of quantum technologies will require science to face.

\ack

JN gratefully acknowledges financial support from the Turku Collegium for Science, Medicine and Technology as well as the Academy of Finland under Project No. 348854. 

\section*{References}

\typeout{}
\bibliographystyle{unsrt}
\bibliography{references}

\end{document}